\documentclass{emulateapj}
\usepackage{colordvi}
\usepackage{epsfig}
\usepackage{graphicx}
\usepackage{amssymb}
\usepackage{enumitem}
\newcommand{\Ha}{\hbox{{\rm H}\kern 0.1em$\alpha$}}
\newcommand{\Hb}{\hbox{{\rm H}\kern 0.1em$\beta$}}
\newcommand{\MgII}{\hbox{{\rm Mg}\kern 0.1em{\sc ii}}}
\newcommand{\CIV}{\hbox{{\rm C}\kern 0.1em{\sc iv}}}
\newcommand{\NeV}{\hbox{[{\rm Ne}\kern 0.1em{\sc v}]}}
\newcommand{\OII}{\hbox{[{\rm O}\kern 0.1em{\sc ii}]}}
\newcommand{\NeIII}{\hbox{[{\rm Ne}\kern 0.1em{\sc iii}]}}
\newcommand{\OIII}{\hbox{[{\rm O}\kern 0.1em{\sc iii}]}}
\newcommand{\NII}{\hbox{[{\rm N}\kern 0.1em{\sc ii}]}}
\newcommand{\SII}{\hbox{[{\rm S}\kern 0.1em{\sc ii}]}}
\newcommand{\sfrtot}{SFR$_{\mathrm{IR+UV}}$~}

\newcommand{\lssfr}{log(sSFR/Gyr$^{-1}$)}
\newcommand{\lmass}{log(M/M$_{\odot}$)}
\newcommand{\av}{$A_{V}$~}

\begin{document}

\title{CANDELS+3D-HST: compact SFGs at \lowercase{z}$\sim$2--3, the progenitors of
  the first quiescent galaxies}

\author{G. Barro\altaffilmark{1},
S. M. Faber\altaffilmark{1},
P. G.~P\'{e}rez-Gonz\'{a}lez\altaffilmark{2,3},
C. Pacifici\altaffilmark{4},
J. R.~Trump\altaffilmark{5},
D. C. Koo\altaffilmark{1}, 
S. Wuyts\altaffilmark{6},
Y. Guo\altaffilmark{1}, 
E. Bell\altaffilmark{7},
A. Dekel\altaffilmark{8},
L. Porter\altaffilmark{9},
J. Primack\altaffilmark{9},
H. Ferguson\altaffilmark{10}, 
M. L. N. Ashby\altaffilmark{11},
K. Caputi\altaffilmark{12},
D. Ceverino \altaffilmark{13},
D. Croton\altaffilmark{14}, 
G. G. Fazio\altaffilmark{11},
M. Giavalisco\altaffilmark{15}, 
L. Hsu\altaffilmark{6},
D. Kocevski\altaffilmark{16}, 
A. Koekemoer\altaffilmark{10}, 
P. Kurczynski\altaffilmark{17},
P. Kollipara\altaffilmark{9},
J. Lee\altaffilmark{4}, 
D. McIntosh\altaffilmark{18},
E. McGrath\altaffilmark{19}, 
C. Moody\altaffilmark{9},
R. Somerville\altaffilmark{17},
C. Papovich\altaffilmark{20},
M. Salvato\altaffilmark{6},
P. Santini\altaffilmark{21},
C. C. Williams\altaffilmark{15},
S. P. Willner\altaffilmark{11},
A. Zolotov\altaffilmark{8},
%Arjen van der Wel\altaffilmark{4},
%alphabetical}
}

\altaffiltext{1}{University of California, Santa Cruz}
\altaffiltext{2}{Universidad Complutense de Madrid}
\altaffiltext{3}{Steward Observatory, University of Arizona}
\altaffiltext{4}{Yonsei University Observatory}
\altaffiltext{5}{Pennsylvania State University}
\altaffiltext{6}{Max-Planck-Institut f\"{u}r extraterrestrische Physik}
\altaffiltext{7}{Department of Astronomy, University of Michigan}
\altaffiltext{8}{The Hebrew University}
\altaffiltext{9}{Santa Cruz Institute for Particle Physics}
\altaffiltext{10}{Space Telescope Science Institute}
\altaffiltext{11}{Harvard-Smithsonian Center for Astrophysics}
\altaffiltext{12}{Kapteyn Astronomical Institute}
\altaffiltext{13}{Universidad Autonoma de Madrid}
\altaffiltext{14}{Swinburne University of Technology}
\altaffiltext{15}{University of Massachusetts}
\altaffiltext{16}{University of Kentucky}
\altaffiltext{17}{Rutgers University}
\altaffiltext{18}{University of Missouri-Kansas City}
\altaffiltext{19}{Colby College}
\altaffiltext{20}{University of Texas A\&M}
\altaffiltext{21}{Osservatorio Astronomico di Roma - INAF}
\slugcomment{Submitted to the Astrophysical Journal} 

%\slugcomment{Last edited: \today} \date{Submitted: \today}
%\pagerange{\pageref{firstpage}--\pageref{lastpage}} \pubyear{2008}
%\maketitle
\label{firstpage}
\begin{abstract}

  We analyze the star-forming and structural properties of 45 massive
  (\lmass$>10$) compact star-forming galaxies (SFGs) at $2<z<3$ to
  explore whether they are progenitors of compact quiescent galaxies
  at $z\sim2$.  The optical/NIR and far-IR {\it Spitzer}/{\it
    Herschel} colors indicate that most compact SFGs are heavily
  obscured. Nearly half (47\%) host an X-ray bright AGN. In contrast,
  only about 10\% of other massive galaxies at that time host
  AGNs. Compact SFGs have centrally-concentrated light profiles and
  spheroidal morphologies similar to quiescent galaxies, and are thus
  strikingly different from other SFGs, which typically are disk-like
  and sometimes clumpy or irregular. Most compact SFGs lie either {\it
    within} the SFR--mass main sequence (65\%) or {\it below} it
  (30\%), on the expected evolutionary path towards quiescent
  galaxies.  These results show conclusively that galaxies become more
  compact before they lose their gas and dust, quenching star
  formation. Using extensive {\it HST} photometry from CANDELS and
  grism spectroscopy from the 3D-HST survey, we model their stellar
  populations with either exponentially declining ($\tau$) star
  formation histories (SFHs) or physically-motivated SFHs drawn from
  semi-analytic models (SAMs). SAMs predict longer formation
  timescales and older ages $\sim$2~Gyr, which are nearly twice as old
  as the estimates of the $\tau$ models. While both models yield good
  SED fits, SAM SFHs better match the observed slope and zero point of
  the SFR--mass main sequence. Contrary to expectations, some low-mass
  compact SFGs (\lmass$=10-10.6$) have younger ages but lower sSFRs
  than that of more massive galaxies, suggesting that the low-mass
  galaxies reach the red sequence faster. If the progenitors of
  compact SFGs are extended SFGs, state-of-the-art SAMs show that
  mergers and disk instabilities are both able to shrink galaxies, but
  disk instabilities are more frequent (60$\%$ versus 40$\%$) and form
  more concentrated galaxies. We confirm this result via
  high-resolution hydrodynamic simulations.

\end{abstract}

\section{Introduction}\label{intro}
The formation history of very massive galaxies is not well
understood. Present-day massive galaxies are known to be a homogeneous
population characterized by red optical colors that follow a tight
correlation with stellar mass (i.e., the red sequence;
\citealt{kauffman03}; \citealt{baldry04}). This population consists
mostly of galaxies with early-type morphologies and passively evolving
stellar populations (e.g., \citealt{djor87};
\citealt{thomas05}). However, a coherent evolutionary picture of their
early star formation histories (SFHs) and the buildup of their stellar
mass over cosmic time is still lacking.

Observations at higher redshifts suggest that the first quiescent
galaxies formed very early during the first 2--3 Gyr of the Universe,
becoming the dominant population among massive galaxies as early as
$z\sim2$ (\citealt{fontana09}; \citealt{ilbert10}; \citealt{brammer11};
\citealt{muzzin13smf}). Recent works have even identified,
photometrically and spectroscopically, a small number of these
galaxies at $z\sim3-4$ (\citealt{guo11}; \citealt{gobat12};
\citealt{stefanon13}), indicating that a fraction of the quiescent
population appear in only $\sim$1~Gyr. What is perhaps even more
surprising is that the first quiescent galaxies were structurally very
different from their local analogs, having effective radii up to a
factor of $\sim$$3-5$ smaller than those of quiescent galaxies at
$z\sim0$. (e.g., \citealt{daddi05}; \citealt{dokkum08};
\citealt{trujillo07}; \citealt{buitrago08}; \citealt{toft07};
\citealt{saracco10}; \citealt{cassata11}; \citealt{szo11}).

Theoretical models are slowly converging on an evolutionary picture
that describes the formation of quiescent galaxies as a two-stage
process (e.g., \citealt{naab07}; \citealt{oser10}). First, an early
phase of highly dissipative {\it in situ} star formation fueled by
cold gas streams (\citealt{haloquench}; \citealt{keres05}; \citealt{dekel06};
\citealt{dekel09a}) or gas-rich mergers (\citealt{hopkins06};
\citealt{hopkins08a}) leads to the formation of a compact remnant
(\citealt{elme08}; \citealt{dekel09b}; \citealt{wuyts10}). Then, a
late phase of size growth dominated by the accretion of smaller
systems slowly increases their radii (\citealt{bournaud07};
\citealt{naab09b}; \citealt{bezanson09}; although see
\citealt{poggianti13}; \citealt{carollo13} for a different
picture). One of the main challenges of this picture is identifying
the mechanism responsible for both the truncation of the
star formation and the structural transformation to understand the
relation between the quiescent population and their star-forming
progenitors. Indeed, observations reveal that these two populations
have significantly different structural properties at
$z\gtrsim2$. While star-forming galaxies (SFGs) exhibit disk-like
morphologies and, in many cases, irregular and clumpy structures
(e.g., \citealt{elme05}; \citealt{elme07}; \citealt{kriek09};
\citealt{guo12b}), quiescent galaxies at that redshift are
spheroid-dominated (\citealt{szo11}; \citealt{bell12}). Furthermore,
quiescent galaxies have smaller sizes than star-forming galaxies of
the same stellar mass (\citealt{williams10}; \citealt{wuyts11a}). This
difference suggests that the observational picture is missing a key
population in the evolutionary sequence that connects star-forming and
quiescent galaxies, namely, objects transitioning from star-forming to
quiescent that simultaneously experience a shrinkage in size.

Naively one would expect that such a connection would be through a
population of massive, compact star-forming galaxies (SFGs). However,
evidence for such galaxies remained elusive even for the deepest HST
optical surveys, which can only probe the rest-frame UV of
$z\gtrsim1.5$ galaxies, and thus tend to miss dust-obscured
galaxies. Now, owing to new IR capabilities of the WFC3 camera, the
analysis of (rest-frame optical) sizes can be extended to higher
redshifts, and evidence has started to accumulate, revealing the
existence of an abundant population of compact but red SFGs at
$z\sim2-3$ (\citealt{wuyts11b}; \citealt{patel13}; \citealt{barro13};
\citealt{stefanon13}; \citealt{williams13}). These galaxies typically
have large stellar masses, heavily obscured star formation and
spheroid-like morphologies. More importantly, they exhibit the small
radii of compact quiescent galaxies implying that structural changes
can occur on timescales comparable to the SF quenching timescale.

\citet[hereafter B13]{barro13} demonstrated that the radii and stellar
mass surface densities of compact SFGs quantitatively matched those of
compact quiescent galaxies. They further showed that the fall in the
number density of compact SFGs since $z\sim3$ is consistent with the
observed increase in the density of compact quiescent galaxies
assuming quenching times for the former of 300~Myr to
1~Gyr. Relatively short star formation timescales are plausible if, as
pointed out in \citet{wuyts11b}, some of these galaxies are offset
from the SFR$-M_{\star}$ correlation (the so-called main sequence;
\citealt{mainseq}; \citealt{elbaz07}) towards the high-SFR upper
envelope. In such extreme cases, gas exhaustion, supernova and AGN
feedback lead to a rapid decline in SFR. The latter can be
particularly relevant in compact galaxies sustaining large SFRs over
very small regions. As shown in, e.g., \citet{stanic12} (see also
\citealt{tremonti07}; \citealt{snewman13}) high SFR surface densities
are associated with strong galactic outflows that can deplete the gas
reservoirs in a short period of time.

Understanding the location and evolution of compact SFGs in relation
to the main sequence of ``normal'' galaxies is critical to assess
whether they represent extreme cases of high star formation efficiency
({\it starburst}; \citealt{daddi10b}; \citealt{rodi10b}) such as SMG
and HyLIRGs (\citealt{smail97}; \citealt{blain02};
\citealt{targett13}), or if they are forming stars gradually over
longer timescales. Life-paths on this diagram allow us to explore the
proposed evolutionary connection with quiescent galaxies, as well as
discriminate between possible formation scenarios that predict
different trends in the structural properties, the morphological type
or the amount of dust extinction along the evolutionary track.

The aim of this paper is to extend the results of B13 and present
further evidence that compact SFGs at $2<z<3$ are the natural
progenitors of compact quiescent galaxies at $z\sim2$. To that end, we
present a detailed analysis of a sample of 45 compact SFGs, selected
from the CANDELS survey in GOODS-S. First, we assemble comprehensive
UV-to-far IR stellar energy distributions (SEDs) for these galaxies,
and we study their observed and rest-frame colors, SFRs, morphologies,
structural properties and AGN activity with respect to other
star-forming and quiescent galaxies at the same redshift. Then, we
model their SEDs using stellar population synthesis (SPS) to: 1)
estimate their stellar ages, 2) study the implications of the assumed
SFH on the predicted tracks in the SFR--M diagram, and compare these
with the observed galaxy distribution; 3) estimate their quenching
times and compare the predicted number of quenched compact SFGs as a
function of time with the observed number density of quiescent
galaxies since $z=3$. Finally, we speculate on the origin of compact
SFGs by studying their possible formation mechanisms using
semi-analytic models (SAMs) and N-body simulations.

The structure of the paper is as follows. In Section~\ref{data} we
describe the datasets, the SED modeling, the procedure to estimate
galaxy properties and the criteria to select compact SFGs. In
Section~\ref{props} we study the SFRs, optical/NIR/far IR colors,
structural properties and AGN activity of these galaxies. In
Section~\ref{sedmodels} we compare the best-fit stellar ages and
formation timescales for compact SFGs obtained with 3 different SFH
models. In Section~\ref{lifepath} we show the evolutionary tracks in
the SFR--M and UVJ diagrams inferred from their SFHs, and we discuss
the implications for the proposed evolutionary sequence from
star-forming to quiescent. In Section~\ref{theory}, we discuss the
formation mechanisms of compact SFGs in the context of theoretical
simulations.

Throughout the paper, we adopt a flat cosmology with $\Omega_{M}$=0.3,
$\Omega_{\Lambda}$=0.7 and H$_{0}=70$~km~s$^{-1}$~Mpc$^{-1}$ and we
quote magnitudes in the AB system.

\begin{figure}[t]
\includegraphics[width=8.7cm,angle=0.]{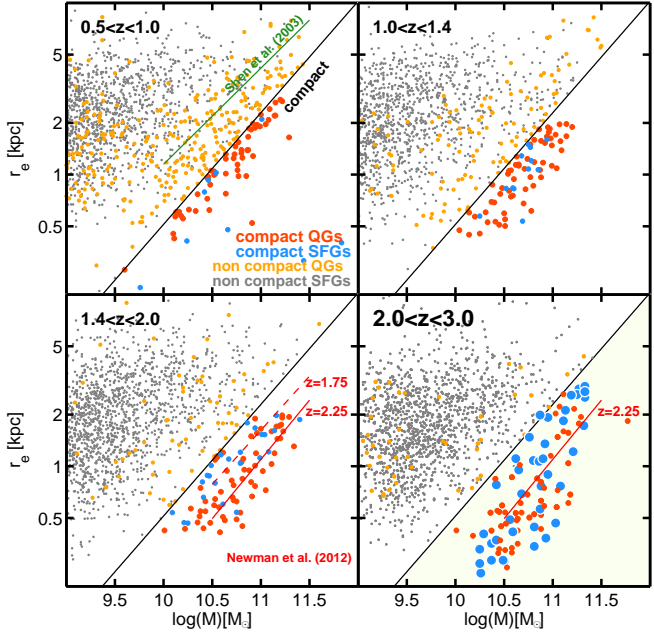}
\caption{\label{sample_size} Evolution of the mass--size distribution
  from $z=3$ to $z=0.5$ for galaxies in GOODS-S. The black line shows
  the compactness selection criterion
  ($\Sigma_{1.5}=10.3~M_{\odot}$kpc$^{-1.5}$). The colored markers
  indicate compact and non-compact SFGs (blue/grey) and quiescent
  galaxies (red/orange). The colored lines show the mass-size relation
  for quiescent galaxies at $z=2.25$ and 1.75 (red) from
  \citet{newman12} and $z=0$ (green) from \citet{shen03}. Note the
  steady increase in radii of these population with time. The 45
  compact SFGs at $2<z<3$ (bottom-right panel) analyzed in this paper
  are highlighted with larger marker sizes.  At $z\gtrsim2$ the
  majority of compact galaxies (highlighted area below the black line)
  are star-forming, as opposed to lower redshifts where these are
  predominantly quiescent.}
\end{figure} 

\section{DATA}\label{data}

The sample of compact SFGs analyzed in this paper is drawn from the
parent catalog presented in B13. In the following, we briefly outline
the datasets on which the catalog is based, the procedures to estimate
stellar properties and SFRs, and the most relevant over B13.

\subsection{Multi-band photometric data in GOODS-S/CANDELS}

The parent galaxy sample is derived from a {\it HST}/WFC3 F160W
($H$-band) selected catalog in the GOODS-S field.  The WFC3/IR
observations in this field cover a total area of
$\sim$173~arcmin$^{2}$ at different depths. The Early Release Science
(ERS, \citealt{windhorst11}) and the CANDELS Wide regions
(\citealt{candelsgro}; \citealt{candelskoe}) cover $\sim$2/3 of the
area at 2-orbit depth ($H_{5\sigma}=27.4$~mag; 115~arcmin$^{2}$), and
the CANDELS Deep region covers the remaining 1/3 at 10-orbit depth
($H_{5\sigma}=28.2$~mag; 55~arcmin$^{2}$).  The galaxies were selected
from a combined mosaic drizzled to a 0.06\arcsec/pixel scale with a
typical point spread function (PSF) of $\sim$0.18~arcsec.  The
multi-wavelength catalog based on the $H$-band selection includes
photometry in 14 passbands ranging from U to 8$\mu$m, with 7
high-resolution bands from {\it HST}/ACS and WFC3
(B$_{435}$,V$_{606}$,i$_{775}$,z$_{850}$,YJH) and the deepest {\it
  Spitzer}/IRAC data from SEDS (\citealt{ashby13}). The merging with
lower resolution data (ground-based and {\it Spitzer}/IRAC) was
computed using TFIT \citep{tfit}.  A comprehensive overview of this
catalog can be found in Guo et al. (2013; see also \citet{galametz13}
for more details).

We also include complementary mid-IR photometry in {\it Spitzer}/MIPS
24 and 70~$\mu$m (30~$\mu$Jy and 1~mJy, $5\sigma$) from \citet{pg08b},
and far-IR from the GOODS-Herschel (\citealt{elbaz11}) and PEP
(\citealt{magnelli13}) surveys, including PACS- 100 and
160~$\mu$m, and SPIRE- 250, 350 and 500~$\mu$m. A description of the
method used to derive consistent mid-to-far IR SEDs is presented in
\citet{pg08,pg10}. X-ray source identifications and total
luminosities ($L$$_{\mathrm{X}}\equiv L_{2-8\mathrm{kev}}$) were
computed for the sources identified in the {\it Chandra} 4~Ms catalog
\citep{chandra4m}.

\begin{figure*}
\centering
\includegraphics[width=18cm,angle=0.,bb=51 298 490 454]{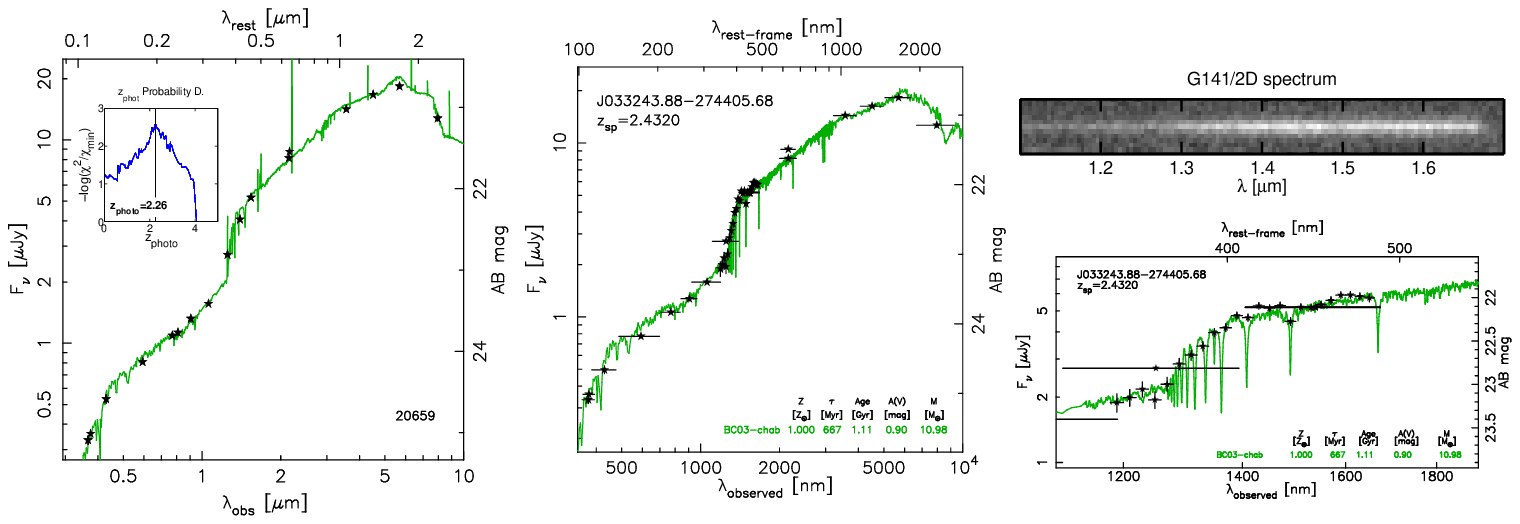}
\caption{\label{composite} Description of the procedure to merge
  multi-wavelength broad-band photometry with G141 grism spectroscopy
  to create a composite SED.  {\it Left panel:} Best fit stellar
  template to the broad-band SED. The photometric redshift probability
  distribution (upper-left inset) and the best-fit value are
  indicated.  {\it Right panel:} Above, the 2D-G141 spectrum of the
  example galaxy. Below, the 1D spectrum collapsed in the spatial
  direction and divided in discrete photometric blocks created by
  convolving the data with square filters of FHWH$=$200\AA. A small
  ($<2\%$) correction is applied to the G141 flux calibration based on
  the comparison to F125W and F140W photometry (black stars).  {\it
    Central-panel:} Best fit stellar template to the merged broad-band
  plus grism SED. The increased spectral resolution of the composite
  SED provides tighter constraints on the SED-modeling increasing the
  precision of photometric redshift and stellar population
  properties.}
\end{figure*}

\subsection{Inferred galaxy properties}\label{galprop}

In order to select a sample of compact SFGs, we first derive
photometric redshifts, stellar masses and star formation rates (SFRs)
for all the galaxies in the $H-$band selected catalog from SED
modeling.  These properties have been used in previous works by
\citet{wuyts11b, wuyts12} and B13. Therefore we describe here only the
most relevant details. In brief, photometric redshifts were estimated
from a variety of different codes available in the literature which
are then combined to improve the individual performance. The technique
is fully described in \citet{dahlen13}, and the catalog will be
released in Dahlen et al. (2014, in prep).  Based on the best
available redshifts (spectroscopic or photometric; see Table 1) we
estimated stellar masses and other stellar population properties (such
age, extinction, UV-based SFR, etc.) using FAST \citep{fast}. The
modeling is based on a grid of \citet[][BC03]{bc03} models that assume
a \citet{chabrier} IMF, solar metallicity, exponentially declining
star formation histories, and a \citet{calzetti} extinction
law. Rest-frame magnitudes based on the best-fit redshifts and stellar
templates were computed using EAZY \citep{eazy}.

We compute SFRs on a galaxy-by-galaxy basis using a {\it ladder} of
SFR indicators as described in \citet{wuyts11a}. The method
essentially relies on IR-based SFR estimates for galaxies detected at
mid- to far-IR wavelengths, and SED-modeled SFRs for the rest. As
shown in \citet{wuyts11a} the agreement between the two estimates for
galaxies with a moderate extinction (faint IR fluxes) ensures the
continuity between the different SFR estimates (see also
Appendix~\ref{sfrcompare}). For IR-detected galaxies the total SFRs,
\sfrtot, were then computed from a combination of IR and rest-frame UV
luminosities (uncorrected for extinction) following \citet{ken98} (see
also \citealt{bell05}):

\begin{equation}
SFR_{\mathrm{UV+IR}}=1.09\times10^{-10}(L_{\mathrm{IR}}+3.3L_{2800})[M_{\odot}/yr]
\end{equation}

The normalization factor corresponds to a \citet{chabrier} IMF, and
$L_{2800}$ is estimated from the best fitting SED template.  Note that
for the analysis of compact SFGs we perform a more exhaustive SED
modeling in \S~\ref{sedmodels}. However, we base our sample selection
in an easy-to-reproduce method using the standard data described
above. We verify that the re-computed stellar properties do not
introduce any significant difference on the sample selection.

The shape of the two-dimensional surface brightness profiles measured
from the {\it HST}/WFC3 F160W image were modeled using GALFIT
\citep{galfit}. The method and the catalog are fully described in
\citet{vdw12}. A single component fit was performed to determine the
circularized, effective (half-light) radius,
r$_{e}$$\equiv$$a_{\mathrm{eff}}\sqrt{(b/a)}$ ($a_{\mathrm{eff}}$ is
the half-light radius along the major axis), and the S\'ersic index,
$n$. Spatially variable point spread functions (PSFs) were created and
processed with TinyTim \citep{tinytim} to replicate the conditions of
the observed data when fitting light profiles. We note that the
circularized radius, although widely use in the literature, may cause
edge-on galaxies to appear smaller (see \S~\ref{strucprop}).

\subsection{Selection of compact SFGs}

Following the criteria of B13, we select a sample of massive
($M_{\star}>10^{10}M_{\odot}$) compact star-forming galaxies at
$2<z<3$, using a threshold in {\it pseudo} stellar mass surface
density, $\Sigma_{1.5}$, of
log$(M/r^{1.5}_{\mathrm{e}})>10.3~M_{\odot}$kpc$^{-1.5}$ (below the
black line in Figure~\ref{sample_size}) and specific SFR,
\lssfr$>-1$. The latter is set slightly above a mass doubling time of
$3\times t_{\mathrm{Hubble}}$ at $z\sim2.5$ to reject the majority of
passively evolving galaxies, whereas the limit in $\Sigma_{1.5}$ is
chosen to select galaxies in the region of the mass size diagram
occupied by quiescent galaxies at $z>2$. Based on these criteria we
identify a total of 45 compact SFGs. The overall properties of these
galaxies are summarized in Table~1.  We note that 4 of these galaxies
are excluded from the analysis of the stellar properties in the
following sections (except \S~\ref{hersprop}) due to AGN continuum
contamination in the SED (see Appendix~\ref{agncont}).

\begin{figure*}[t]
\centering
\includegraphics[width=8.5cm,angle=0.]{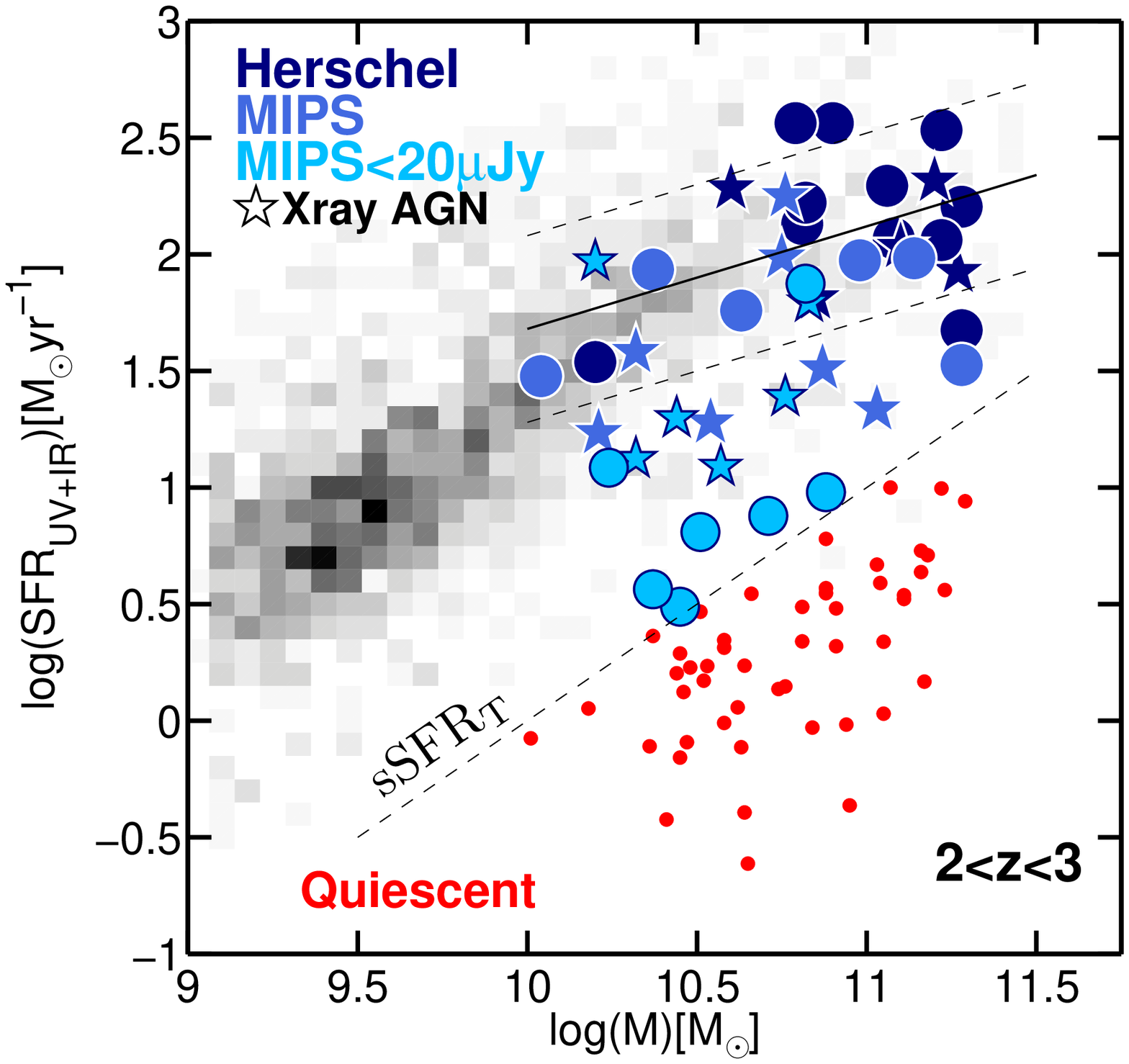}
\includegraphics[width=8.4cm,angle=0.,bb=50 100 575 630]{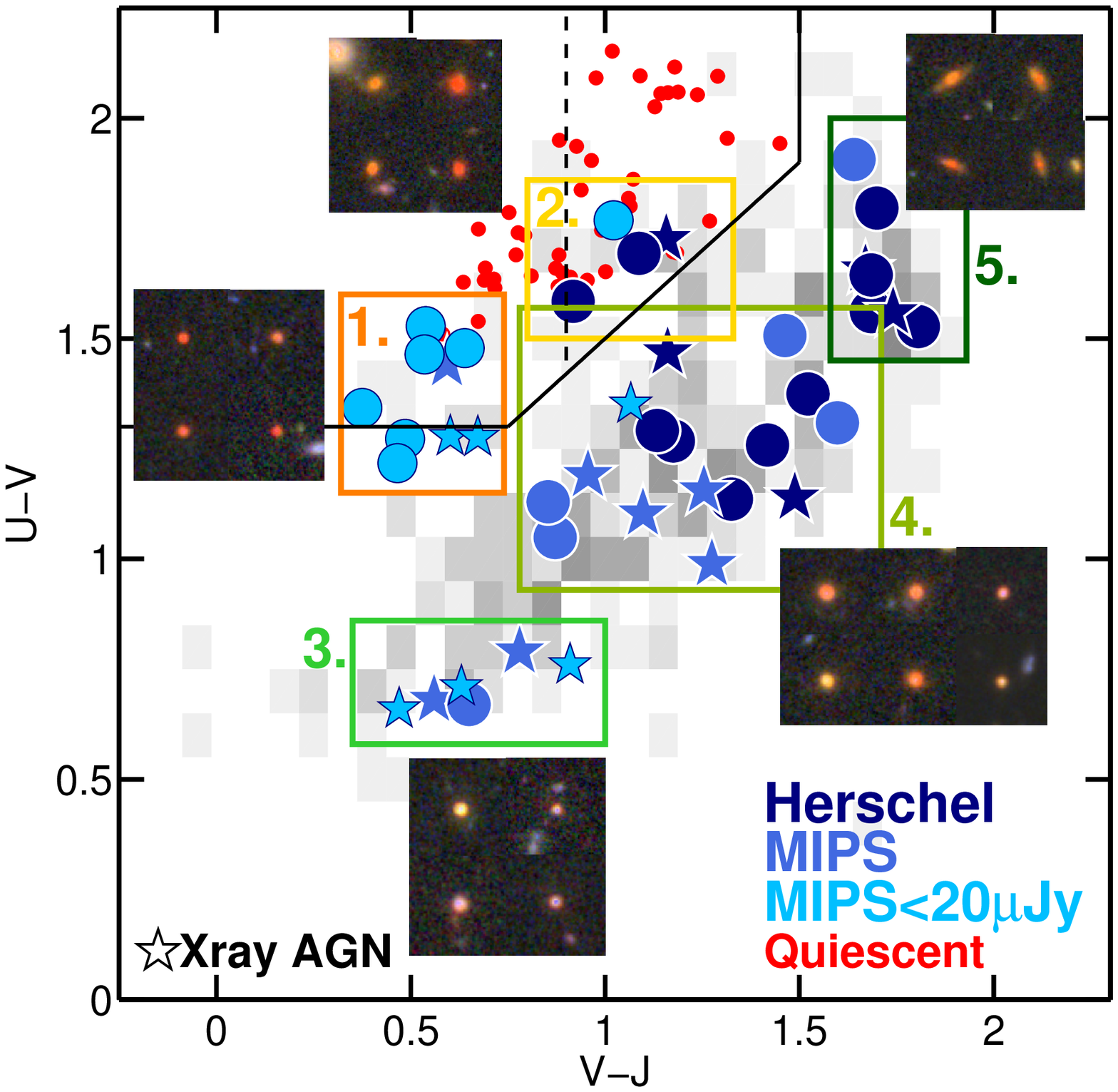}
\caption{\label{sample} {\it Left panel:} SFR--mass diagram for all
  galaxies at $2<z<3$ in the parent catalog. The boxed gray scale
  shows the main sequence of SFGs. The red markers show the quiescent
  population (\lssfr$<-1$; dashed black line). Compact SFGs are shown
  in blue colors; cyan indicates galaxies undetected in the far-IR
  (S$_{24\mu m}<20\mu$Jy); the other two (light-to-dark) colors
  indicate detections in {\it Spitzer}/MIPS~24~$\mu$m and {\it
    Herschel} (PACS or SPIRE). Far-IR emission is usually associated
  with a higher level of dust obscuration and SFR activity. The stars
  indicate that the galaxy hosts an AGN. The black line (dashed) shows
  the best-fit ($\pm$3$\sigma$) to the main sequence of massive
  (\lmass$>10$) SFGs. Only 2 compact SFGs present SFRs 3$\sigma$ above
  the main sequence ({\it starburst} galaxies;
  \citealt{daddi10b}). {\it Right panel:} UVJ diagram for galaxies
  more massive than \lmass$>10$ (boxed gray scale). The symbols
  indicate the same as in the left panel. For the purpose of further
  discussion, we divide the diagram in 5 regions (see e.g.,
  Figure~\ref{mediansfhs}). Regions 1 and 2 (orange/yellow) overlap
  partially with the UVJ-quiescent area, including mainly galaxies
  with lower SFRs. Regions 3 to 5 (green) contain SFGs with increasing
  levels of extinction and larger stellar masses from left to
  right. The 5\arcsec$\times$5\arcsec (ACS/WFC3) zJH color stamps show
  representative examples of galaxies in each region. Compact SFGs
  present undisturbed spheroidal morphologies, except the most
  obscured ones which appear to be edge-on and disk-like.}
\end{figure*}

To illustrate the motivation of the selection criteria,
Figure~\ref{sample_size} shows the mass--size relation for galaxies
more massive than \lmass$>9$ in 4 redshift intervals ranging from
$0.5<z<3.0$. The sub-populations of compact SFGs and quiescent
galaxies are highlighted in blue and red, respectively.  As discussed
in B13 and other previous works, the region limited by the
$\Sigma_{1.5}$ threshold encloses most of the quiescent population at
$z\gtrsim1.4$, which appears to follow a very tight mass-size
correlation with a nearly constant slope and an increasing zero point
towards larger sizes as a function of time (see e.g.,
\citealt{cassata11}; \citealt{newman12}).  Interestingly, at
$z\gtrsim2$, this region becomes more densely populated with compact
SFGs (large blue markers) which are not nearly as abundant at lower
redshifts (Figure~2 of B13). The changing nature of the galaxies
populating the compact region, from star-forming to quiescent around
$z\sim2$, jointly with the remarkable similarity of their structural
properties (S\'ersic, $r_{e}$ and $\Sigma_{1.5}$) and the rapid
increase in the number density of compact quiescent galaxies
(\citealt{cassata13}) was interpreted in B13 as an indication of the
evolutionary connection between the two populations.

\subsection{3D-HST NIR grism spectra of compact SFGs}\label{compositesed}

In order to improve the spectral characterization of the compact SFGs,
we combined the broad-band photometry SEDs with HST/WFC3 G141 grism
spectroscopy from the 3D-HST survey \citep{3dhst}. The grism provides
continuous wavelength coverage from $\lambda=$1.1 to 1.6~$\mu$m with
medium resolution ($R\sim130$; 47~\AA/pixel), yielding $5\sigma$
continuum detections for sources brighter than $H_{F140W}=23$.  This
allows us to improve the spectral resolution around the age-sensitive
features in the rest-frame optical of galaxies at $z\gtrsim2$
($\lambda_{\mathrm{rest}}\sim4000$~\AA; e.g., \citealt{kriek11};
\citealt{whitaker13}). The source catalog and one-dimensional
flux-calibrated spectra reduced with the aXe software \citep{axe} were
drawn from \citet{trump13}. All compact SFGs fall within the area
covered by the G141 observations. However, the grism spectra of nearby
objects can sometimes overlap with the main extraction causing
partial, or severe, flux contamination \citep{3dhst}.  Nevertheless,
the aXe reduction provides a contamination estimate that can be used
to determine the un-contaminated spectral range. Based on that
determination, we extracted good quality spectra ($<1\%$
contamination) for 36 of the 45 compact SFGs.

Figure~\ref{composite} illustrates the procedure to merge a G141
spectrum with the broad-band SED. Briefly, we extract the 1D spectrum
in the wavelength range 1.1$-$1.7~$\mu$m at a native resolution of
46.5~\AA/pixel. Then, we convolve it with a square filter transmission
of FWHM$=$200\AA~for the purpose of combining it with the broad-band
photometry. The spectra are already flux calibrated, but we perform an
additional re-calibration by comparing to the broad-band photometry in
F140W and F125W. This results in small variations of less than 2\%
level. For the faintest objects in the sample, we bin the spectra by a
factor 2$-$3 to increase the SNR at the expense of lowering the
spectral resolution.

At $2<z<3$ the G141 spectra can yield detections of emission lines due
to \OIII~or \OII, or absorption lines in the Balmer series or the
G-band.  However, due to the low spectral resolution and the
additional broadening caused by the intrinsic galaxy shape, high-EW
emission lines are more easily detected (e.g., \citealt{trump11b};
\citealt{fumagalli12}) than absorption lines, which are only
identified with high significance in bright galaxies
(\citealt{dokkum10b}) or in stacked spectra (e.g.,
\citealt{whitaker13}; \citealt{bedregal13};
\citealt{krogager13}). Among our compact SFGs only 3 galaxies show
emission lines, and another 6 (with confirmed spectroscopic redshift)
show evidence of absorption lines at the appropriate rest-frame
wavelengths (Barro et al. 2014, in prep). Nonetheless, the grism data
provide a solid detection of the stellar continuum (top-right panel of
Figure~\ref{composite}) which, when combined with the broad-band
photometry, increase the spectral resolution of the SED, improving the
quality of the photometric redshifts and the estimated stellar
properties (\citealt{3dhst}; \citealt{bedregal13}). Roughly half of
the sample (23/45; see Table 1) have previous spectroscopic
redshifts. For those galaxies, the overall accuracy of the photometric
redshifts based on the composite SEDs is better than 1\% ($\Delta
z/(1+z)=0.7$\%).

\section{Properties of compact SFGs at $2<\lowercase{z}<3$}\label{props}

In this section, we review and expand the analysis of the properties
of compact SFGs at $2<z<3$ presented in B13, and in \S~\ref{sedmodels}
and \S~\ref{lifepath} and we model their stellar populations to verify
if the proposed evolutionary connection with compact quiescent
galaxies is consistent with the estimated ages, quenching times, and
evolutionary tracks on the SFR--M diagram.

\begin{figure*}[t]
\centering
\includegraphics[width=17cm,angle=0.,bb=107 45 938 583]{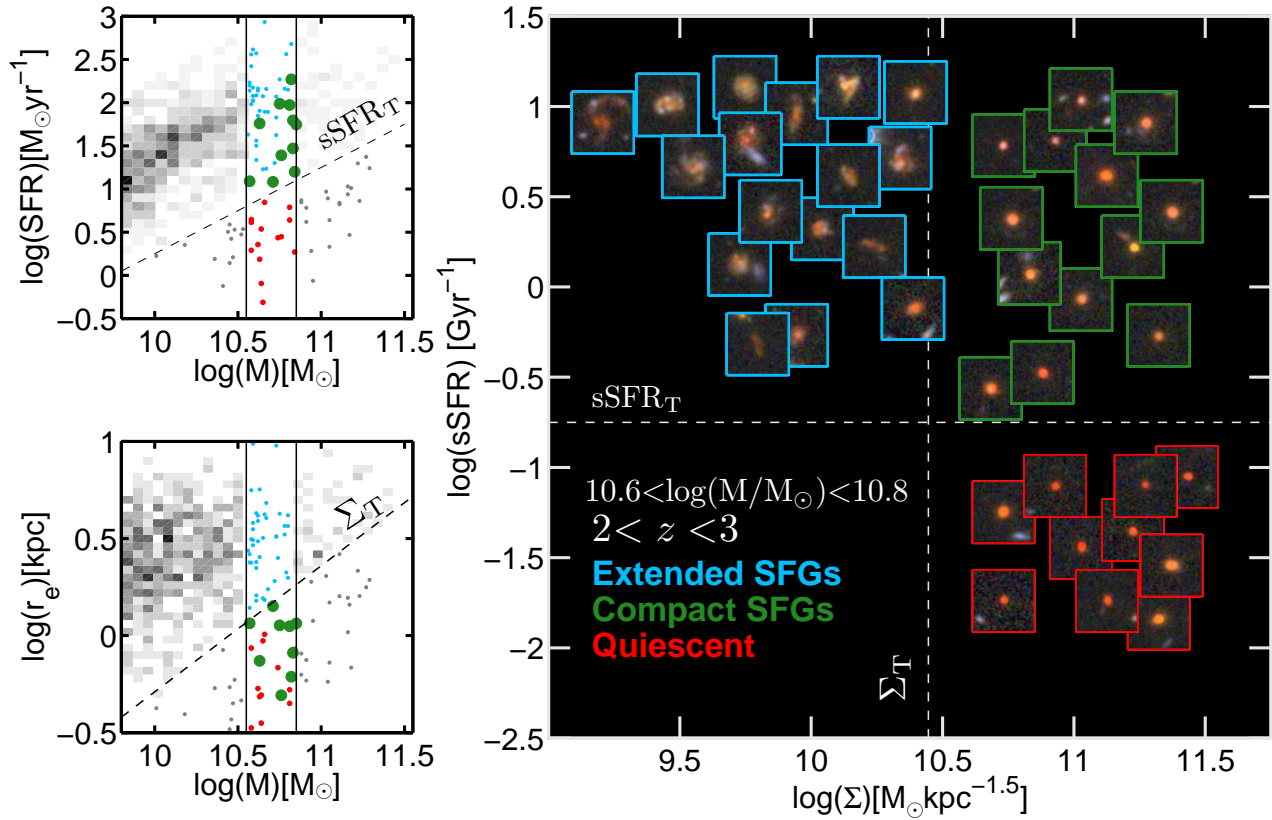}
\caption{\label{postageplot} {\it Left panels:} Distribution of
  galaxies at $2<z<3$ (boxed gray scale) in the SFR--M (upper-panel)
  and M--size (lower-panel) diagrams. The colored markers indicate
  compact SFGs (green), other non-compact SFGs (cyan) and quiescent
  galaxies (red) in a thin slice of stellar mass
  ($10.6<$\lmass$<10.8$; vertical black lines). The dashed lines show
  the selection thresholds for star-forming (above \lssfr=-1) and
  compact (below log($\Sigma_{T}/M_{\odot}\mathrm{kpc}^{1.5}$=10.3)
  galaxies. Compact SFGs have nearly {\it normal}, main sequence,
  SFRs, but small effective radii typical of the quiescent
  population. {\it Right panel:} sSFR--$\Sigma$ diagram showing the galaxies
  in the highlighted mass slice of the left panels.  The location of
  compact SFGs in upper-right quadrant and the lack of galaxies in the
  lower-left quadrant suggest that the formation of compact quiescent
  galaxies follows an evolutionary sequence from extended (upper-left)
  to compact (upper-right) SFGs, likely due to a strongly
  dissipational process (see \S~\ref{theory}), and then to quiescence
  (bottom-right) by simply shutting down SFR. The
  5\arcsec$\times$5\arcsec zJH color postages illustrate that compact
  SFGs not only share the stellar masses and effective radii of
  quiescent galaxies, but they also present similar spheroidal
  morphologies, significantly different from those of extended SFGs
  which appear to be disk-like and sometimes irregular or clumpy.}
\end{figure*}

\subsection{Distribution in the SFR--M plane}

The left panel of Figure~\ref{sample} shows SFR versus stellar mass
for all SFGs (boxed gray scale) and quiescent galaxies (red markers)
at $2<z<3$ highlighting the location of compact SFGs (blue markers;
the shade of blue indicates the strength of the far-IR detection). A
correlation between SFR--M, usually referred as the ``main sequence''
of star-forming galaxies (\citealt{mainseq}; \citealt{elbaz11}), is
visible across the whole mass range. In agreement with previous
studies of massive galaxies (\lmass$>10$), we find that the majority
of compact SFGs present dust obscured star formation
(\citealt{papovich07}; \citealt{pg08b}; \citealt{bauer11}), based on
their detection at mid-to-far IR wavelengths: 71\% are detected in
MIPS 24~$\mu$m, and 44\%/13\% are detected in {\it Herschel}
PACS/SPIRE, respectively. All {\it Herschel} detections are also
recovered in the deeper 24~$\mu$m data (S$_{24\mu
  m;3\sigma}\sim$20$\mu$Jy vs. $\sim$1mJy in the {\it Herschel}
bands). In addition, the fraction of IR-detections increases towards
the most massive galaxies indicating that IR-luminosity, and in
general dust attenuation ($A(V)\propto
L_{\mathrm{IR}}$/$L_{\mathrm{UV}}$; \citealt{barro11b}), both
correlate with stellar mass in SFGs, i.e., the most massive,
star-forming galaxies are more obscured (\citealt{brammer11};
\citealt{wuyts11b}).

The fraction of compact SFGs among all massive SFGs is
$\sim$20\%. This number however depends on the stellar mass,
increasing from 10\% to 30\% and 37\% at
\lmass$=$[$10-10.6$],[$10.6-11.2$] and [$>11.2$].  This is because, at
the high-mass end, the number of non-compact SFGs and the scatter in
their mass--size distribution (i.e., the range in $\Sigma_{1.5}$)
decreases, leading to an increase in the relative number of compact
vs. extended SFGs (below and above the black line in
Figure~\ref{sample_size}).

In agreement with previous works, we find a flattening in the slope of
the main sequence at the high-mass end (\citealt{bauer11};
\citealt{whitaker12b}). The black line in Figure~\ref{sample} shows
the best fit ($\pm$3$\sigma$) to a single power-law ($\alpha=0.44$)
for all massive (\lmass$>10$) SFGs.  With respect to this fit, most
compact SFGs are found either on the main sequence ($\sim$65\%) or
below it ($\sim$30\%), which is consistent with the idea that at least
some of these galaxies are in transit to the red sequence. We note
also that compact SFGs below the main sequence have a stellar mass
distribution more skewed towards smaller values (\lmass$=10-10.6$)
than those in the main sequence (see also \S~\ref{shortlived} for
further discussion). Only 2 compact SFGs present SFRs slightly above
the 3$\sigma$ upper-limit of the main sequence (we reject another 2
due to AGN contamination; see next section). Such galaxies, usually
called {\it starburst} (\citealt{daddi10b}; \citealt{rodi11}), are
thought to be in a short-lived, high star formation efficiency phase
(high gas-to-SF ratio), possibly triggered by an external mechanism,
such as mergers or galaxy interactions. If compact SFGs are the
precursors of the quiescent population, the small {\it starburst}
fraction suggests that quenching is not usually preceded by a strong
peak in the SFR, or, alternatively, the duty cycle of the {\it
  starburst} phase is very short compared to the duration of the
star-forming phase. Incidentally, we do not find evidence for tidal
features or disturbed appearances in either of these 2 galaxies, but
we note that these are not necessarily representative of the whole
{\it starburst} population, as is the case in, e.g.,
\citet{kartaltepe12}.

\subsection{Extinction properties}\label{uvjprop}

The right panel of Figure~\ref{sample} shows the rest-frame $U-V$
vs. $V-J$ color (hereafter UVJ) for compact SFGs and other massive
star-forming and quiescent galaxies at $2<z<3$. The UVJ diagram is
an alternative diagnostic to distinguish between reddened star-forming
and quiescent galaxies according to their SEDs. This method has been
shown to be very successful in breaking the dust /age degeneracy using
the $V-J$ color as a proxy for dust attenuation (\citealt{wuyts07};
\citealt{williams10}; \citealt{whitaker11}).  The UVJ colors of
compact SFGs are consistent with their distribution on the SFR--M
diagram, and support the idea of their transitory nature from
star-forming to quiescent. Roughly 70\% of the compact SFGs,
predominantly those with far-IR detections, present red $V-J$ colors
characteristic of dusty SFGs, while the remaining $\sim$30\% appear to
lie within (or close to) the quiescent region. Within the latter, we
also find evidence for differences in the attenuation level as a
function of the $V-J$ color. Those with lower extinctions, to the left
of $\sim V-J=0.75$, were identified in \citet{whitaker11} as recently
quenched galaxies, following a nearly vertical color track (i.e.,
maintaining a low extinction) starting as low-mass, un-extinguished,
galaxies. If that is the case, those nearly quiescent galaxies at
$V-J=0.75$ could indicate the arrival point on the red sequence for
more dusty compact SFGs. We further investigate the possible
evolutionary tracks of compact SFGs in the UVJ diagram as a function
of their stellar mass and SFH in \S~\ref{uvjevol}.  For the purpose of
further discussion in the following sections, we divide the UVJ
diagram into 5 regions, roughly corresponding to the following overall
properties: 1) low-SFR, low-extinction; 2) low-SFR, higher-extinction;
3, 4 and 5) star-forming with increasing stellar mass and extinction.

\begin{figure}[t]
\centering
\includegraphics[width=9cm,angle=0.]{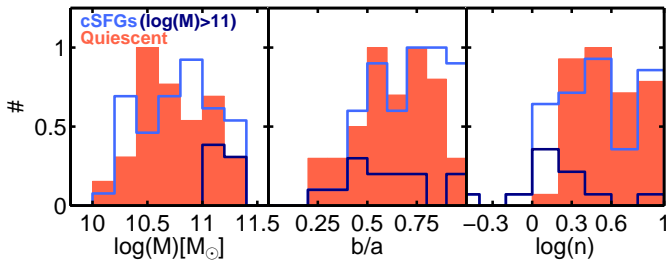}
\caption{\label{summaryprops} {\it Left to Right panels:} Distribution
  of stellar masses, axis-ratios and S\'ersic indices for compact SFGs
  (light blue) and quiescent galaxies (red). Both populations have
  consistent stellar mass distributions and similarly high values of
  the axis ratio and the S\'ersic index, typical of a
  spheroid-dominated population. The dark blue line indicates the
  distribution of the most massive (\lmass$>$11) and larger
  ($r_{e}>2$~kpc) compact SFGs, which appear to have more disk-like
  morphologies and a flatter distribution of axis ratios.}
\end{figure}

\subsection{Structural properties and visual appearance}\label{strucprop}

Compact SFGs are selected to have high stellar mass surface densities
similar to those of quiescent galaxies at $z\sim2$.  This means that,
for a given slice in stellar mass, compact SFGs exhibit the remarkably
small effective radii of quiescent galaxies while having the SFRs of
normal, main sequence, galaxies (Figure~\ref{postageplot}). As a
result, they occupy a distinct region in the sSFR--$\Sigma$ diagram
which, combined with the lack of extended-quiescent galaxies, suggest
that, at these redshifts, the quenching of star formation takes place
in the most compact (higher $\Sigma_{1.5}$) galaxies (see e.g.,
\citealt{cheung12}, and \citealt{fang13} for an extension of this
result to lower redshifts), and thus compact SFGs are the immediate
progenitors of the quiescent population. Under the assumption that
galaxies grow both in stellar mass and size during their star-forming
phase, the natural precursors of massive, compact SFGs are larger
SFGs, suggesting that the evolutionary sequence in
Figure~\ref{postageplot} goes from extended to compact SFGs
(left-to-right) and then to quiescence by simply shutting down star
formation. This sequence implies also the need for a mechanism to
shrink the size and to change the structure of extended SFGs
transforming them into compact SFGs. We explore these mechanisms in
\S~\ref{theory}.

The evolution from compact SFGs to compact quiescent galaxies is
supported by the histograms in Figure~\ref{summaryprops} which show
that compact SFGs span roughly the same range in stellar mass
(\lmass$=10-11.5$) as quiescent galaxies, while having high S\'ersic
indices ($n=3.4$) and axis-ratios ($b/a=0.75$), characteristic of that
population (e.g., \citealt{szo12}; \citealt{bell12}). We also find a
trend with stellar mass such that the most massive (\lmass$>11$) and
larger ($r_{e}>2$~kpc) compact SFGs are more akin to edge-on disks,
with lower Sersic indices and a flatter distribution of axis ratios
(dark-blue line in Figure~\ref{summaryprops}). This however is also
true also massive compact quiescent galaxies at $z\sim2$
(\citealt{vdw11a}; \citealt{bruce12}; \citealt{chang13}), indicating
that the structural similarities between the two populations are also
preserved at different stellar masses.

In terms of their visual appearance, compact SFGs show undisturbed
spheroidal morphologies, very similar to compact quiescent galaxies,
but strikingly different from non-compact SFGs, which are
predominantly disk-like or irregular (right panel of
Figure~\ref{postageplot}). Only the most massive and dust-obscured
galaxies, mainly in region 5 of the UVJ diagram, present different
morphologies, more similar to edge-on disks or patchy galaxies with
diffuse light profiles nearly undetected in the rest-frame UV images
(see also \citealt{patel12}). Indeed, their selection as compact
galaxies may be partially due to an inclination effect. Signatures of
mergers or interactions are uncommon among compact SFGs and, in
general, among spheroidal galaxies at $z\sim2$
\citep{kaviraj12,kaviraj13a}. The few examples of disturbed
morphologies within our sample appear to be in the most unobscured
galaxies in region 3 of the UVJ. Interestingly, 3 of these galaxies
host the most luminous X-ray galaxies ($L_{\mathrm{X}}>10^{44}$~erg/s)
in the sample.

\begin{figure*}[t]
\centering
\includegraphics[width=17cm,angle=0.]{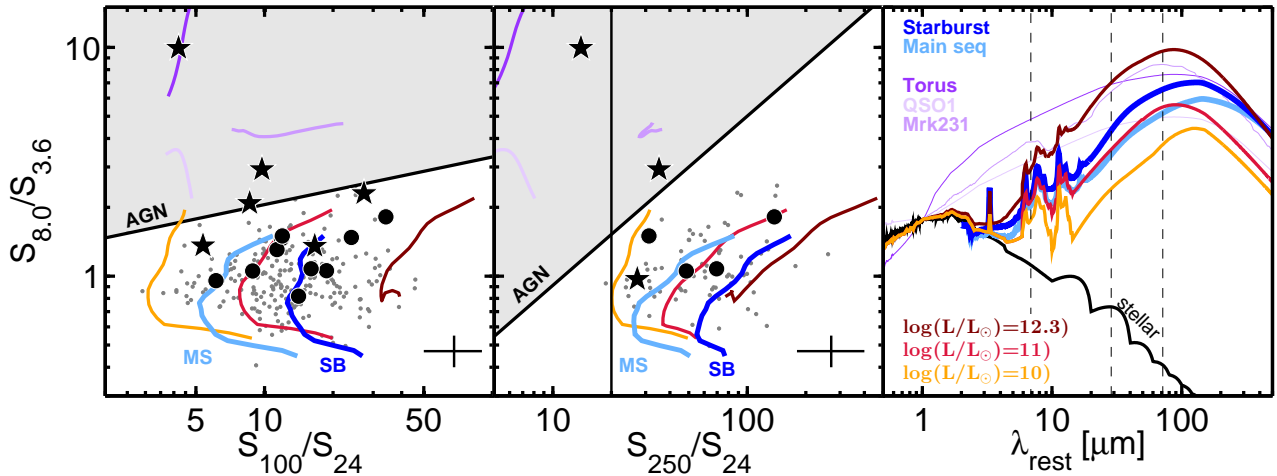}
\caption{\label{tempdust} Far-IR ({\it Spitzer}/{\it Herschel})
  color-color diagrams for compact (black) and non-compact (gray) SFGs
  at $2<z<3$ detected in PACS~100~$\mu$m (left panel) and
  SPIRE~250~$\mu$m (central panel). Stars indicate X-ray
  detections. The gray shaded area indicates the selection region for
  galaxies with a significant AGN contribution to the IR-emission
  (\citealt{kirkpatric13}). Also from \citet{kirkpatric13}, the
  vertical line at $S_{250}/S_{24}=30$ indicates the minimum color for
  a typical star-forming SED. Galaxies to the right of that line are
  expected to have significant star formation regardless of their AGN
  emission. The solid lines indicate evolutionary tracks (bottom-up
  from $z=2$ to 3) for 3 groups of IR-SEDs show in the right panel,
  namely: AGN dominated (purple), SFGs with increasing
  L$_{\mathrm{TIR}}$ (orange to brown) drawn from the library of
  \citet{ce01}, and the {\it main sequence}/{\it starburst} templates
  of \citet{elbaz11}. The dashed lines in the right panel indicate
  rest-frame wavelength probed by MIPS~24~$\mu$m, PACS~100~$\mu$m and
  SPIRE~250~$\mu$m for a galaxy at $z=2.5$. Overall, compact SFGs
  present colors consistent with star formation and IR-SEDs ranging
  between the {\it main sequence} and {\it starburst} templates. Only
  3 X-ray detected AGNs appear to have some AGN contribution to their
  IR-SEDs. These galaxies are also flagged as AGN dominated based on
  their optical/NIR SEDs (Appendix~\ref{agncont}).}
\end{figure*}

\subsection{Far IR colors and SEDs}\label{hersprop}

The emission at mid-to-far IR wavelengths is typically associated with
dust heated by star formation. However, if part of this emission
originates from a different source, such as an AGN or an evolved
stellar population, it could lead us to overestimate the IR-based
SFR. The latter case appears to be relevant only for galaxies with low
SFRs (\citealt{salim09}; \citealt{fumagalli13}), however, an obscured
AGN can have a significant contribution to the IR-emission even in
strongly SFGs (\citealt{daddi07,daddi07b}). The shape of the IR SED,
probed by {\it Spitzer}/{\it Herschel} colors, provides an effective
diagnostic tool to identify the power source of the dust heating
(e.g., \citealt{kirkpatric13}). Dust heated by star formation has
colder temperatures ($T_{\mathrm{dust}}=15-50$~K), and thus emits at
longer wavelengths than dust heated by an AGN
($T_{\mathrm{dust}}=150$~K), which is a more intense heating
source. Figure~\ref{tempdust} shows the S$_{250}$/S$_{24}$ and
S$_{100}$/S$_{24}$ {\it Spitzer}/{\it Herschel} colors versus the
S$_{8.0}$/S$_{3.6}$ IRAC color for 19 (7) compact SFGs detected in
PACS (SPIRE) compared with other massive SFGs at $2<z<3$.  In the
presence of an AGN, the emission at shorter wavelengths leads to bluer
{\it Herschel} colors and redder IRAC colors, typically within the
gray shaded areas. IRAC colors have been widely used in the literature
as an AGN selection criterion (e.g., \citealt{lacy04};
\citealt{stern05}; \citealt{donley07}), but combined with their {\it
  Herschel} colors, also provide additional information on the heating
source (SF vs. AGN) or the nature of the star formation (main sequence
vs. starburst). To illustrate these differences, Figure~\ref{tempdust}
shows color tracks of IR templates with increasing levels of AGN
activity (purple) and IR luminosity (red).

The overall {\it Herschel} colors and the distribution with respect to
other SFGs indicates that the IR emission in compact SFGs is mainly
fueled by star formation. Only 2/6 X-ray detected compact SFGs (22603
and 9834) appear to have a significant contribution from the AGN to
the IR emission, whereas the other 4 present colors consistent, or
slightly above, the star-forming range. The first 2 galaxies and
another one of the second group, were already excluded from our
analysis on the basis of AGN contamination in the stellar SED, but are
shown in this section to illustrate the effects of AGN emission in the
IR-colors.

We also find an excellent agreement between the main sequence/{\it
  starburst} classification derived from the SFR--M diagram and the
{\it Herschel} colors. Only 2 galaxies (25998 and 14876), the same
ones above the main sequence in Figure~\ref{sample}, appear to have a
high S$_{100}$/S$_{24}$$>$20 ratio characteristic of {\it starburst}
galaxies. The remaining compact SFGs lie roughly between the tracks of
the main sequence and {\it starburst} templates
(\citealt{elbaz11}; \citealt{magdis12}), with IR luminosities ranging
from $L_{\mathrm{TIR}}=10^{10.8-11.6}L_{\odot}$.

\subsection{AGN identification from X-rays}\label{xrayprop}

Using the {\it Chandra} 4~Ms catalog, we find that roughly $\sim20\%$
of all massive ($M_{\star}>10^{10}M_{\odot}$) galaxies at $2<z<3$ host
an X-ray detected AGN (see also \citealt{wang12}). Interestingly, the
majority of these luminous ($L_{X}>10^{43}$~erg/s) AGNs are found in
compact hosts. In this sample, which covers a slightly wider area of
GOODS-S than B13, 47\% (21/45) of the compact SFGs are X-ray detected
AGN, 6 of which are also selected using the IRAC power-law criteria
(PLG; \citealt{donley07,donley08}; \citealt{caputi13}). Note also
that, at the depth of the 4~Ms {\it Chandra} survey, only the most
luminous AGNs can be detected at $z>2$. Thus the intrinsic fraction
could be higher if we were able to detect lower-luminosity AGNs. For
comparison, only 9\% and 17\% of the non-compact SFGs and compact
quiescent galaxies host an AGN. In the context of the evolutionary
sequence, the large fraction of AGNs among compact SFGs suggests that
the transformation from extended to centrally concentrated compact SFG
triggers a phase of black hole and stellar bulge growth which could
signal the building of the M$_{\star}-\sigma$ relation
(\citealt{cisternas11a}; \citealt{mullaney12b}).

Compact SFGs with and without AGNs (stars in Figure~\ref{sample})
appear to have a similar distribution in the UVJ and SFR--M
diagrams, which provides no conclusive evidence on the role of the AGN
in the quenching of the star formation. We note however, that the high
AGN fraction among compact SFGs makes the quenching scenario more
likely at high redshift than in the local Universe, where AGNs are
more frequent among older star-forming galaxies or post-starburst
(\citealt{davies07}; \citealt{wild10}; Yesuf et al. 2013, in prep.).

\section{Stellar population modeling of compact SFGs}\label{sedmodels}

In \S~\ref{compositesed} we described the method to create composite
SEDs combining broad-band photometry in 17 bands with WFC3/G141 grism
spectroscopy. Here we make use of these detailed SEDs to estimate the
stellar properties of compact SFGs from SED-fitting. In particular, we
focus on deriving stellar ages and formation timescales (see also
Appendix for a discussion on SFRs), and we study the differences
arising from the use of 3 different parametrizations of the SFH,
namely: single (\S~\ref{singletau}) and delayed
(\S~\ref{delayedtau}) exponentially declining ($\tau$) models, and a
library of SFHs derived from semi-analytic models (SAMs) of galaxy
formation (\S~\ref{samtau}). There are different possible definitions
of galaxy age that are frequently used in the literature (e.g., SFR-
or mass-weighted ages, \citealt{wuyts11a}). Here, we refer to the
best-fit age as the time since the onset of star formation,
$t=t_{\mathrm{obs}}-t_{\mathrm{form}}$.

\subsection{Single $\tau$ models}\label{singletau}

The main modeling assumptions used with single (and also delayed)
$\tau$ models are: the BC03 stellar library, a \citet{chabrier} IMF
($M\in[0.1-100]M_{\odot}$), with a \citet{calzetti} attenuation law
ranging from $A_{V}=0-4$, and solar metallicity. In addition, the
$e-$folding times are required to be larger than 300~Myr to obtain a
better agreement between the \sfrtot and the best-fit SFR from the
models (see Appendix 1 for detailed discussion). No other constraints
are imposed on the best-fit ages.

The best-fit ages estimated from single $\tau$ models range from
$t=0.3-1.1$~Gyr, with a median value of $t=850$~Myr. Based on those
ages, the formation redshifts range between $z_{\mathrm{form}}=3.5$
and $z_{\mathrm{form}}=4.2$.  The majority of compact SFGs present
best-fit $e$-folding times close to the minimum threshold, with a
median value of $\tau=400$~Myr, and $90\%$ of the galaxies presenting
values lower than $\tau=800$~Myr. The distribution of $t/\tau$ values
peaks around $\sim$2.6, as expected from the star-forming nature of
the sample. For comparison, quiescent galaxies are often selected with
a threshold of $t/\tau>6$ (e.g., \citealt{fontana09}), which roughly
corresponds to \lssfr$=-1.5$.

\begin{figure}[t]
\centering
\includegraphics[width=8.5cm,angle=0.]{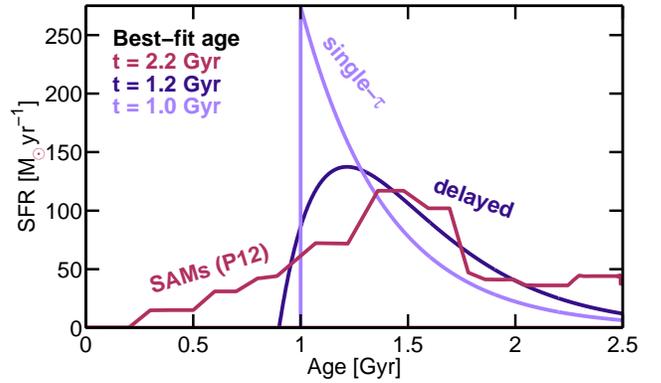}
\caption{\label{history} Example of the 3 different SFHs used to model
  the stellar populations of compact SFGs. The purple lines show the
  evolution of single and delayed $\tau$ models with the same
  $\tau=300$~Myr. The main difference between them is an early phase
  ($t\ll\tau$) of increasing SFR in the latter. The magenta line
  illustrates a non-parametric SFH drawn from a SAM
  \citep{paci12}. SAM SFHs present fluctuations of the SFR on short
  timescales inherited from the accretion and merging histories of the
  SAMs. Due to having a gradual increase of the SFR at early times,
  SAM SFHs estimate older stellar ages than $\tau$ models}
\end{figure}

\subsubsection{Short-lived compact SFGs}\label{shortlived}

We verify that the overall $\chi^{2}$ for constrained ($\tau>300$~Myr)
SED fits is fully consistent with the values obtained imposing no
restrictions on the $e$-folding time. For the majority of galaxies a
weak constraint favors solutions with smaller $\tau$ and younger ages
for similar values of log(t/$\tau$)$\sim$constant, i.e., within the
well-known degeneracy in age-$\tau$ (see, e.g., Figure~11 of
\citealt{shards}).

Nevertheless, for the (low-sSFR) galaxies in region 1 of the UVJ
diagram (Figure~\ref{sample}b) the constrained $\tau$ models
overestimate the rest-frame UV luminosity
($\lambda_{\mathrm{rest}}<3000$~\AA) providing a poor fit in that
spectral range. Forcing a maximally-old age to reduce the
UV-luminosity worsens the $\chi^{2}$, suggesting that shorter
formation timescales (i.e., shorter $\tau$) are required to reproduce
the SEDs of these galaxies. Indeed, the unconstrained $\tau$ models
provide the best-fit for typical $e$-folding times of $\tau=10-30$~Myr
and a median age of $t\sim1$~Gyr. As mentioned in \S~\ref{uvjprop},
region 1 of Figure~\ref{sample}b was identified in \citet{whitaker12}
as the arrival point on the red sequence for recently quenched
galaxies (see also \citealt{mcintosh13}). The small values of $\tau$
needed to fit these galaxies suggest that this is not only the arrival
point for recently quenched galaxies, but specifically for galaxies
with short assembly histories, i.e, a ``fast track'' to the red
sequence. Meanwhile, other compact SFGs, well reproduced with
constrained $\tau$ models, may follow a different route to the
quiescent region of the UVJ. We further discuss this possibility in
\S~\ref{samtau} and \S~\ref{uvjevol}.

\subsection{Delayed $\tau$ models}\label{delayedtau}

Several recent papers have addressed the issue of how the use of
different SFHs affects the best-fit stellar properties
(\citealt{lee09}; \citealt{pforr12}). While it is not yet clear what
is the preferred functional form, there is a general agreement on the
limitations of declining $\tau$ models to recover the stellar
properties of young star-forming galaxies at $z\gtrsim3$
(\citealt{finlator11}; \citealt{papovich11};
\citealt{schaerer11}). The problem arises because some galaxies might
undergo increasing, rather than decreasing, SFHs during the early
phases of their lives. For such galaxies there is often a better
agreement between the properties derived from SED-modeling and
observed estimates of the stellar age and SFR using {\it inverted}
(exponentially increasing) $\tau$ models or the delayed models,
SFR$\sim~t\times\mathrm{exp}(-t/\tau)$ (\citealt{maraston10};
\citealt{gonzalez12}; \citealt{curtislake12}). The main difference
between single and delayed $\tau$ models is an early phase
($t\ll$$\tau$) of increasing SFR in the latter, while at
intermediate-to-late times both models present the same exponential
decline with time (Figure~\ref{history}).

\begin{figure}
\centering
\includegraphics[width=8.cm,angle=0.]{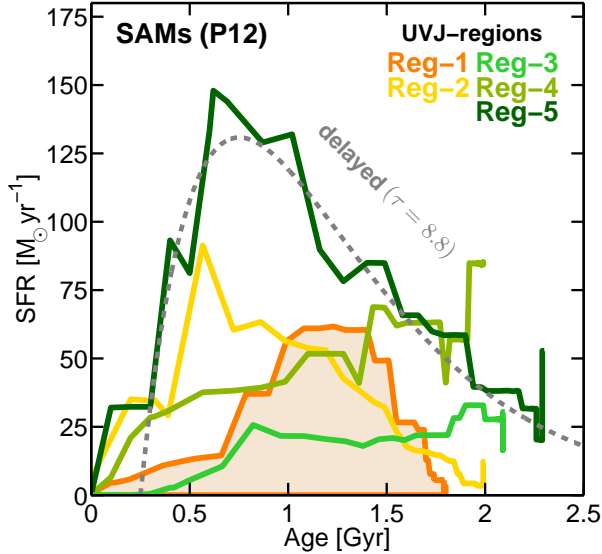}
\caption{\label{mediansfhs} Averaged SAM SFHs for compact SFGs in
  different regions of the UVJ diagram. The colors indicate the
  regions in Figure~\ref{sample}b. Galaxies in region 1 (orange shaded
  area) appear to have the shorter formation timescales, i.e., they
  assembled the bulk of their stellar mass in a shorter period of
  time. Galaxies in region 5 show a SFH similar to a delayed model
  with $\tau=600$~Myr (dashed gray line). However, the majority of
  compact SFGs (regions 1, 3, 4) have a more gradual increase of the
  SFR at early times and a longer plateau phase (SFR$\sim$constant)
  than the delayed models.}
\end{figure}

\begin{figure*}[t]
\centering
\includegraphics[width=8cm,angle=0.]{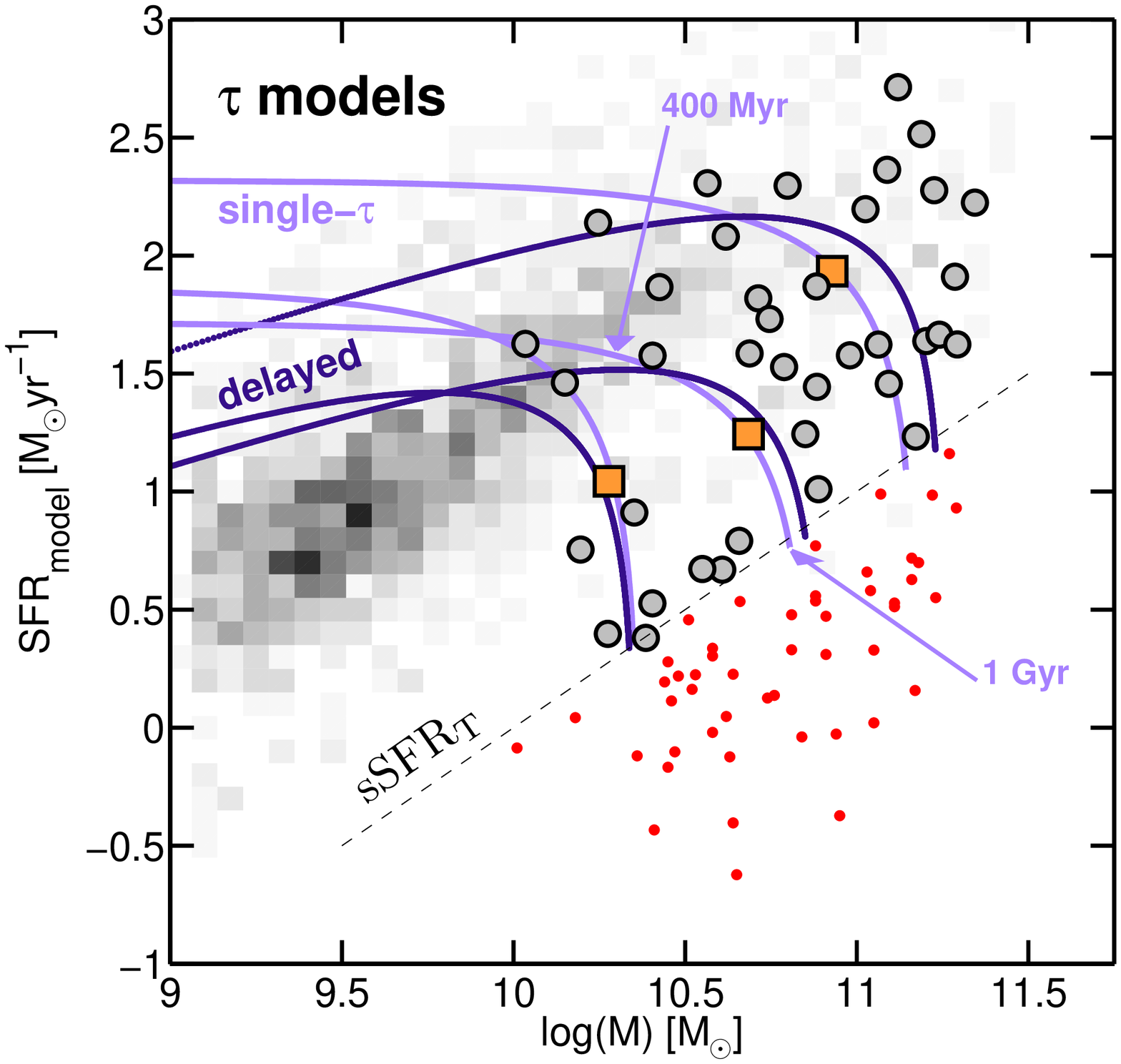}
\hspace{0.5cm}
\includegraphics[width=8cm,angle=0.]{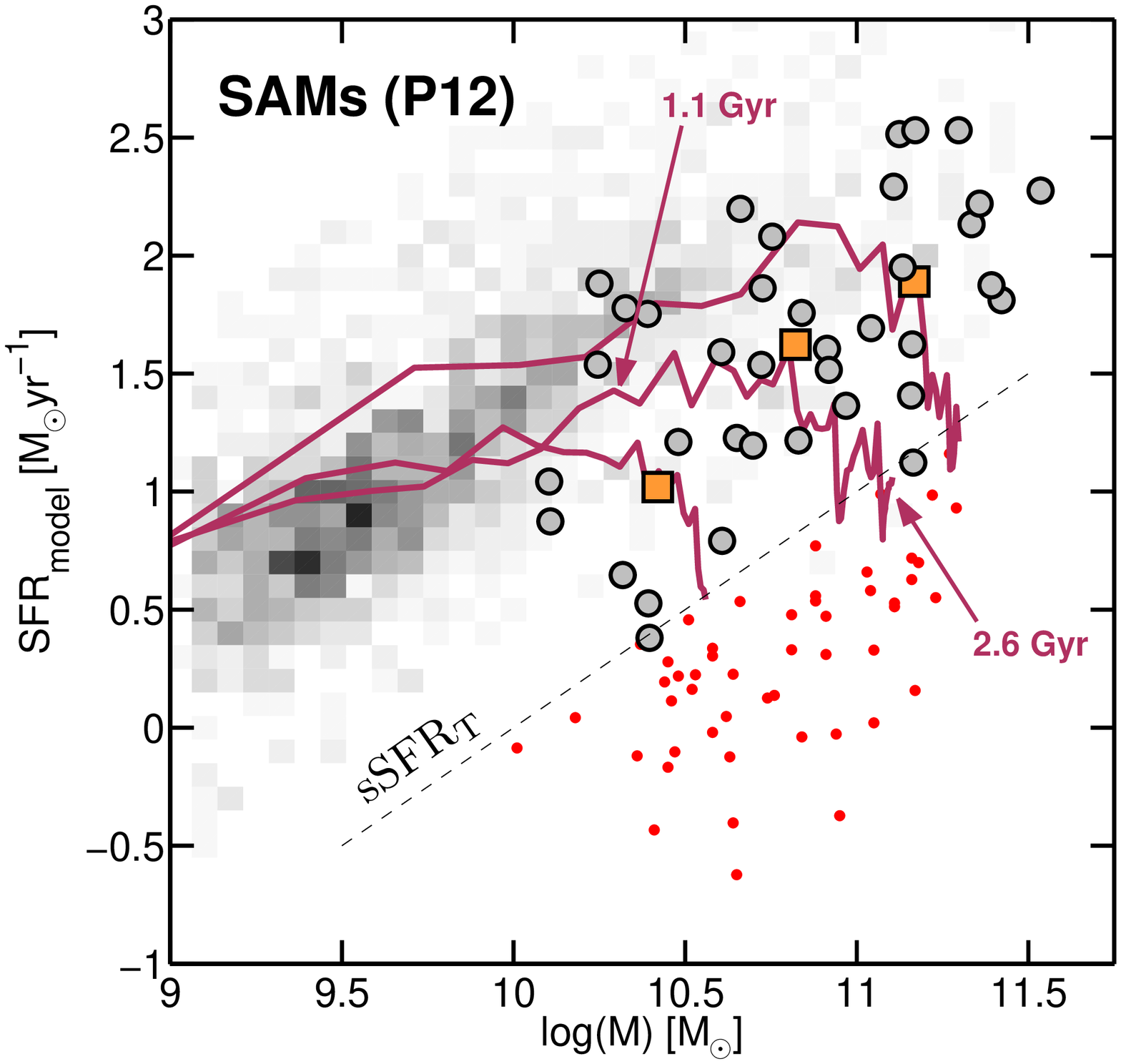}
\caption{\label{mainseq} Distribution of the compact SFGs in the
  SFR--M diagram (grey circles) determined from the best-fit stellar
  masses and SFRs to single $\tau$ models (left-panel) and SAM SFHs
  (right-panel). The distribution of other (non-compact) SFGs (boxed
  gray scale) and quiescent galaxies (red) at $2<z<3$ is the same as
  in Figure~\ref{sample}. The purple lines in the left-panel show the
  evolutionary tracks for 3 compact SFGs (orange squares) based on the
  best-fit SFHs to single and delayed $\tau$ models. In both cases,
  the slopes of the tracks appear to be shallower than the observed
  main sequence. The magenta lines in the right-panel show the
  evolutionary tracks for the same 3 galaxies based on SAM SFHs.
  Qualitatively, these tracks are similar to those of the delayed
  models. However, the SAM tracks present a steeper slope closer to
  that of the main sequence, and thus predict a longer duration of the
  star-forming phase. They also predict older stellar ages, which
  indicates that galaxies evolve in these tracks at a slower pace than
  in the tracks of the $\tau$ models, as indicated by the  (see also
  Figure~\ref{ssfrtime}).}
\end{figure*}

We find however, that none of the compact SFGs are in the increasing
SFR phase at the epoch of observation, i.e., they are predominantly in
intermediate-to-late evolutionary stages ($t>\tau$), for which the
single and delayed $\tau$ models have similar behavior, and thus
provide similar best-fit stellar properties. The only significant
difference is that stellar ages are $\sim30\%$ older in the delayed
models, with values ranging from $t=0.5-1.3$~Gyr, and a median value
of $t=1.1$~Gyr. The median $e$-folding time is the same in both
models, $\tau=400$~Myr, and the distribution is similarly skewed
towards the minimum threshold, with $90\%$ of the galaxies having
$\tau<1$~Gyr. The stellar masses are also fully consistent, with only
a small offset ($\Delta$\lmass$=0.05$~dex) towards larger stellar
masses in the delayed models. The typical scatter of the comparison is
smaller than 0.1~dex. Thus we conclude that, for compact SFGs, the use
of single or delayed $\tau$ models produce similar best-fit results,
with the only noticeable difference of slightly older stellar ages in
the latter.

\subsection{SFHs from semi-analytic models of galaxy formation}\label{samtau}

An alternative option to parametric SFHs is to use a library of
physically motivated SFHs drawn from theoretical models of galaxy
formation.  The advantage of this approach is that the range of
possible SFHs is more diverse, including increasing and decreasing
phases, as well as sudden bursts of star formation caused by galaxy
interactions. This method has been successfully used in
\citet{finlator07} to analyze a sample of $z\sim3$ galaxies using
hydrodynamic simulations and, recently, in \citet[][hereafter
  P12]{paci12} to reproduce evolutionary paths of low and high mass
galaxies at $z=0-1$ using a template library derived from
semi-analytical modeling (SAM) of a dark matter simulation. In this
section we follow the latter approach to analyze the SFH of compact
SFGs.

The details of the modeling procedure are described in P12. Briefly,
the template library is based on SFHs and chemical enrichment
histories for galaxies drawn from the Millenium cosmological
simulation \citep{springel05} as processed by the semi-analytical
recipes of \citet{delucia07}. The initial library is expanded in two
ways: adding galaxies at randomly drawn stages of their evolution
(i.e., not only at the default redshift given by the model); varying
the SFR and metallicity of the galaxies in the last 10~Myr (defined as
{\it current} SFR in P12) before the time of observation. The effect
of these variations is similar to the addition of a recent burst or a
sudden truncation of the SFR. The stellar populations are modeled
using the latest version of the \citet{bc03} models with a
\citet{chabrier} IMF and the two-phase dust attenuation recipe (birth
clouds and ambient ISM) of \citet{cf00}.

\subsubsection{Overall results for SAM SFHs}\label{overallsams}

On average, we find that the evolution of the best-fit SAM SFHs for
compact SFGs is similar to that of a delayed $\tau$ model with a long
$e$-folding time, particularly at mid and late times
(Figure~\ref{history}). The key difference is that SAM SFHs present
more gradual increase of the SFR during the rising phase at the onset
of star formation. As a result, the best-fit stellar ages tend to be
older than those of the $\tau$ models. The typical stellar ages for
compact SFGs obtained with SAM SFHs range between $t=1.6-2.4$~Gyr,
with a median value of $t=2$~Gyr. This is a factor of $\sim$2 older
than the estimates of the $\tau$ models. Attending to these values,
their formation redshifts would increase from $z_{\mathrm{form}}=3-4$
in the $\tau$ models, to $z_{\mathrm{form}}=6-7$, suggesting that
these galaxies are nearly maximally old. The age of the Universe at
$z\sim2.5$ is 2.6~Gyr. Despite the longer duration of the increasing
SFR phase in SAM SFHs, we again find very few galaxies ($<2\%$) with
rising SFRs at the epoch of observation. The majority of compact SFGs
present, on average, either declining or roughly constant SFRs.

Figure~\ref{mediansfhs} shows the averaged best-fit SFHs for compact
SFGs in different regions of the UVJ diagram from
Figure~\ref{sample}b. In \S~\ref{shortlived} we showed that fits of
the SEDs of (low-sSFR) galaxies in region 1 with $\tau$ models
required short formation timescales ($\tau<300$~Myr). The best-fit SAM
SFHs support this result by showing that these galaxies have, on
average, a bell-shaped SFH with a FHWM of only $\sim$500~Myr (shaded
orange area). The average SFHs of compact SFGs in other regions show
longer duration. As a result, galaxies in region 1 are also younger
(1.8~Gyr) than the average of other compact SFGs (2.1~Gyr). The fact
that younger compact SFGs have lower sSFRs than the older ones (see
also Figure~\ref{ssfrtime}) implies that, in this case, age correlates
with sSFR, contrary to intuition. We note, however, that such
correlations only apply when using constant or declining SFHs, and
when comparing galaxies with the same formation timescale
($\tau$). Figure~\ref{mediansfhs} also illustrates that, although some
SAM SFHs may resemble the evolution of a long-$\tau$ delayed
model, in general, they display a broad range of trends in SFR
vs. time, presenting phases of increasing, nearly constant, or rapidly
decreasing SFR.

\section{Evolutionary tracks of compact SFGs: linking progenitors to their descendants}\label{lifepath}

In this section we study the evolutionary tracks of compact SFGs in
the SFR--M and UVJ diagrams as inferred from their best-fit SFHs.
In particular, we study the predictions of 3 different SFH models for
the slope of the SFR--M correlation at early times
($t\ll t_{\mathrm{obs}}$), and the duration of the main sequence phase
for the compact SFGs. Then, we compare the galaxy number densities
estimated from the forward extrapolation of these SFHs with the
observed number density of quiescent galaxies, and thus verify the
proposed evolutionary sequence between the 2 populations. Finally, we
study the distribution of compact SFGs in the UVJ diagram as a
function of stellar mass and extinction, and we discuss the
implications for their formation timescales and their history of dust
production and destruction.

\subsection{Evolutionary tracks in the SFR-M diagram}\label{sfrpath}

\begin{figure}[t]
\centering
\includegraphics[width=8cm,angle=0.]{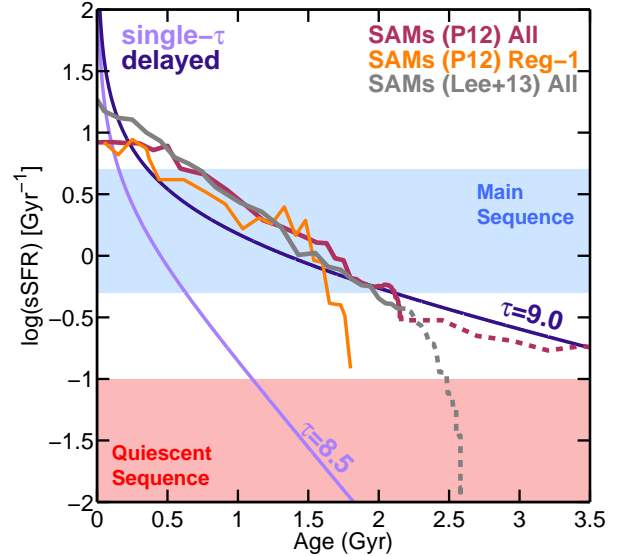}
\caption{\label{ssfrtime} Evolution of the sSFR as a function of time
  for different SFHs. The light-purple line shows a single
  $\tau$ model with a fast decline, $\tau=300$~Myr (the typical value
  for compact SFGs). The dark-purple line shows a delayed model with
  slow decline, $\tau=1$~Gyr. The solid magenta/black lines show the
  averaged, best-fit SAM SFHs for all compact SFGs using the libraries
  of \citealt{paci12} and \citet{lee13}, respectively. Both libraries
  find similar SFHs, but different forward evolution (dashed
  lines). The library of \citet{lee13}, consisting exclusively of
  simulated galaxies that are quenched by $z\sim2$, predicts shorter
  quenching times. The orange line shows the SAM SFH of the (low-sSFR)
  compact SFGs in region 1 of the UVJ (Figure~\ref{sample}b), which
  present an abrupt decay, compared to the average evolution of
  compact SFGs.}
\end{figure}

The left panel of Figure~\ref{mainseq} shows the distribution of
compact SFGs in the SFR--M diagram based on SFRs and stellar masses
(gray circles) derived from single $\tau$ models. The values derived
from delayed models are very similar and thus are not shown. However,
the predicted evolution on the diagram is not the same for each
model. The light and dark blue lines illustrate the differences in the
evolutionary tracks (from the onset of star formation until they
become quiescent) of 3 compact SFGs fitted with single and delayed
$\tau$ models, respectively. While both models present the same
exponential decline of the SFR at late ($t\gg t_{\mathrm{obs}}$)
times, the predicted slope of the SFR--M correlation at early times
($t\ll t_{\mathrm{obs}}$) is different. For delayed models, the slope,
log(SFR)$=\alpha$log(M), is $\alpha\sim0.50$, whereas for single
$\tau$ models the SFR is nearly independent of the stellar mass,
$\alpha\sim$0.  Neither of these, however, appear to follow the
steeper observed slope of the main sequence at $z\gtrsim2$ and
\lmass$<10$ ($\alpha\sim0.6$, e.g., \citealt{santini09};
\citealt{whitaker12b}) suggesting that these SFHs do not adequately
reproduce the early phases of galaxy growth.  If the SFR zero point of
the main sequence keeps increasing at $z>2$, it could explain a
flatter slope for an evolutionary track that follows the main sequence
as a function of time. However, the evolution of the SFR zero point
since $z\sim4$ to $z\sim2$ is not strong enough to reproduce the flat
evolutionary track of the single $\tau$ models
($\Delta$sSFR$\sim-0.2$~dex; \citealt{stark09}; \citealt{karim11};
\citealt{bouwens12}; \citealt{gonzalez12}). This suggest that an
increasing SFR is more appropriate to reproduce the early phases of
stellar mass growth (e.g., \citealt{maraston10}).  In that regard, the
tracks of the SAM SFHs (right panel of Figure~\ref{mainseq}) produce
the best results, following more closely the observed slope of the
main sequence at $2<z<3$ (see also \citet{paci13} for a similar result
at $z\sim1$), presenting a steeper slope ($\alpha=0.58$) than the
delayed models. This difference arises from a more gradual increase of
the SFR at early times ($t\lesssim500$~Myr) in SAM SFHs
(Figure~\ref{mediansfhs}).

\begin{figure}[t]
\centering
\includegraphics[width=7.5cm,angle=0.]{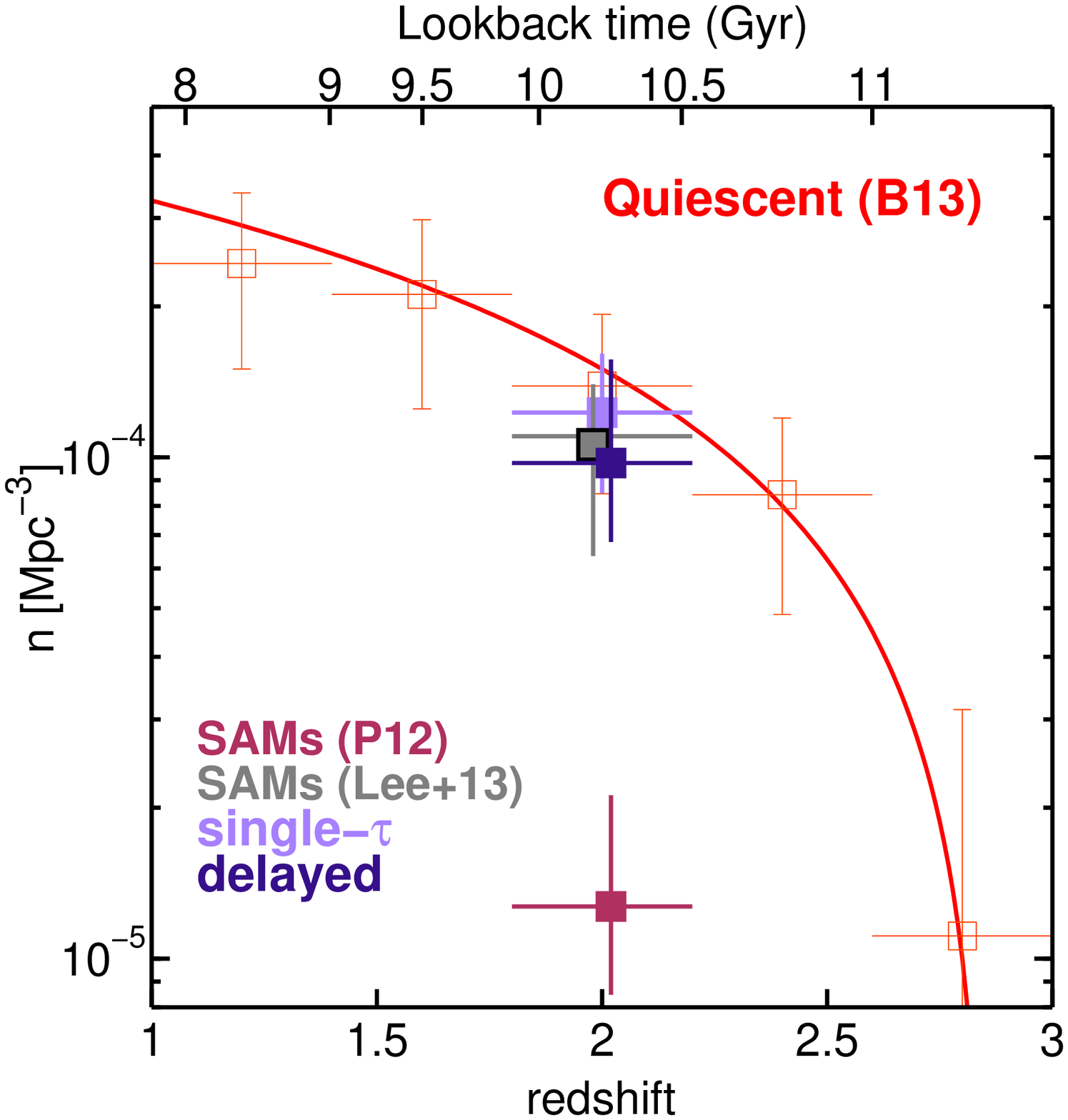}
\caption{\label{elapsed} Comparison of the observed number density of
  compact quiescent galaxies from \citet{barro13} (red squares and
  best-fit line) with the predicted number of compact SFGs that would
  be quenched by that time. We estimate this number extrapolating the
  best-fit SFHs to: single (light-purple) and delayed (dark-purple)
  $\tau$ models, and SAM SFHs from \citet[][magenta]{paci12} and
  \citet[][black]{lee13}. All the SFHs except those based on the P12
  library, which decline more slowly (dashed magenta line in
  Figure~\ref{ssfrtime}), match the observations. This implies that to
  reproduce the emergence of the quiescent population using SAM SFHs,
  compact SFGs must experience an abrupt decline of the SFR.}
\end{figure}

Figure~\ref{ssfrtime} show the evolution of sSFR versus time for the
different SFHs. The shaded regions indicate the approximate location
of the star-forming main sequence and the quiescent sequence, defined
a range in sSFR. The longer formation timescales of SAM SFHs compared
to the $\tau$ models lead to a longer duration of the main sequence
phase, $t_{\mathrm{MS}}=1.5$~Gyr (magenta/gray lines). To obtain a
similarly long main sequence phase with delayed models, would require
$e$-folding times of 1~Gyr (dark-purple). However, the majority of
compact SFGs are best fitted with much shorter timescales
($\sim$300~Myr), which result in an average duration of the main
sequence of $t_{\mathrm{MS}}<600$~Myr (light-purple line).

Not surprisingly, if we extrapolate the SFHs to estimate the quenching
times, i.e., the time since the epoch of observation until the galaxy
becomes quiescent (\lssfr$=-1$),
$t_{\mathrm{q}}=t_{\mathrm{obs}}-t_{\mathrm{quiescent}}$, we also
obtain significantly larger average values for the P12 SAM SFHs
($t_{\mathrm{q}}=2.5$~Gyr) compared to single
($t_{\mathrm{q}}=300$~Myr) and delayed ($t_{\mathrm{q}}=600$~Myr)
$\tau$ models. We note however, that this trend is strongly dependent
on the choice of the library of SAM SFHs.  The P12 library was created
for the analysis of {\it local} galaxies, and therefore favors
long-lived SFHs, similar to those of main-sequence galaxies at
$z=0$. If we build a library of SEDs including only galaxies that have
low sSFR (\lssfr$<-1$) at $z=2$, the predicted quenching times became
substantially shorter ($t_{\mathrm{q}}=400$~Myr; magenta vs. gray
dashed lines in Figure~\ref{ssfrtime}). For this purpose, we used a
slightly different SAM \citep{lee13} because it allowed a simpler
selection of the library of quenched galaxies by $z=2$. As shown by
the good agreement between the solid gray and magenta lines in
Figure~\ref{ssfrtime}, the best-fit SFHs from this library and that of
P12 are fully consistent. Only the forward evolution of the SFH
differs. Individual examples of short SFHs can be found in the P12
library. For example, the orange line shows the rapid decline in the
sSFR of compact SFGs in region 1 ($t\sim1.8$~Gyr). However, if the
galaxy presents a high current SFR, the P12 library usually favors a
long-lived forward evolution over a short one.

Although this test shows that the extrapolated SFHs have limited
predictive power (i.e., the results depend on the model choice), we
can test if, assuming any of the previous SFHs, it is possible to
reproduce the emergence of the quiescent population in terms of
quenched compact SFGs . In B13 we showed that the number densities of
these two populations are in good agreement for quenching times
between $t_{\mathrm{q}}=300-800$~Myr (see also
\citealt{williams13}). Figure~\ref{elapsed} shows that the single and
delayed $\tau$ models predict number densities that are also
consistent, or slightly lower, than the observed value. For SAM SFHs
however there is a strong dichotomy depending on which template
library we use. While the default (long-lived) templates under-predict
the number density by more than an order of magnitude, the short-lived
templates are in good agreement with the observations.  If, as argued
above, SAM SFHs are better at describing the evolutionary paths of
compact SFGs, this implies that, to reproduce the emergence of the
quiescent population, these galaxies would have to end their lives
with a sharp truncation of the SFR on a short timescale compared to
their average age (t$_{\mathrm{q}}\sim400$~Myr over $t=2$~Gyr; see
also \citet{stefanon13} for a similar argument).  This could indicate
the action of a strong quenching mechanism triggered (or enhanced) by
the transformation from the extended to the compact phase.

\subsection{Evolutionary paths in the UVJ diagram}\label{uvjevol}

\begin{figure}[t]
\centering
\includegraphics[width=8cm,angle=0.]{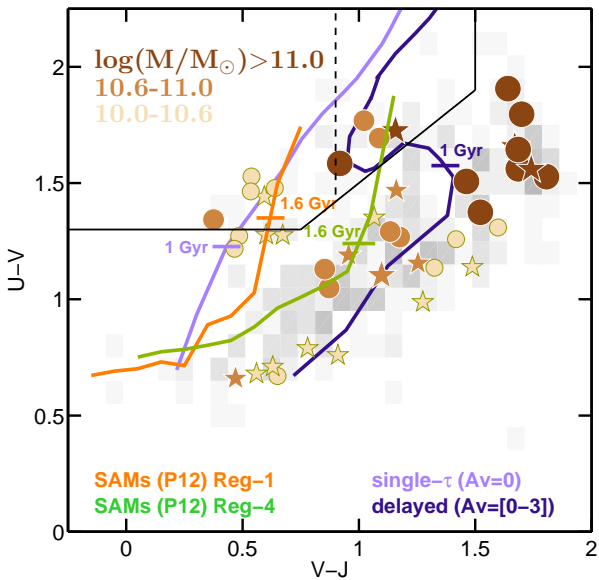}
\caption{\label{uvjmass} Distribution of the compact SFGs in the UVJ
  diagram for different bins of stellar mass. The boxed gray scale
  show other massive (\lmass$>10$) SFGs at the same redshift. The
  low-sSFR, low-extinction galaxies in region 1 (see
  Figure~\ref{sample}b) have lower stellar masses than the majority of
  compact SFGs. This suggest that there are different tracks to the
  quiescent region with different dust formation histories. To
  illustrate this idea, the light-purple line shows the color-track of
  a dust-free single $\tau$ model, while the dark-purple line shows a
  delayed model with variable dust-extinction modeled after the SFH
  (ranging from Av=0 to 2, and then 0.5 at \lssfr$=-1$). Similarly,
  the orange (green) line shows the SAM SFH color-track of a compact
  SFG in region 1 (region 4) of the UVJ. For the SAM SFHs we also
  model the evolution of the dust-extinction after the SFH.}
\end{figure}

Figure~\ref{uvjmass} shows the UVJ diagram for compact SFGs
color-coded by stellar mass. As discussed in \S~\ref{uvjprop}, we find
a strong correlation between stellar mass and extinction ($\propto
V-J$; see also \citealt{wuyts11b}; \citealt{brammer11}). We also find
that nearly all compact SFGs in region 1 belong in the lowest stellar
mass bin (\lmass$=10-10.6$). As shown in the previous sections, these
galaxies appear to have shorter formation timescales and younger ages
than other compact SFGs. Recent studies of the stellar populations of
quiescent galaxies at $z\sim2$ report a similar correlation between
age and stellar mass (\citealt{whitaker12,whitaker13};
\citealt{newman13}). In those cases, however, the older galaxies are
also more quiescent (i.e., with lower sSFRs and redder colors),
whereas for compact SFGs, this correlation is reversed. The more
massive (older) galaxies are forming stars more actively than the
lower-mass ones.  This result suggests that at least some low-mass
compact SFGs arrive onto the red sequence before the more massive
ones, thus populating faster the intermediate-to-low mass end of the
quiescent stellar mass function (e.g., \citealt{tomczak13}).

We speculate that the age difference leads to different evolutionary
tracks for compact SFGs in the UVJ diagram depending on their stellar
mass.  The more massive, long-lived SFGs undergo active star formation
for a longer duration, thus gradually increasing the amount of dust
(produced in supernovae and late phases of stellar evolution) and
reaching higher levels of extinction.  On the contrary, low-mass
compact SFGs have shorter formation timescales and thus exhibit a
smaller range of extinctions before shutting down star
formation. Regardless of the final stellar mass, however, the amount
of dust must decrease prior to quenching to be able to match the low
extinction observed in quiescent galaxies.

To illustrate these possibilities, Figure~\ref{uvjmass} shows the
evolutionary tracks for different SFHs with variable levels of dust
extinction. For simplicity, we model the evolution of the extinction
after the SFHs (i.e., the \av$\propto$SFR). This parametrization
reproduces, qualitatively, the observed correlation between SFR (mass)
and obscuration, and leads to lower \av after quenching. Although this
model is not exhaustive (see e.g., \citealt{zahid13} for detailed
modeling based on similar hypothesis) it provides a simple way to
study plausible evolutionary tracks in the UVJ diagram. The
light-purple line shows the simplest case of a single $\tau$ model
with no extinction. The dark-purple line shows a delayed model where
the dust-extinction ranges from \av$=0$ to 3~mag. The orange and green
lines show the color tracks based on the best-fit SAM SFHs for a
compact SFG in region 1 and another one in region 4. In this case, we
normalize the extinction to match the best-fit \av of the galaxies at
the time of observation. The difference between these tracks is that
the low-mass galaxy (orange) has a shorter star-forming phase, and
thus reaches lower extinction levels and bluer V$-$J colors than the
more massive galaxy (green). Each evolutionary track indicates the
point where the galaxies are 1.6~Gyr old. At that age, the galaxy in
region 1 is nearly quiescent, and un-obscured, while the galaxy in
region 4 is still star-forming and has an \av$=1.6$~mag. For
comparison, the evolutionary tracks for single and delayed $\tau$
models with $\tau=300$~Myr reach the quiescent region earlier at ages
of $\sim$1~Gyr.

Overall, the evolutionary tracks reproduce well the observed location
in the UVJ diagram of the compact SFGs and their quiescent
descendants. The only exceptions are the most extinguished galaxies in
region 5, whose location on the UVJ diagram is, at least partially,
the result of inclination effects (see \S~\ref{uvjprop}). We note also
that a diminishing extinction level following the shutdown of star
formation in the most massive galaxies is consistent with the
observations of faint IR detections in UVJ-quiescent galaxies reported
in previous works (\citealt{pg08b}; \citealt{brammer11}). Similarly
recent spectroscopic observations indicate that recently quenched
galaxies span a broad range of rest-frame U-V colors, suggesting they
can indeed arrive on the red sequence through a more dusty track
(\citealt{bezanson13}).

\begin{figure*}[t]
\centering
\includegraphics[width=8.3cm,angle=0.]{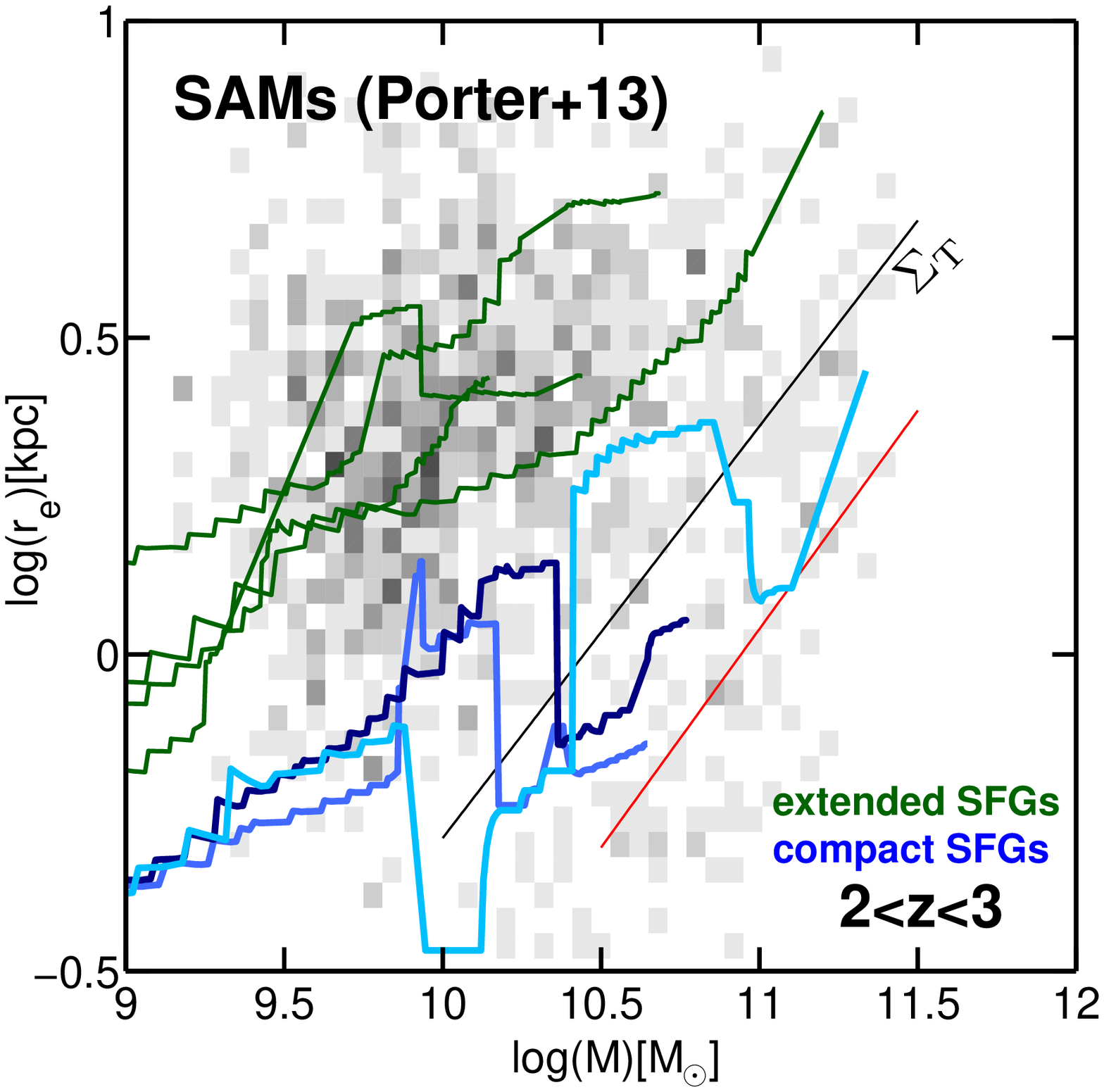}
\includegraphics[width=8.3cm,angle=0.]{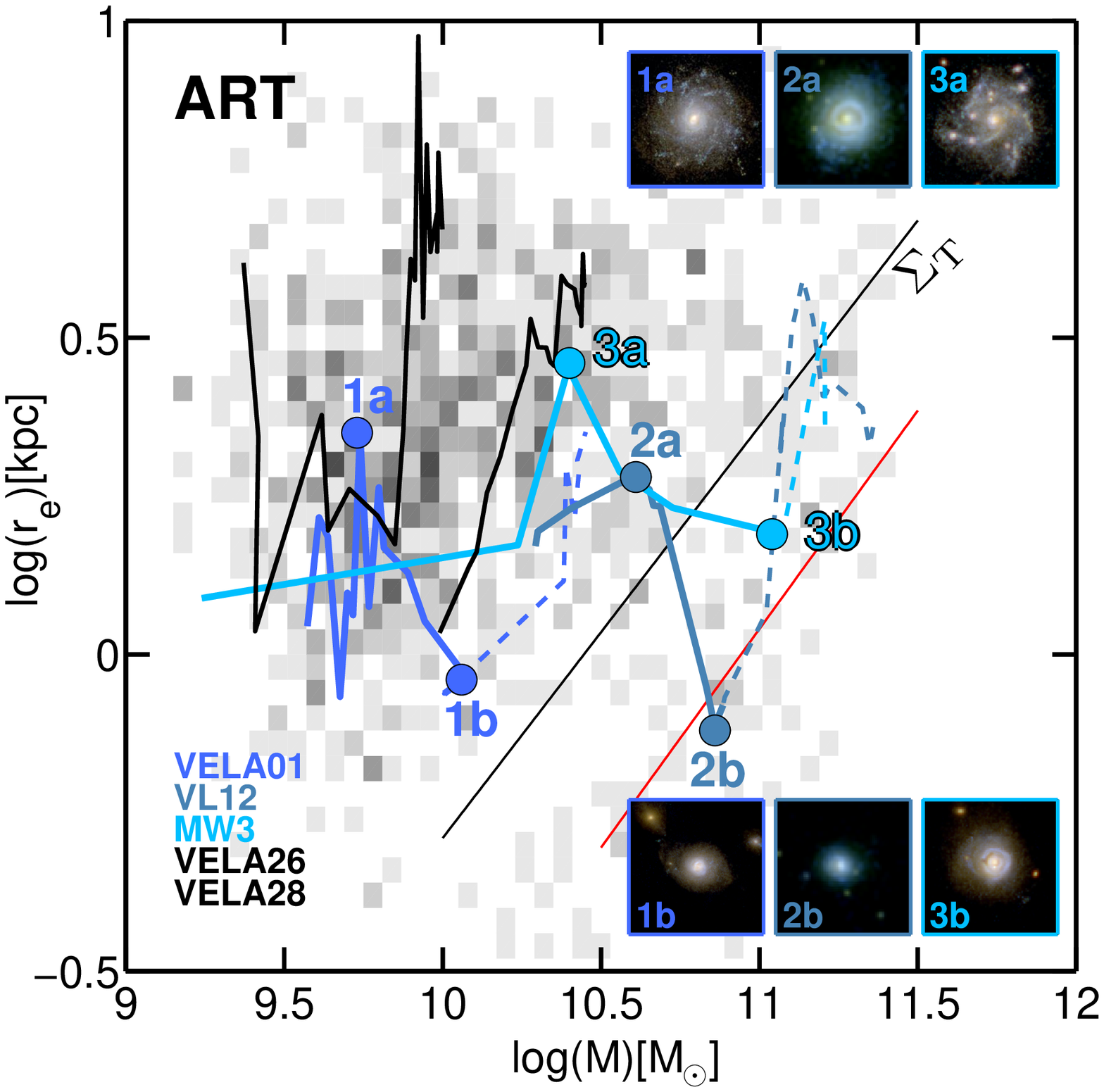}
\caption{\label{simulplot} Comparison of the observed galaxy
  distribution in the mass-size diagram at $2<z<3$ (boxed gray scale)
  with the predictions from theoretical models (colored tracks). The
  black line indicates the selection criteria,
  log($\Sigma_{T}/M_{\odot}\mathrm{kpc}^{1.5}$=10.3),for compact
  galaxies (below the line), and the red line the average mass-size
  relation for quiescent galaxies at $z\sim2$ \citep{newman12}.  {\it
    Left panel:} The green lines show the evolutionary tracks (since
  the onset of star formation until \lssfr$<-1$) of 3 non-compact SFGs
  in the SAMs of Porter et al. (2013a). The evolution of these
  galaxies is consistent with the mass-size correlation for the bulk
  of SFGs. The blue tracks show the evolution of 3 galaxies selected
  to satisfy the compact SFG criterion at $2<z<3$. All these galaxies
  become compact as a result of DIs that cause a sharp contraction in
  the $r_{e}$ by a factor of $\sim2$. {\it Right panel:} The blue
  lines show the evolutionary tracks of 3 galaxies drawn from the
  sample of hydrodynamic simulations of \citet{ceverino12}. All these
  galaxies experience a significant shrinkage due to DIs. For
  comparison, the black lines show 2 other galaxies from that sample
  that have fluctuations in size, but maintain an overall size-growing
  trend. The 5\arcsec$\times$5\arcsec~kpc stamps illustrate the visual
  appearance of these galaxies at the stages of maximum $r_{e}$
  (disk-like and clumpy; 1a,2a,3a) and minimum $r_{e}$ (compact
  spheroid; 1b,2b,3b).}
\end{figure*}

\section{The formation of compact SFGs}\label{theory}

The remarkably small sizes of compact SFGs is among the strongest
evidence in favor of their evolutionary connection to the quiescent
population at $z\sim2$. In B13 we speculated that such small sizes
could be the result of strongly dissipational processes that reduce
the effective radius of SFGs with more extended light profiles.  Gas
rich major mergers (\citealt{springel05b}; \citealt{robertson06}) or
disk instabilities (DI; \citealt{dekel09b}; \citealt{ceverino10})
triggered by strong processes of gas accretion from the halo
(\citealt{keres05}; \citealt{dekel09a}) are plausible
mechanisms. Recent results pointing out the paucity of major mergers
at $z=2-3$ (\citealt{williams11}) and the low incidence of merging
signatures on compact SFGs and quenched galaxies
(\citealt{kaviraj13b}) seem to favor internal, self-regulated
mechanisms like DI.  However, direct evidence of the mechanisms
responsible for the formation of compact SFGs remains to be found.

Cosmological simulations of galaxy formation provide a perfect
framework to test different scenarios and make predictions about which
galaxies are likely to experience significant structural
transformations. In this section we present the results of 2 sets of
theoretical models, namely: the recent SAMs of Porter et al. (2013a)
and a set of hydrodynamical simulations described in
\citet{ceverino10}. Note that the SAMs discussed in this section are
different from those used in \S~\ref{samtau}. The latter were
preferred for the analysis of the stellar populations, as they were
described and exhaustively tested for that purpose in
\citet{paci12,paci13}.  However, the SAMs of Porter et
al. (2013a) include new physical prescriptions critical for the
formation of compact SFGs, and thus are more suitable for the study of
structural evolution. We are now working to incorporate the new SAMs
into the code used for stellar population analysis.

\subsection{Compact SFGs in the SAMs of Porter et al. (2013)}

A full description of the semi-analytic recipes and the accuracy of
the models to reproduce observational trends at low and high redshifts
is presented in Porter et al. (2013abc). Briefly, these SAMs are an
extension of those presented in \citet{somerville08,somerville12},
which included prescriptions for radiative cooling, star formation,
supernova and black hole feedback as well as chemical enrichment and
galaxy mergers. A novel feature of the new version is the addition of
dissipational processes, critical for the formation of spheroidal
galaxies. In particular, these SAMs include a new treatment of
gravitational perturbations that allows the formation of star-forming
clumps and the triggering of disk instabilities. These result in the
inward migration of gas and clumps to the center, making the galaxy
more compact (\citealt{ceverino12}; \citealt{dekel13a}). DIs are
modeled using the \citet{toomre} instability criterion. If a galaxy
exceeds the critical threshold, a fraction of the total stellar mass
is removed from the disk and added to the central (bulge) component,
thereby reducing the half-light (and half-mass) radius. Although a
detailed characterization of the DI requires high-resolution
simulations (see next section), the Toomre criterion allows one to
quickly test a plausible formation scenario for compact SFGs which is
consistent with the observational picture of unstable clumpy disks
(\citealt{elmegreen05}; \citealt{genzel08}; \citealt{fs09}).

The left panel of Figure~\ref{simulplot} shows the predicted
evolutionary tracks in the mass-size diagram for 3 simulated galaxies
that satisfy the compact SFG selection criteria at $z=2-3$ (blue
lines). For comparison, we also show the tracks for a group of
non-compact SFGs at the same redshift (green lines). All compact SFGs
experience a significant shrinkage in effective radius of a factor of
$\sim$2, increasing the mass surface density beyond the compactness
threshold (dashed line). According to Porter et al. (2013c), all
galaxies in the compact region are the result of a dissipational
processes, either DIs (60\%) or major mergers (40\%). Overall, the
remnants of DIs are 10\% more compact than those of mergers, and thus
are more efficient at populating the region of highest $\Sigma$.  In
the SAMs, the contraction enhances the strength of the quenching
mechanisms responsible for shutting down star formation. In
particular, most compact SFGs quench on timescales of a few hundred
Myrs as a result of quasar mode feedback.  These numbers are
consistent with the quenching timescales estimated in
\S~\ref{sfrpath}.

An interesting prediction of the models is that the precursors of
compact SFGs are usually among the smallest galaxies for a given
stellar mass, even before the dissipational process, i.e., they are on
the lower envelope of the mass-size relation for SFGs
(Figure~\ref{simulplot}). They also present the highest sSFRs at a
given mass, which combined with their small sizes, makes them more
susceptible to the Toomre instability.

\subsection{Compact SFGs in Hydrodynamic simulations}

Hydrodynamic simulations provide a high-resolution view of the
processes shaping galaxy structure (gas inflow, DIs, galaxy
interactions, etc.), that is unavailable in the SAMs.  Unfortunately,
such detailed simulations require large computational efforts, which
are only possible for small samples of $\sim$30--50~galaxies.  Here we
focus on a handful of galaxies drawn from the larger sample of
\citet{ceverino10,ceverino13} and \citet{dekel13a}, computed with the
Adaptive Refinement Tree (ART) code (\citealt{arm}) using spatial
resolutions ranging between 17--75~pc (see \citet{ceverino09} for more
details about the code).  To ensure a fair comparison between models
and observations, the structural properties of the simulated galaxies
are measured using GALFIT on images processed to emulate the
properties of the CANDELS data.  This includes degrading the
high-resolution simulations down to 0.06''/px, convolving with the PSF
of the F160W image, and adding random noise to emulate the SNR of the
real data (Mozena et al. 2013, in prep.).

The right panel of Figure~\ref{simulplot} shows the evolutionary
tracks of the simulated galaxies in the mass--size diagram.  The blue
lines depict the only 3 galaxies in the sample that present (at some
point) a decrease in effective radii larger than a factor of 2, over a
period of time of at least 300~Myr (i.e., we reject rapid fluctuations
that last only for 1 or 2 simulation time steps). The shrinkage is
large enough to be significant, and it is similar to the decrease in
$r_{e}$ experienced by compact SFGs in the SAMs. For comparison, the
black lines illustrate the evolution of other simulated galaxies
showing milder size fluctuations. Interestingly, none of the 3 compact
galaxies experienced a major merger.  In all cases, the shrinkage
appears to be the result of DIs. As a result, the star-forming clumps
(clearly seen in the high-resolution images of Figure~\ref{simulplot};
Moody et al. 2013, in prep.), along with large amounts of gas, migrate
inward due to dynamical friction and tidal torques generated by the
unstable disk(e.g., \citealt{dekel09b}; \citealt{bournaud11}). In the
center, the gas is turned into stars very efficiently causing a
significant increase in the inner stellar mass ($\sim$0.3--0.5~dex) at
the same time that the size of the galaxy decreases. In the compact
stage, the galaxies are more centrally concentrated and show no
evidence of star-forming clumps. In terms of their light profiles,
during and after the contraction, we find that the S\'ersic index
increases from low ($n=1-2$) to high ($n>3$), spheroid-like values
(see also Ceverino et al. 2013; Zolotov et al. 2013, in prep.). The
whole transformation process from extended, clumpy disk to compact
spheroid occurs on timescales of a few hundred Myr, i.e., of the order
of the dynamical timescale of the disks (\citealt{ceverino10};
\citealt{cacciato12}).  This is consistent with the scenario proposed
in B13 in which new compact SFGs are being formed at a similar pace at
which the existing ones are turning quiescent.

Based on analytic calculations, \citet{dekel13b} showed that the
process of disk contraction due to DIs is expected to happen for
nearly half of the massive SFGs at $z\gtrsim2$. Such a high fraction
results from a combination of high SFRs and continuous gas accretion
common at these redshifts.  The ratio of the timescales of these
processes, $w=t_{\mathrm{SF}}/t_{\mathrm{infall}}$ ({\it wetness}
parameter), controls the DI process. For $w>1$ the galaxy triggers a
dissipative wet inflow, becoming more compact and centrally
concentrated, whereas for lower values it forms the new stars in an
inside-out growing disk. The duration of the DI phase determines if
the galaxy shrinks enough to become a compact SFGs or if it grows in
size again after the contraction (i.e., the $w<1$ mode). The answer to
this question depends on the process that regulate star formation
(internal quenching) and gas infall (halo quenching). For example, the
version of the simulations used in this paper does not include the
effects of radiation feedback, either from AGN or star formation, and
thus galaxies are unlikely to experience a severe suppression of SFR
\citep{dekel13a} after the shrinkage. In fact, all of them tend to
re-grow an extended disk component. This is illustrated by the dashed
lines in Figure~\ref{simulplot}, which show a rapid increase in
effective radii of the compact SFGs shortly after the contraction. The
relevance of these hydrodynamical simulations for compact SFG
formation is thus the physics and duration of the event, not the total
evolutionary picture.

\section{Summary}
\noindent
1.~We analyze the SFRs, structural properties, and stellar populations of
45 massive (log$(M/M_{\odot})>10$) compact SFGs at $2<z<3$ in GOODS-S
to extend the results of \citet{barro13} and present further evidence
that these galaxies are the natural progenitors of the compact,
quiescent galaxies at $z\sim2$.

\begin{itemize}[leftmargin=*]
\item Compact SFGs present heavily-obscured star formation based on
  their {\it Spitzer}/{\it Herschel} far-IR colors (71\% and 44\%/13\%
  are detected in MIPS and PACS/SPIRE). As a group, they exhibit a
  higher fraction of X-ray detected AGN (47\%) than more extended SFGs
  (12\%) or quiescent galaxies (9\%) at the same redshift and stellar
  mass. Such a high fraction implies that compact SFGs are typically
  in an active phase of black hole growth.
\item The distribution of compact SFGs in the SFR--M diagram
  extends from the {\it normal} main sequence (65\%) to nearly the
  region occupied by quenched galaxies (30\%). This result is
  consistent with the notion that these galaxies are in transit to
  become compact, quiescent galaxies. Only 2/45 compact SFGs lie above
  the SFR--M main sequence and they present far-IR SEDs typical of a
  {\it starburst} galaxy. Such a small fraction suggests that either
  the transition from star-forming to quiescent does not require an
  abrupt burst of star formation, or the {\it starburst} duty cycle is
  very short, $\sim$5\% the duration of the main-sequence phase.
\item The transiting nature of compact SFGs is further supported by
  their location in the UVJ diagram, where most compact SFGs lie
  within the dusty, reddened region occupied by other massive SFGs,
  while those with lower sSFRs scatter off towards the quiescent
  region.
\item The radii, Sersic indices, axial ratios, and spheroidal
  morphologies of compact SFGs match well to those of the quiescent
  galaxies and indicate that they can directly evolve into the red
  population simply by shutting down their star formation.
\item The structural properties of compact SFGs are strikingly
  different from those of non-compact SFGs, which have disk-like
  morphologies and usually clumpy or irregular appearance.  Under the
  assumption that galaxies grow both in stellar mass and size during
  the star-forming phase, the progenitors of massive
  (log$(M/M_{\odot})>10$), compact SFGs are probably among these
  extended SFGs. This suggests the need for a transformation mechanism
  to link the 2 populations that shrinks the size and transforms the
  structure of extended galaxies from disk to spheroid (see point 3.).
\end{itemize}

\noindent
2.~We study the stellar populations of compact SFGs using 2 sets of
SFHs: 1) exponentially-declining (single and delayed) $\tau$ models,
and 2) a library of physically-motivated SFHs extracted from
SAMs. Qualitatively, both sets of SFHs produce similar results and
trends; however,they differ in their quantitative predictions.

\begin{itemize}[leftmargin=*]
\item SAM SFHs predict longer formation timescales and older ages,
  $t=2^{+0.4}_{-0.2}$~Gyr ($z_{\mathrm{form}}=6-7$), which are nearly
  a factor of 2 older than the estimates of single,
  $t=0.9^{+0.2}_{-0.5}$~Gyr, and delayed, $t=1.1^{+0.2}_{-0.6}$~Gyr,
  $\tau$ models ($z_{\mathrm{form}}=3-4$).
\item Both sets of SFHs provide good SED fits. However, the predicted
  evolutionary tracks in the SFR--M diagram from SAM SFHs provide a
  better description of how compact SFGs evolve as a function of time,
  tracking more closely the observed slope of the main sequence and by
  remaining on it for $\sim$70\% of their lifetime.
\item We find that some low-mass, compact SFGs (\lmass$=10-10.6$) have
  younger ages and undergo active star formation for a shorter
  duration than more massive SFGs, suggesting that they arrive earlier
  onto the red sequence. Thus, we speculate that compact SFGs follow
  different evolutionary tracks on the UVJ diagram depending on
  their stellar mass. These tracks differ in the amount of dust
  extinction, whose amount vary according to the SFR: more massive
  galaxies increase their dust content gradually during a longer
  star-forming phase, thus reaching higher levels of extinction,
  whereas lower mass galaxies have shorter formation timescales and
  thus evolve (faster) on a lower extinction track.

\item Both sets of SFHs are able to reproduce the number density of
  compact quiescent galaxies at $z=2$ from compact SFGs quenching on
  timescales of $t_{\mathrm{q}}=300-600$~Myr. In $\tau$ models, these
  quenching times are the result of a rapid assembly,
  $\tau\sim300$~Myr, whereas SAM SFHs predict a gradual assembly
  followed by an abrupt decay, $4\times$ shorter than the duration of
  the star-forming main sequence phase ($t_{\mathrm{MS}}=1.5$~Gyr).
\end{itemize}

\noindent
3.~To analyze the proposed evolutionary sequence from extended to compact
SFGs, we study the formation mechanisms of compact SFGs in theoretical
models of galaxy formation.
\begin{itemize}[leftmargin=*]
\item The recent SAMs of Porter et al. (2013a) suggest that compact
  SFGs form only in strongly dissipational processes, such as major
  mergers or disk instabilities (DIs). However, DIs are more frequent
  (60\% vs. 40\%) and form the majority of the most compact remnants.
\item Compact SFGs formed in DIs are among the smallest SFGs of a
  given stellar mass before the contraction, and, after the DI, they
  experience a contraction in $r_{\mathrm{e}}$ by a factor of 2.
\item We verify these SAM predictions by using the high-resolution
  hydrodynamic simulations of \citet{ceverino12} where we find 3
  representative examples of compact SFGs formed in DIs. These
  galaxies experience a similar size shrinkage while changing their
  morphology from clumpy disk to centrally concentrated spheroid. This
  contraction is also in good agreement with recent predictions from
  \citet{dekel13b} and Zolotov et al. (2013) for the of effect DIs.

\end{itemize}

\section*{Acknowledgments}

We thank David Elbaz for very useful discussions. Support for Program
number HST-GO-12060 was provided by NASA through a grant from the
Space Telescope Science Institute, which is operated by the
Association of Universities for Research in Astronomy, Incorporated,
under NASA contract NAS5-26555. GB acknowledges support from NSF grant
AST-08-08133. PGP-G acknowledges support from grant AYA2012-31277-E.
This work has made use of the Rainbow Cosmological Surveys Database,
which is operated by the Universidad Complutense de Madrid (UCM),
partnered with the University of California Observatories at Santa
Cruz (UCO/Lick,UCSC). 

%Table with properties.
%\placetable{redshifts}
%\begin{landscape}
\begin{deluxetable*}{cccccccccccccccc}
%\tablecolumns{8}
%\rotate{90}
\setlength{\tabcolsep}{0.008in} 
\tablewidth{0pt}
\tabletypesize{\scriptsize}
\tablecaption{\label{redshifts} Properties of compact SFGs}
\tablehead{
\colhead{ID}  & \colhead{R.A.} & \colhead{DEC} & \colhead{$z_{\mathrm{best}}$} & \colhead{$z$-REF} &
\colhead{X-ray} & \colhead{$f_{24\mu m}$} & \colhead{$f_{100\mu m}$}  & \colhead{$f_{250\mu m}$}  &
\colhead{SFR}   &  \colhead{Mass}  &\colhead{r$_{eff}$} &
\colhead{U-V}   &  \colhead{V-J} &  \colhead{Reg} & \colhead{G141}\\
\colhead{(1)} & \colhead{(2)}  & \colhead{(3)} & \colhead{(4)}   & \colhead{(5)}   & 
\colhead{(6)}  & \colhead{(7)}  & \colhead{(8)}   & \colhead{(9)}  &
\colhead{(10)} & \colhead{(11)} & \colhead{(12)} &
\colhead{(13)} & \colhead{(14)}  & \colhead{(15)} & \colhead{(16)}}
\startdata
     21937 & 53.00658860 & -27.72416860 &  2.726 &    9 & $86^{-}$ &    62 &     - &     - &    17 & 10.21 &  0.29 &  0.68 &  0.56 & 3 & A \\ 
     14781 & 53.03332780 & -27.78257480 &  2.619 &    5 & $137^{-}$ &    81 &  2206 &     - &   192 & 10.60 &  1.00 &  1.14 &  1.49 & 4 & A(\OII) \\ 
     23382 & 53.16229880 & -27.71213490 &  2.433 &  1,2 &   534 &    80 &     - &    2175 &    84 & 11.27 &  1.77 &  1.73 &  1.16 & 2 & A \\ 
     21662 & 53.05885220 & -27.72632930 &  2.180 &    - &     - &   155 &  1838 &     - &   341 & 11.22 &  2.70 &  1.56 &  1.69 & 5 & B \\ 
     22069 & 53.10207860 & -27.72256120 &  2.610 &    - &     - &     - &     - &     - &     3 & 10.45 &  0.33 &  1.53 &  0.54 & 1 & B \\ 
     22539 & 53.18736680 & -27.71918680 &  2.315 &  1,3 &     - &     - &     - &     - &    10 & 10.88 &  0.43 &  1.34 &  0.38 & 1 & A \\ 
     23896 & 53.10081420 & -27.71598590 &  2.303 &  1,2 &   326 &    49 &     - &     - &    32 & 10.87 &  1.40 &  1.19 &  0.96 & 4 & A \\ 
     25998 & 53.13757210 & -27.70010390 &  2.453 &  1,2 &     - &   140 &  3359 &     - &   365 & 10.90 &  0.95 &  1.27 &  1.18 & 4 & A \\ 
     26056 & 53.06325870 & -27.69964260 &  2.402 &  1,9 &   215 &   109 &     - &     - &    97 & 10.75 &  1.13 &  1.16 &  1.25 & 4 & B \\ 
     25879 & 53.03444580 & -27.69821010 &  2.474 &    8 & $138^{-}$ &    37 &     - &     - &    19 & 10.54 &  0.47 &  1.10 &  1.10 & 4 & B \\ 
     19298 & 53.01265220 & -27.74724370 &  2.573 &    9 &    93 &    12 &     - &     - &    25 & 10.76 &  0.49 &  1.28 &  0.67 & 1 & B \\ 
     20659 & 53.18283880 & -27.73491130 &  2.432 & 1,2,4 &     - &    72 &     - &     - &    94 & 10.98 &  0.52 &  1.05 &  0.87 & 4 & A \\ 
      9834 & 53.14882690 & -27.82112070 &  2.576 &    7 & $490^{+}$ &   588 &  5731 & 20669 &   391 & 11.06 &  2.43 &  1.42 &  1.16 & 4 & B \\ 
      4150 & 53.05557910 & -27.87400810 &  2.560 &    3 &     - &    78 &     - &     - &   160 & 11.28 &  2.57 &  1.58 &  0.92 & 2 & - \\ 
      9290 & 53.18622010 & -27.82519980 &  2.030 &    - &     - &     - &     - &     - &     6 & 10.51 &  0.54 &  1.27 &  0.49 & 1 & A \\ 
      1883 & 53.16977920 & -27.90078740 &  2.673 &    4 &     - &    33 &     - &     - &    86 & 10.37 &  0.83 &  0.67 &  0.65 & 3 & C \\ 
      2644 & 53.16450380 & -27.89038860 &  2.123 &    4 &   544 &   323 &     - &     - &    62 & 10.83 &  0.82 &  0.66 &  0.47 & 3 & A \\ 
       536 & 53.08917740 & -27.93046510 &  2.611 &    7 &   294 &    31 &     - &     - &    38 & 10.32 &  0.41 &  0.79 &  0.78 & 3 & B \\ 
       580 & 53.08732350 & -27.92954880 &  2.680 &    - &   290 &    47 &     - &     - &    22 & 11.03 &  0.82 &  1.44 &  0.59 & 1 & C \\ 
     18475 & 53.10810690 & -27.75397980 &  2.728 &  1,6 &   359 &     - &     - &     - &    20 & 10.44 &  0.33 &  1.28 &  0.60 & 1 & C \\ 
      3643 & 53.07600440 & -27.87816150 &  2.793 &    8 &   254 &     - &     - &     - &    13 & 10.32 &  0.33 &  0.71 &  0.63 & 3 & B \\ 
     20790 & 53.17444810 & -27.73329980 &  2.576 &    2 &   564 &     7 &     - &     - &    93 & 10.20 &  0.38 &  0.76 &  0.91 & 3 & B \\ 
      1086 & 53.13717980 & -27.91583650 &  2.570 &    - &     - &     - &     - &     - &     4 & 10.37 &  0.42 &  1.46 &  0.54 & 1 & B \\ 
     19143 & 53.02794410 & -27.74866340 &  2.300 &    - &   123 &     - &     - &     - &    12 & 10.57 &  1.16 &  1.35 &  1.07 & 4 & B \\ 
     26659 & 53.08446670 & -27.70418600 &  2.510 &    - &     - &     - &     - &     - &     8 & 10.71 &  1.53 &  1.48 &  0.64 & 1 & B \\ 
     11701 & 53.09402810 & -27.80412630 &  2.560 &    - & $310^{+}$ &   147 &   792 &     - &   442 & 11.16 &  1.27 &  1.66 &  1.67 & 5 & B \\ 
     15614 & 53.14889600 & -27.77750460 &  2.070 &    - &     - &    64 &  1193 &     - &    35 & 10.20 &  0.32 &  1.14 &  1.32 & 4 & A(\OIII) \\ 
     26231 & 53.06500180 & -27.70001300 &  2.500 &    - &     - &     - &     - &     - &    12 & 10.24 &  0.45 &  1.22 &  0.47 & 1 & B \\ 
     26211 & 53.06595180 & -27.70185220 &  2.110 &    - &     - &   153 &  1738 &     - &   135 & 10.81 &  1.12 &  1.69 &  1.09 & 2 & C \\ 
     25952 & 53.12113620 & -27.69807510 &  1.970 &    1 &     - &    91 &     - &     - &    57 & 10.63 &  0.74 &  1.13 &  0.85 & 4 & A \\ 
      3280 & 53.06061510 & -27.88237230 &  2.150 &    - &     - &   228 &  2724 &  7101 &   167 & 10.82 &  0.61 &  1.29 &  1.13 & 4 & - \\ 
     26612 & 53.07743450 & -27.70465270 &  2.080 &    - &     - &    40 &     - &     - &    30 & 10.76 &  0.75 &  1.31 &  1.60 & 4 & B \\ 
     22603 & 53.10701590 & -27.71822560 &  2.291 &  1,5 & $351^{+}$ &   554 &  2323 &  7729 &   554 & 11.10 &  2.65 &  1.54 &  1.47 & 4 & B \\ 
     15432 & 53.14614830 & -27.77988200 &  2.640 &    - &   482 &    48 &     - &     - &    65 & 10.85 &  1.16 &  1.47 &  1.16 & 4 & B \\ 
     10973 & 53.18582930 & -27.80996560 &  2.583 &    4 &   593 &    47 &     - &     - &   316 & 10.76 &  0.39 &  0.99 &  1.27 & 4 & B \\ 
      7670 & 53.14817910 & -27.83916220 &  2.150 &    - &     - &   110 &  1755 &  7616 &   115 & 11.22 &  2.87 &  1.53 &  1.81 & 5 & C \\ 
     24367 & 53.14392700 & -27.67773850 &  2.420 &    - &   475 &    78 &  1290 &     - &   207 & 11.20 &  2.67 &  1.56 &  1.74 & 5 & B \\ 
     18562 & 53.02739130 & -27.75388610 &  2.040 &    - &     - &    67 &     - &     - &    34 & 11.28 &  2.97 &  1.91 &  1.64 & 5 & B \\ 
     14876 & 53.11879020 & -27.78281820 &  2.309 &  1,2 &     - &   207 &  7068 & 28531 &   365 & 10.79 &  1.83 &  1.26 &  1.42 & 4 & B \\ 
     22883 & 53.14216250 & -27.70742850 &  2.150 &    - &     - &   176 &     - &     - &    75 & 10.82 &  1.77 &  1.77 &  1.02 & 2 & B \\ 
     22200 & 53.05424860 & -27.72164870 &  2.307 &    1 &     - &   174 &  1549 &  8467 &   197 & 11.06 &  1.89 &  1.64 &  1.68 & 5 & B \\ 
     21901 & 53.12859620 & -27.72429960 &  2.020 &    - &     - &    91 &  1278 &     - &    47 & 11.28 &  2.77 &  1.80 &  1.70 & 5 & B(\OIII) \\ 
     26788 & 53.09951770 & -27.70616110 &  2.260 &    - &     - &   179 &  1094 &     - &   117 & 11.08 &  2.17 &  1.37 &  1.52 & 4 & B(\OIII) \\ 
      7579 & 53.14454840 & -27.83969560 &  2.050 &    - &     - &   128 &     - &     - &    96 & 11.14 &  2.64 &  1.51 &  1.46 & 4 & B \\ 
     15377 & 53.02228390 & -27.77890030 &  2.650 &    - & $111^{+}$ &   178 &  1531 &     - &   108 & 11.10 &  1.79 &  1.63 &  1.22 & 2 & - \\ 
\enddata                                                                    
\tablecomments{\\ 
(1) General ID in Guo et al. (2013).\\
(2,3) R.A and Declination J2000.\\
(4) Photometric or spectroscopic redshift.\\
(5) Spectroscopic redshift from different references: 1) Barro, in prep ({\it Keck} MOSFIRE), 2) Kurk et al. (2013) [GMASS-2578,2443,2467,1989,2043], 3)
Kriek et al. (2008), 4) Ballestra et al. (2010), 5) Silverman et al. 2010, 6) Wuyts et al. (2009), 7) Vanzella et al. (2008),
8) Stern et al in prep., 9) Szokoly et al. (2004).\\
(6) X-ray ID in Xue et al. (2011). The superscript index indicate degree of contamination on the SED from AGN emission based on the comparison to SED-fits 
that include AGN templates (See \S~\ref{xrayprop}). {\it +:} Severe. {\it -:} Mild.\\
(7,8,9) Far-IR fluxes in {\it Spitzer}/MIPS~24~$\mu$m and {\it Herschel}/PACS~100~$\mu$m and SPIRE~250~$\mu$m.\\
(10) Total star formation rate (SFR$_{UV+IR}$ [M$_{\odot}$yr$^{-1}$]), see \S~\ref{galprop}.\\
(11) Stellar mass (\lmass) determined from SED fitting using \citet{bc03}, see \S~\ref{galprop}.\\
(12) Circularized, effective (half-light) radius (kpc) measured with GALFIT, see \S~\ref{galprop}.\\
(13,14) Rest-frame colors estimated from the best-fit stellar template using {\it EAZY}.\\ 
(15) Location of the galaxy in the regions of the $UVJ$ diagram indicated in Figure 2.\\
(16) Qualitative flag for the G141 spectra: A) Highest signal-to-noise spectra, showing a continuum break 
and (occasionally) absorption lines, or a emission line (\OII or \OIII/\Hb, indicated); B) Low signal-to-noise, absent continuum break, weak emission lines;
C) Significant flux contamination or truncated spectrum.\\
\\}                                                                         
\end{deluxetable*}                                                          
%\end{landscape}     

%\clearpage

\appendix
\section{Testing AGN contamination in the SEDs}\label{agncont}

\begin{figure}
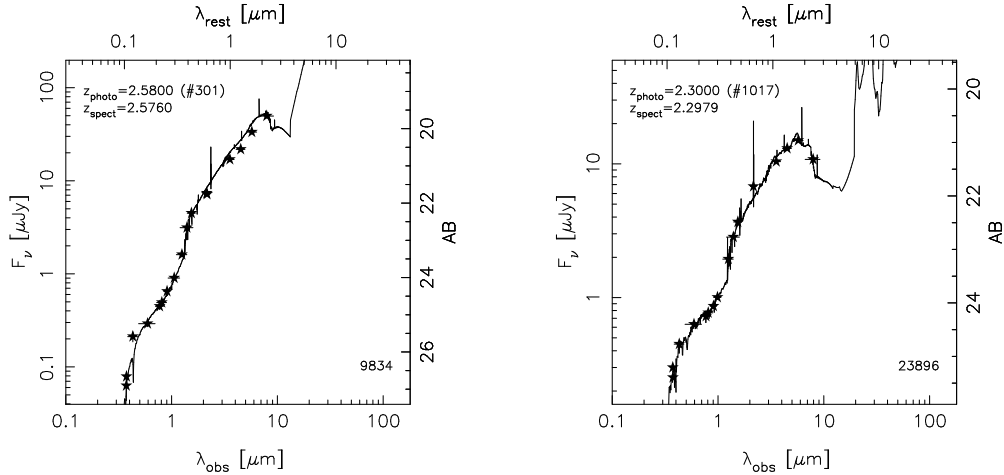

\centering
\includegraphics[width=6.5cm,angle=-90.,bb=80 30 591 507]{fig14a.eps}
\hspace{1cm}
\includegraphics[width=6.5cm,angle=-90.,bb=80 30 591 507]{fig14b.eps}
\caption{\label{agnsed} SEDs of compact SFGs hosting an AGN, and their
  best-fit optical-to-NIR templates. {\it Left panel:} Fitting the SED
  requires a hybrid model combining an stellar component (60\%) plus
  an AGN component (40\%). The AGN contribution is more prominent in
  the rest-frame NIR (the 1.6~$\mu$m bump is slightly shifted
  redward), but the hybrid template suggests that there is also some
  contribution of the AGN to the rest-frame UV luminosity (Hsu et
  al. 2013, in prep., will present the analysis of the SEDs of the
  AGNs in GOODS-S using hybrid templates). {\it Right panel:} The SED
  is well fitted with a pure stellar component template.}
\end{figure}

We fit the SEDs of the 21/45 X-ray detected compact SFGs galaxies with
the AGN templates of \citet{salvato09, salvato11} to identify a
possible AGN contamination in the stellar SED. We find that 4
galaxies, best-fitted with Type I AGN templates, present strong
continuum contamination from the non-stellar component and IRAC
power-law slopes. Another 3 are fitted with composite Type 2/stellar
templates suggesting that a minor AGN contamination could be
possible. To test this effect we compare the stellar masses of the 21
galaxies derived from stellar templates with those of
\citet{santini12b} calculated using a combination of stellar and
nuclear \citep{silva04} components. In agreement with the previous
analysis, we find a significant offset in the stellar masses
($\Delta$M$_{\star}$=0.3~dex) of the 4 aforementioned Type I AGNs,
while for the other 3 galaxies the contamination is small, and the
stellar masses differ by less than 0.1~dex. The left panel of
Figure~\ref{agnsed} illustrates the SED of a compact SFG with a
significant ($\sim$40\%) AGN contribution to the SED. The right panel
shows the SED of another X-ray detected compact SFG whose SED is well
fitted by a pure stellar component. Based on the results of the SED
fitting to hybrid stellar+AGN templates, we exclude the 4 strongly
contaminated galaxies (indicated in Table~\ref{redshifts}) from the
analysis in Sections~\ref{props} and \ref{sedmodels}.

\section{SFR$_{\mathrm{IR+UV}}$ vs. SFR-model consistency}\label{sfrcompare}

In the following we assess the agreement between \sfrtot and SFR-model
for different SFHs. While \sfrtot is, arguably, the best SFR indicator
for heavily obscured galaxies, SFR-model is tied to all other stellar
properties (age, quenching time, etc.), and thus a fully consistent
analysis of the galaxy properties requires a good match between both
SFR indicators.

\subsection{Constrained $\tau$ models}

As shown in \citet{wuyts11a}, the motivation for constraining the
$e$-folding time to be larger than $\tau=300$~Myr (i.e., rejecting
quasi-instantaneous bursts) is to obtain a better agreement between
\sfrtot and SFR-model. Otherwise, the SED fits tend to favor solutions
with shorter formation timescales, which lead to SFR-model
systematically lower than \sfrtot.

\begin{figure}
\centering
\includegraphics[width=16cm,angle=0.]{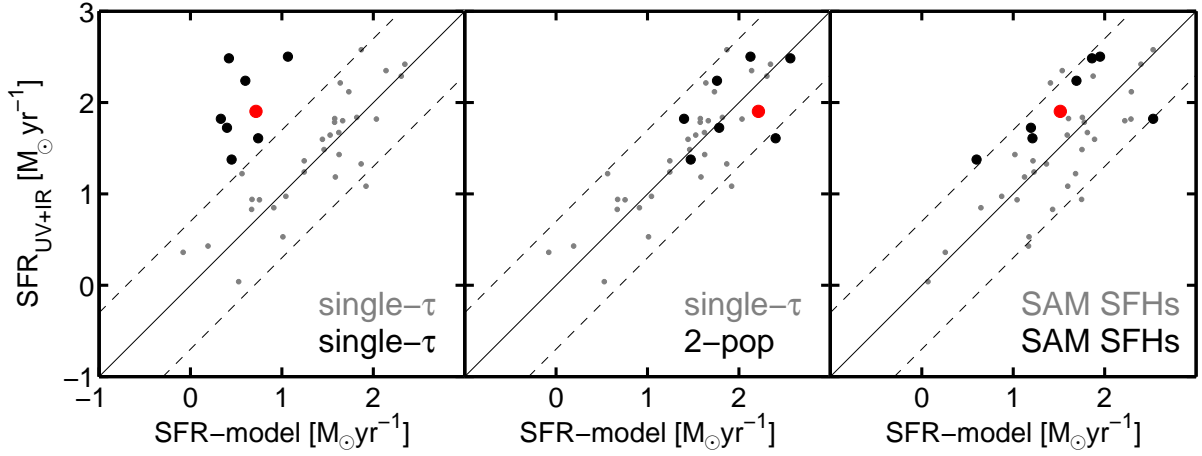}
\caption{\label{sfrtest} Comparison between SFR$_{\mathrm{IR+UV}}$ and
  the SFR-model for compact SFGs. Each panel illustrates the
  comparison for SFR-model derived from different SFHs. {\it Left
    panel:} Single $\tau$ models; the overall agreement between SFR
  estimates is good, but there are some outliers (black) for which
  SFR-model$\ll$\sfrtot. These appear to be either extremely obscured
  galaxies with high SFRs or galaxies with lower SFRs for which the
  last (obscured) burst of star formation is shutting down. In both
  cases the bets-fit $\tau$ model is too old (low sSFR) to match the
  observed \sfrtot. {\it Central-panel:} 2-population $\tau$ models;
  Using a composite stellar population consisting of a relatively old
  underlying population plus a young, obscured burst provides a better
  agreement between the two independent SFR indicators.  {\it Right
    panel:} SAM SFHs; using these models, the agreement in the SFRs is
  better than for single $\tau$ models and similar to the 2-population
  case. For most of these galaxies the SAM SFHs features a burst of
  star formation prior to the epoch of observation, similarly to the
  2-population $\tau$ SFHs. The red marker highlights one of the
  outliers in region 2 of the UVJ; Figure~\ref{region5} shows the
  best-fit templates for this galaxy based on different SFHs.}
\end{figure}

The left panel of Figure~\ref{sfrtest} shows the comparison of \sfrtot
and SFR-model for compact SFGs based on constrained single $\tau$
models. Overall, the agreement is quite good, and the scatter is
consistent with the typical uncertainties in SFRs
($rms$[SFR]$\sim$0.39~dex; e.g., \citealt{daddi07}). There are,
however, a group of galaxies ($\sim20\%$) which present significantly
lower values of SFR-model compared to \sfrtot
($\Delta$SFR$~>1$~dex). These discrepancies have been reported in
previous works, where the authors attributed to cases of high dust
obscuration in which reddening saturates as extinction indicator
(\citealt{santini09}; \citealt{reddy10}). In such cases, the
\citet{calzetti} attenuation law fails to provide a realistic
approximation, and the additional reddening is compensated by the
models aging the stellar populations, i.e., they are fitted with old
stellar populations instead of young and dusty. We find the same
systematic outliers if we use delayed models instead of single
$\tau$. These outliers are preferentially located in regions 2 (dusty
+ intermediate age) and 5 (highly obscured) of the UVJ diagram
(Figure~\ref{sample_size}b).

The case of galaxies in region 5 ($V-J>$1.5) has been discussed in
detail in \citet{patel12}, where the authors show that, in many cases,
the high dust obscuration is largely driven by inclination effects. In
agreement with this result, we find that compact SFGs in region 5 have
small S\'ersic indices and lower axis ratios, characteristic of
edge-on disks (see \S~\ref{props}). The outliers in region 2 are mainly
IR-detected galaxies located near the quiescent region of the UVJ
diagram. Similar cases have been reported in other works (e.g.,
\citealt{brammer11}; \citealt{patel13}), which suggest that the IR
luminosity could be powered by either an obscured AGN, or the final
stages of star formation in quenching galaxy. In our sample, 1 out of
the 4 galaxies is an X-ray detected AGN, but neither that galaxy or
any of the other 3 present present IRAC power-law SEDs or {\it
  Spitzer}/{\it Herschel} colors indicative of obscured AGNs activity.
The outliers are always among the largest ($r_{\mathrm{e}}>1.5$~kpc),
most massive compact SFGs, and their rest-frame UV emission appears to
confined in a small central region. This could be signaling the fading
of a nuclear burst, or perhaps that optical and IR emission are
associated with physically distinct regions of the galaxy (e.g., an
obscured surrounding component).

In order to reconcile the \sfrtot and SFR-model for the outliers, we
first tried to use a different extinction law, namely \citet{cf00},
which allows for a steeper (less gray) wavelength dependence than
Calzetti, and thus increases the attenuation in the rest-frame
UV. However, the best-fit ages obtained with approach were only
marginally younger, and thus the values of SFR-model were still
severely under-predicted. As an alternative approach to account for
missing star formation, we modeled the SED of the outliers using two
single $\tau$ models, that represent a intermediate-age, moderate
extinction, underlying population, plus a young recent burst with
younger age and higher obscuration (see e.g., Bell et al. 2003,
\citealt{pg08}). In addition, we introduced the far-IR fluxes as a
prior, by requiring that the absorbed UV-luminosity of the young burst
matched the IR-emission from the heated dust within a factor of
$\sim$2 (see e.g., \citealt{dacunha08,dacunha10} for a similar
approach). As shown in the central panel of Figure~\ref{sfrtest}, the
2-population approach provides a better agreement between \sfrtot and
SFR-model for the outliers, and therefore we adopted the best-fit
stellar properties and SFRs estimated with this method for the
analysis in \S~\ref{lifepath}. Nevertheless, the most significant
difference affected the SFR-model ($\Delta$SFR of up to 1.5~dex),
whereas the masses and stellar ages (of the older component) changed
by $\Delta M\sim0.15$~dex and $\Delta t\sim0.2$~dex, respectively.

The success of the 2-population approach to reconcile the apparent
inconsistency between optical and far-IR SEDs could indicate that, at
least for some galaxies, the SFHs are not as smooth as single $\tau$
models. In fact, in the hierarchical picture of galaxy formation, the
SFHs are likely to be more stochastic due to discrete accretion
events. The addition of a second population to the SFH provides a
first order approximation to reflect at least the most recent of such
events.

\begin{figure}
\centering
\includegraphics[width=6.1cm,angle=90.,bb=577 500 101 34]{fig16a.eps}
\includegraphics[width=6.1cm,angle=90.,bb=577 500 101 34]{fig16b.eps}
\includegraphics[width=5.6cm,angle=0.]{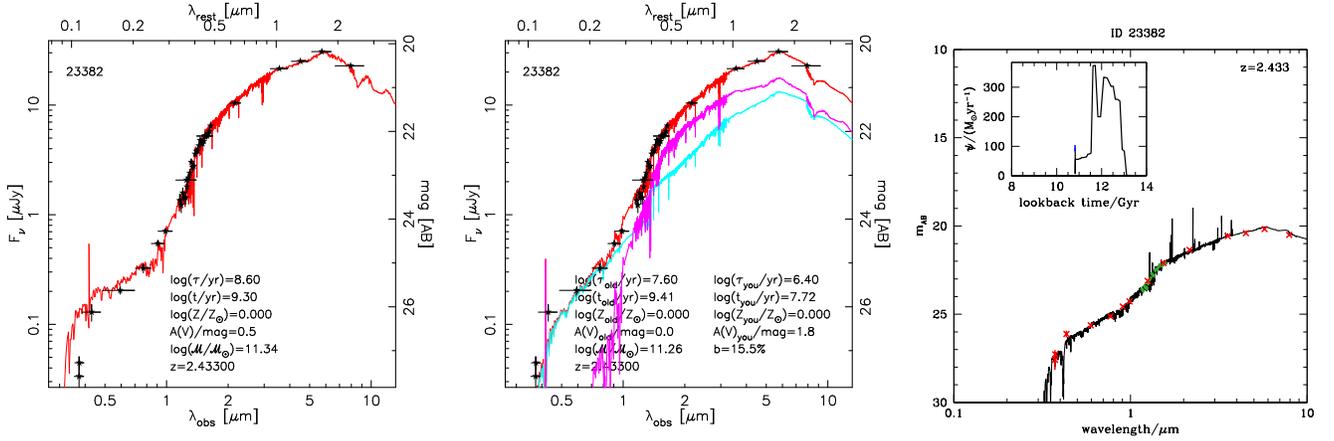}
\caption{\label{region5} Best-fit stellar templates an inferred
  stellar properties for one of the outliers in the \sfrtot
  vs. SFR-model comparison ($\Delta SFR>0.7$~dex), marked in red in
  Figure~\ref{sfrtest}. {\it Left panel:} The best-fit BC03 model based
  on a single $\tau$ SFH suggests that the galaxy in nearly quiescent
  (\lssfr=-1), and moderately obscured, which is inconsistent with the
  high SFR values derived from the IR-luminosity, based on MIPS and
  {\it Herschel} detections. {\it Central-panel:} Using 2 single
  $\tau$ models provides a better match between SFR-model and \sfrtot
  from a composite stellar population of an old, low-attenuation
  component (magenta) and a young, obscured component (cyan) with
  high-SFR that would be responsible for the observed
  IR-emission. {\it Right panel:} Best-fit stellar template based on
  the SAM SFHs of \citet{paci12}. According to this SFH, the galaxy
  was at the peak of his SFR 1~Gyr before the epoch of
  observation. Now the SFR is declining slowly, which provides a good
  match to \sfrtot without any other additional requirements.}
\end{figure}

\subsection{SAMs}

Interestingly, using SAM SFHs, which include such stochastically, the
agreement between \sfrtot and SFR-model is better than for single
$\tau$ models. Although the overall scatter is slightly higher than in
the single $\tau$ models ($rms$[SFR]$\sim$0.43~dex), there are fewer
systematic outliers (right panel of Figure~\ref{sfrtest}). Attending
to the SFHs of the outliers, we find that, in most cases ($\sim$60\%),
the good agreement between \sfrtot and SFR-model is the result of a
short burst of star formation in the last 10~Myr (right panel of
Figure~\ref{region5}), similarly to the what we obtained using the 2
population approach. For the rest of the outliers, the values of
SFR-model are higher because their overall SFHs present a longer
duration of the actively star-forming phase and a slower decline of
the SFR (i.e., a shallower slope) than the $\tau$ models (see e.g.,
Figure~\ref{history} or Figure~\ref{mediansfhs}). Consequently, the
SFR remains high for a longer period of time.

\bibliographystyle{aa}
\bibliography{../../referencias}

\begin{thebibliography}{179}
\expandafter\ifx\csname natexlab\endcsname\relax\def\natexlab#1{#1}\fi

\bibitem[{{Ashby} {et~al.}(2013){Ashby}, {Willner}, {Fazio}, {Huang}, {Arendt},
  {Barmby}, {Barro}, {Bell}, {Bouwens}, {Cattaneo}, {Croton}, {Dav{\'e}},
  {Dunlop}, {Egami}, {Faber}, {Finlator}, {Grogin}, {Guhathakurta},
  {Hernquist}, {Hora}, {Illingworth}, {Kashlinsky}, {Koekemoer}, {Koo},
  {Labb{\'e}}, {Li}, {Lin}, {Moseley}, {Nandra}, {Newman}, {Noeske}, {Ouchi},
  {Peth}, {Rigopoulou}, {Robertson}, {Sarajedini}, {Simard}, {Smith}, {Wang},
  {Wechsler}, {Weiner}, {Wilson}, {Wuyts}, {Yamada}, \& {Yan}}]{ashby13}
{Ashby}, M.~L.~N., {Willner}, S.~P., {Fazio}, G.~G., {et~al.} 2013, \apj, 769,
  80

\bibitem[{{Baldry} {et~al.}(2004){Baldry}, {Glazebrook}, {Brinkmann},
  {Ivezi{\'c}}, {Lupton}, {Nichol}, \& {Szalay}}]{baldry04}
{Baldry}, I.~K., {Glazebrook}, K., {Brinkmann}, J., {et~al.} 2004, \apj, 600,
  681

\bibitem[{{Barro} {et~al.}(2013){Barro}, {Faber}, {P{\'e}rez-Gonz{\'a}lez},
  {Koo}, {Williams}, {Kocevski}, {Trump}, {Mozena}, {McGrath}, {van der Wel},
  {Wuyts}, {Bell}, {Croton}, {Ceverino}, {Dekel}, {Ashby}, {Cheung},
  {Ferguson}, {Fontana}, {Fang}, {Giavalisco}, {Grogin}, {Guo}, {Hathi},
  {Hopkins}, {Huang}, {Koekemoer}, {Kartaltepe}, {Lee}, {Newman}, {Porter},
  {Primack}, {Ryan}, {Rosario}, {Somerville}, {Salvato}, \& {Hsu}}]{barro13}
{Barro}, G., {Faber}, S.~M., {P{\'e}rez-Gonz{\'a}lez}, P.~G., {et~al.} 2013,
  \apj, 765, 104

\bibitem[{{Barro} {et~al.}(2011){Barro}, {P{\'e}rez-Gonz{\'a}lez}, {Gallego},
  {Ashby}, {Kajisawa}, {Miyazaki}, {Villar}, {Yamada}, \&
  {Zamorano}}]{barro11b}
{Barro}, G., {P{\'e}rez-Gonz{\'a}lez}, P.~G., {Gallego}, J., {et~al.} 2011,
  \apjs, 193, 30

\bibitem[{{Bauer} {et~al.}(2011){Bauer}, {Conselice}, {P{\'e}rez-Gonz{\'a}lez},
  {Gr{\"u}tzbauch}, {Bluck}, {Buitrago}, \& {Mortlock}}]{bauer11}
{Bauer}, A.~E., {Conselice}, C.~J., {P{\'e}rez-Gonz{\'a}lez}, P.~G., {et~al.}
  2011, \mnras, 417, 289

\bibitem[{{Bedregal} {et~al.}(2013){Bedregal}, {Scarlata}, {Henry}, {Atek},
  {Rafelski}, {Teplitz}, {Dominguez}, {Siana}, {Colbert}, {Malkan}, {Ross},
  {Martin}, {Dressler}, {Bridge}, {Hathi}, {Masters}, {McCarthy}, \&
  {Rutkowski}}]{bedregal13}
{Bedregal}, A.~G., {Scarlata}, C., {Henry}, A.~L., {et~al.} 2013, ArXiv
  e-prints

\bibitem[{{Bell} {et~al.}(2005){Bell}, {Papovich}, {Wolf}, {Le Floc'h},
  {Caldwell}, {Barden}, {Egami}, {McIntosh}, {Meisenheimer},
  {P{\'e}rez-Gonz{\'a}lez}, {Rieke}, {Rieke}, {Rigby}, \& {Rix}}]{bell05}
{Bell}, E.~F., {Papovich}, C., {Wolf}, C., {et~al.} 2005, \apj, 625, 23

\bibitem[{{Bell} {et~al.}(2012){Bell}, {van der Wel}, {Papovich}, {Kocevski},
  {Lotz}, {McIntosh}, {Kartaltepe}, {Faber}, {Ferguson}, {Koekemoer}, {Grogin},
  {Wuyts}, {Cheung}, {Conselice}, {Dekel}, {Dunlop}, {Giavalisco},
  {Herrington}, {Koo}, {McGrath}, {de Mello}, {Rix}, {Robaina}, \&
  {Williams}}]{bell12}
{Bell}, E.~F., {van der Wel}, A., {Papovich}, C., {et~al.} 2012, \apj, 753, 167

\bibitem[{{Bezanson} {et~al.}(2013){Bezanson}, {van Dokkum}, {van de Sande},
  {Franx}, \& {Kriek}}]{bezanson13}
{Bezanson}, R., {van Dokkum}, P., {van de Sande}, J., {Franx}, M., \& {Kriek},
  M. 2013, \apjl, 764, L8

\bibitem[{{Bezanson} {et~al.}(2009){Bezanson}, {van Dokkum}, {Tal},
  {Marchesini}, {Kriek}, {Franx}, \& {Coppi}}]{bezanson09}
{Bezanson}, R., {van Dokkum}, P.~G., {Tal}, T., {et~al.} 2009, \apj, 697, 1290

\bibitem[{{Birnboim} \& {Dekel}(2003)}]{haloquench}
{Birnboim}, Y. \& {Dekel}, A. 2003, \mnras, 345, 349

\bibitem[{{Blain} {et~al.}(2002){Blain}, {Smail}, {Ivison}, {Kneib}, \&
  {Frayer}}]{blain02}
{Blain}, A.~W., {Smail}, I., {Ivison}, R.~J., {Kneib}, J.-P., \& {Frayer},
  D.~T. 2002, \physrep, 369, 111

\bibitem[{{Bournaud} {et~al.}(2011){Bournaud}, {Dekel}, {Teyssier}, {Cacciato},
  {Daddi}, {Juneau}, \& {Shankar}}]{bournaud11}
{Bournaud}, F., {Dekel}, A., {Teyssier}, R., {et~al.} 2011, \apjl, 741, L33

\bibitem[{{Bournaud} {et~al.}(2007){Bournaud}, {Jog}, \& {Combes}}]{bournaud07}
{Bournaud}, F., {Jog}, C.~J., \& {Combes}, F. 2007, \aap, 476, 1179

\bibitem[{{Bouwens} {et~al.}(2012){Bouwens}, {Illingworth}, {Oesch}, {Franx},
  {Labb{\'e}}, {Trenti}, {van Dokkum}, {Carollo}, {Gonz{\'a}lez}, {Smit}, \&
  {Magee}}]{bouwens12}
{Bouwens}, R.~J., {Illingworth}, G.~D., {Oesch}, P.~A., {et~al.} 2012, \apj,
  754, 83

\bibitem[{{Brammer} {et~al.}(2008){Brammer}, {van Dokkum}, \& {Coppi}}]{eazy}
{Brammer}, G.~B., {van Dokkum}, P.~G., \& {Coppi}, P. 2008, \apj, 686, 1503

\bibitem[{{Brammer} {et~al.}(2012){Brammer}, {van Dokkum}, {Franx},
  {Fumagalli}, {Patel}, {Rix}, {Skelton}, {Kriek}, {Nelson}, {Schmidt},
  {Bezanson}, {da Cunha}, {Erb}, {Fan}, {F{\"o}rster Schreiber}, {Illingworth},
  {Labb{\'e}}, {Leja}, {Lundgren}, {Magee}, {Marchesini}, {McCarthy},
  {Momcheva}, {Muzzin}, {Quadri}, {Steidel}, {Tal}, {Wake}, {Whitaker}, \&
  {Williams}}]{3dhst}
{Brammer}, G.~B., {van Dokkum}, P.~G., {Franx}, M., {et~al.} 2012, \apjs, 200,
  13

\bibitem[{{Brammer} {et~al.}(2011){Brammer}, {Whitaker}, {van Dokkum},
  {Marchesini}, {Franx}, {Kriek}, {Labb{\'e}}, {Lee}, {Muzzin}, {Quadri},
  {Rudnick}, \& {Williams}}]{brammer11}
{Brammer}, G.~B., {Whitaker}, K.~E., {van Dokkum}, P.~G., {et~al.} 2011, \apj,
  739, 24

\bibitem[{{Bruce} {et~al.}(2012){Bruce}, {Dunlop}, {Cirasuolo}, {McLure},
  {Targett}, {Bell}, {Croton}, {Dekel}, {Faber}, {Ferguson}, {Grogin},
  {Kocevski}, {Koekemoer}, {Koo}, {Lai}, {Lotz}, {McGrath}, {Newman}, \& {van
  der Wel}}]{bruce12}
{Bruce}, V.~A., {Dunlop}, J.~S., {Cirasuolo}, M., {et~al.} 2012, ArXiv e-prints

\bibitem[{{Bruzual} \& {Charlot}(2003)}]{bc03}
{Bruzual}, G. \& {Charlot}, S. 2003, \mnras, 344, 1000

\bibitem[{{Buitrago} {et~al.}(2008){Buitrago}, {Trujillo}, {Conselice},
  {Bouwens}, {Dickinson}, \& {Yan}}]{buitrago08}
{Buitrago}, F., {Trujillo}, I., {Conselice}, C.~J., {et~al.} 2008, ArXiv
  e-prints

\bibitem[{{Cacciato} {et~al.}(2012){Cacciato}, {Dekel}, \&
  {Genel}}]{cacciato12}
{Cacciato}, M., {Dekel}, A., \& {Genel}, S. 2012, \mnras, 421, 818

\bibitem[{{Calzetti} {et~al.}(2000){Calzetti}, {Armus}, {Bohlin}, {Kinney},
  {Koornneef}, \& {Storchi-Bergmann}}]{calzetti}
{Calzetti}, D., {Armus}, L., {Bohlin}, R.~C., {et~al.} 2000, \apj, 533, 682

\bibitem[{{Caputi}(2013)}]{caputi13}
{Caputi}, K.~I. 2013, \apj, 768, 103

\bibitem[{{Carollo} {et~al.}(2013){Carollo}, {Bschorr}, {Renzini}, {Lilly},
  {Capak}, {Cibinel}, {Ilbert}, {Onodera}, {Scoville}, {Cameron}, {Mobasher},
  {Sanders}, \& {Taniguchi}}]{carollo13}
{Carollo}, C.~M., {Bschorr}, T.~J., {Renzini}, A., {et~al.} 2013, \apj, 773,
  112

\bibitem[{{Cassata} {et~al.}(2011){Cassata}, {Giavalisco}, {Guo}, {Renzini},
  {Ferguson}, {Koekemoer}, {Salimbeni}, {Scarlata}, {Grogin}, {Conselice},
  {Dahlen}, {Lotz}, {Dickinson}, \& {Lin}}]{cassata11}
{Cassata}, P., {Giavalisco}, M., {Guo}, Y., {et~al.} 2011, \apj, 743, 96

\bibitem[{{Cassata} {et~al.}(2013){Cassata}, {Giavalisco}, {Williams}, {Guo},
  {Lee}, {Renzini}, {Ferguson}, {Faber}, {Barro}, {McIntosh}, {Lu}, {Bell},
  {Koo}, {Papovich}, {Ryan}, {Conselice}, {Grogin}, {Koekemoer}, \&
  {Hathi}}]{cassata13}
{Cassata}, P., {Giavalisco}, M., {Williams}, C.~C., {et~al.} 2013, \apj, 775,
  106

\bibitem[{{Ceverino} {et~al.}(2010){Ceverino}, {Dekel}, \&
  {Bournaud}}]{ceverino10}
{Ceverino}, D., {Dekel}, A., \& {Bournaud}, F. 2010, \mnras, 404, 2151

\bibitem[{{Ceverino} {et~al.}(2012){Ceverino}, {Dekel}, {Mandelker},
  {Bournaud}, {Burkert}, {Genzel}, \& {Primack}}]{ceverino12}
{Ceverino}, D., {Dekel}, A., {Mandelker}, N., {et~al.} 2012, \mnras, 420, 3490

\bibitem[{{Ceverino} \& {Klypin}(2009)}]{ceverino09}
{Ceverino}, D. \& {Klypin}, A. 2009, \apj, 695, 292

\bibitem[{{Ceverino} {et~al.}(2013){Ceverino}, {Klypin}, {Klimek},
  {Trujillo-Gomez}, {Churchill}, {Primack}, \& {Dekel}}]{ceverino13}
{Ceverino}, D., {Klypin}, A., {Klimek}, E., {et~al.} 2013, ArXiv e-prints

\bibitem[{{Chabrier}(2003)}]{chabrier}
{Chabrier}, G. 2003, \pasp, 115, 763

\bibitem[{{Chang} {et~al.}(2013){Chang}, {van der Wel}, {Rix}, {Holden},
  {Bell}, {McGrath}, {Wuyts}, {H{\"a}u{\ss}ler}, {Barden}, {Faber}, {Mozena},
  {Ferguson}, {Guo}, {Galametz}, {Grogin}, {Kocevski}, {Koekemoer}, {Dekel},
  {Huang}, {Hathi}, \& {Donley}}]{chang13}
{Chang}, Y.-Y., {van der Wel}, A., {Rix}, H.-W., {et~al.} 2013, ArXiv e-prints

\bibitem[{{Charlot} \& {Fall}(2000)}]{cf00}
{Charlot}, S. \& {Fall}, S.~M. 2000, \apj, 539, 718

\bibitem[{{Chary} \& {Elbaz}(2001)}]{ce01}
{Chary}, R. \& {Elbaz}, D. 2001, \apj, 556, 562

\bibitem[{{Cheung} {et~al.}(2012){Cheung}, {Faber}, {Koo}, {Dutton}, {Simard},
  {McGrath}, {Huang}, {Bell}, {Dekel}, {Fang}, {Salim}, {Barro}, {Bundy},
  {Coil}, {Cooper}, {Conselice}, {Davis}, {Dominguez}, {Kassin}, {Kocevski},
  {Koekemoer}, {Lin}, {Lotz}, {Newman}, {Phillips}, {Rosario}, {Weiner}, \&
  {Willmer}}]{cheung12}
{Cheung}, E., {Faber}, S.~M., {Koo}, D.~C., {et~al.} 2012, ArXiv e-prints

\bibitem[{{Cisternas} {et~al.}(2011){Cisternas}, {Jahnke}, {Inskip},
  {Kartaltepe}, {Koekemoer}, {Lisker}, {Robaina}, {Scodeggio}, {Sheth},
  {Trump}, {Andrae}, {Miyaji}, {Lusso}, {Brusa}, {Capak}, {Cappelluti},
  {Civano}, {Ilbert}, {Impey}, {Leauthaud}, {Lilly}, {Salvato}, {Scoville}, \&
  {Taniguchi}}]{cisternas11a}
{Cisternas}, M., {Jahnke}, K., {Inskip}, K.~J., {et~al.} 2011, \apj, 726, 57

\bibitem[{{Curtis-Lake} {et~al.}(2012){Curtis-Lake}, {McLure}, {Pearce},
  {Dunlop}, {Cirasuolo}, {Stark}, {Almaini}, {Bradshaw}, {Chuter}, {Foucaud},
  \& {Hartley}}]{curtislake12}
{Curtis-Lake}, E., {McLure}, R.~J., {Pearce}, H.~J., {et~al.} 2012, \mnras,
  422, 1425

\bibitem[{{da Cunha} {et~al.}(2008){da Cunha}, {Charlot}, \&
  {Elbaz}}]{dacunha08}
{da Cunha}, E., {Charlot}, S., \& {Elbaz}, D. 2008, \mnras, 388, 1595

\bibitem[{{da Cunha} {et~al.}(2010){da Cunha}, {Charmandaris},
  {D{\'{\i}}az-Santos}, {Armus}, {Marshall}, \& {Elbaz}}]{dacunha10}
{da Cunha}, E., {Charmandaris}, V., {D{\'{\i}}az-Santos}, T., {et~al.} 2010,
  \aap, 523, A78

\bibitem[{{Daddi} {et~al.}(2007{\natexlab{a}}){Daddi}, {Alexander},
  {Dickinson}, {Gilli}, {Renzini}, {Elbaz}, {Cimatti}, {Chary}, {Frayer},
  {Bauer}, {Brandt}, {Giavalisco}, {Grogin}, {Huynh}, {Kurk}, {Mignoli},
  {Morrison}, {Pope}, \& {Ravindranath}}]{daddi07b}
{Daddi}, E., {Alexander}, D.~M., {Dickinson}, M., {et~al.} 2007{\natexlab{a}},
  \apj, 670, 173

\bibitem[{{Daddi} {et~al.}(2007{\natexlab{b}}){Daddi}, {Dickinson}, {Morrison},
  {Chary}, {Cimatti}, {Elbaz}, {Frayer}, {Renzini}, {Pope}, {Alexander},
  {Bauer}, {Giavalisco}, {Huynh}, {Kurk}, \& {Mignoli}}]{daddi07}
{Daddi}, E., {Dickinson}, M., {Morrison}, G., {et~al.} 2007{\natexlab{b}},
  \apj, 670, 156

\bibitem[{{Daddi} {et~al.}(2010){Daddi}, {Elbaz}, {Walter}, {Bournaud},
  {Salmi}, {Carilli}, {Dannerbauer}, {Dickinson}, {Monaco}, \&
  {Riechers}}]{daddi10b}
{Daddi}, E., {Elbaz}, D., {Walter}, F., {et~al.} 2010, \apjl, 714, L118

\bibitem[{{Daddi} {et~al.}(2005){Daddi}, {Renzini}, {Pirzkal}, {Cimatti},
  {Malhotra}, {Stiavelli}, {Xu}, {Pasquali}, {Rhoads}, {Brusa}, {di Serego
  Alighieri}, {Ferguson}, {Koekemoer}, {Moustakas}, {Panagia}, \&
  {Windhorst}}]{daddi05}
{Daddi}, E., {Renzini}, A., {Pirzkal}, N., {et~al.} 2005, \apj, 626, 680

\bibitem[{{Dahlen} {et~al.}(2013){Dahlen}, {Mobasher}, {Faber}, {Ferguson},
  {Barro}, {Finkelstein}, {Finlator}, {Fontana}, {Gruetzbauch}, {Johnson},
  {Pforr}, {Salvato}, {Wiklind}, {Wuyts}, {Acquaviva}, {Dickinson}, {Guo},
  {Huang}, {Huang}, {Newman}, {Bell}, {Conselice}, {Galametz}, {Gawiser},
  {Giavalisco}, {Grogin}, {Hathi}, {Kocevski}, {Koekemoer}, {Koo}, {Lee},
  {McGrath}, {Papovich}, {Peth}, {Ryan}, {Somerville}, {Weiner}, \&
  {Wilson}}]{dahlen13}
{Dahlen}, T., {Mobasher}, B., {Faber}, S.~M., {et~al.} 2013, \apj, 775, 93

\bibitem[{{Davies} {et~al.}(2007){Davies}, {M{\"u}ller S{\'a}nchez}, {Genzel},
  {Tacconi}, {Hicks}, {Friedrich}, \& {Sternberg}}]{davies07}
{Davies}, R.~I., {M{\"u}ller S{\'a}nchez}, F., {Genzel}, R., {et~al.} 2007,
  \apj, 671, 1388

\bibitem[{{De Lucia} \& {Blaizot}(2007)}]{delucia07}
{De Lucia}, G. \& {Blaizot}, J. 2007, \mnras, 375, 2

\bibitem[{{Dekel} \& {Birnboim}(2006)}]{dekel06}
{Dekel}, A. \& {Birnboim}, Y. 2006, \mnras, 368, 2

\bibitem[{{Dekel} {et~al.}(2009{\natexlab{a}}){Dekel}, {Birnboim}, {Engel},
  {Freundlich}, {Goerdt}, {Mumcuoglu}, {Neistein}, {Pichon}, {Teyssier}, \&
  {Zinger}}]{dekel09a}
{Dekel}, A., {Birnboim}, Y., {Engel}, G., {et~al.} 2009{\natexlab{a}}, \nat,
  457, 451

\bibitem[{{Dekel} \& {Burkert}(2013)}]{dekel13b}
{Dekel}, A. \& {Burkert}, A. 2013, ArXiv e-prints

\bibitem[{{Dekel} {et~al.}(2009{\natexlab{b}}){Dekel}, {Sari}, \&
  {Ceverino}}]{dekel09b}
{Dekel}, A., {Sari}, R., \& {Ceverino}, D. 2009{\natexlab{b}}, \apj, 703, 785

\bibitem[{{Dekel} {et~al.}(2013){Dekel}, {Zolotov}, {Tweed}, {Cacciato},
  {Ceverino}, \& {Primack}}]{dekel13a}
{Dekel}, A., {Zolotov}, A., {Tweed}, D., {et~al.} 2013, \mnras, 435, 999

\bibitem[{{Diamond-Stanic} {et~al.}(2012){Diamond-Stanic}, {Moustakas},
  {Tremonti}, {Coil}, {Hickox}, {Robaina}, {Rudnick}, \& {Sell}}]{stanic12}
{Diamond-Stanic}, A.~M., {Moustakas}, J., {Tremonti}, C.~A., {et~al.} 2012,
  \apjl, 755, L26

\bibitem[{{Djorgovski} \& {Davis}(1987)}]{djor87}
{Djorgovski}, S. \& {Davis}, M. 1987, \apj, 313, 59

\bibitem[{{Donley} {et~al.}(2008){Donley}, {Rieke}, {P{\'e}rez-Gonz{\'a}lez},
  \& {Barro}}]{donley08}
{Donley}, J.~L., {Rieke}, G.~H., {P{\'e}rez-Gonz{\'a}lez}, P.~G., \& {Barro},
  G. 2008, \apj, 687, 111

\bibitem[{{Donley} {et~al.}(2007){Donley}, {Rieke}, {P{\'e}rez-Gonz{\'a}lez},
  {Rigby}, \& {Alonso-Herrero}}]{donley07}
{Donley}, J.~L., {Rieke}, G.~H., {P{\'e}rez-Gonz{\'a}lez}, P.~G., {Rigby},
  J.~R., \& {Alonso-Herrero}, A. 2007, \apj, 660, 167

\bibitem[{{Elbaz} {et~al.}(2007){Elbaz}, {Daddi}, {Le Borgne}, {Dickinson},
  {Alexander}, {Chary}, {Starck}, {Brandt}, {Kitzbichler}, {MacDonald},
  {Nonino}, {Popesso}, {Stern}, \& {Vanzella}}]{elbaz07}
{Elbaz}, D., {Daddi}, E., {Le Borgne}, D., {et~al.} 2007, \aap, 468, 33

\bibitem[{{Elbaz} {et~al.}(2011){Elbaz}, {Dickinson}, {Hwang},
  {D{\'{\i}}az-Santos}, {Magdis}, {Magnelli}, {Le Borgne}, {Galliano},
  {Pannella}, {Chanial}, {Armus}, {Charmandaris}, {Daddi}, {Aussel}, {Popesso},
  {Kartaltepe}, {Altieri}, {Valtchanov}, {Coia}, {Dannerbauer}, {Dasyra},
  {Leiton}, {Mazzarella}, {Alexander}, {Buat}, {Burgarella}, {Chary}, {Gilli},
  {Ivison}, {Juneau}, {Le Floc'h}, {Lutz}, {Morrison}, {Mullaney}, {Murphy},
  {Pope}, {Scott}, {Brodwin}, {Calzetti}, {Cesarsky}, {Charlot}, {Dole},
  {Eisenhardt}, {Ferguson}, {F{\"o}rster Schreiber}, {Frayer}, {Giavalisco},
  {Huynh}, {Koekemoer}, {Papovich}, {Reddy}, {Surace}, {Teplitz}, {Yun}, \&
  {Wilson}}]{elbaz11}
{Elbaz}, D., {Dickinson}, M., {Hwang}, H.~S., {et~al.} 2011, \aap, 533, A119

\bibitem[{{Elmegreen} {et~al.}(2008){Elmegreen}, {Bournaud}, \&
  {Elmegreen}}]{elme08}
{Elmegreen}, B.~G., {Bournaud}, F., \& {Elmegreen}, D.~M. 2008, \apj, 688, 67

\bibitem[{{Elmegreen} \& {Elmegreen}(2005{\natexlab{a}})}]{elme05}
{Elmegreen}, B.~G. \& {Elmegreen}, D.~M. 2005{\natexlab{a}}, \apj, 627, 632

\bibitem[{{Elmegreen} \& {Elmegreen}(2005{\natexlab{b}})}]{elmegreen05}
---. 2005{\natexlab{b}}, \apj, 627, 632

\bibitem[{{Elmegreen} {et~al.}(2007){Elmegreen}, {Elmegreen}, {Ravindranath},
  \& {Coe}}]{elme07}
{Elmegreen}, D.~M., {Elmegreen}, B.~G., {Ravindranath}, S., \& {Coe}, D.~A.
  2007, \apj, 658, 763

\bibitem[{{Fang} {et~al.}(2013){Fang}, {Faber}, {Koo}, \& {Dekel}}]{fang13}
{Fang}, J.~J., {Faber}, S.~M., {Koo}, D.~C., \& {Dekel}, A. 2013, \apj, 776, 63

\bibitem[{{Finlator} {et~al.}(2007){Finlator}, {Dav{\'e}}, \&
  {Oppenheimer}}]{finlator07}
{Finlator}, K., {Dav{\'e}}, R., \& {Oppenheimer}, B.~D. 2007, \mnras, 376, 1861

\bibitem[{{Finlator} {et~al.}(2011){Finlator}, {Oppenheimer}, \&
  {Dav{\'e}}}]{finlator11}
{Finlator}, K., {Oppenheimer}, B.~D., \& {Dav{\'e}}, R. 2011, \mnras, 410, 1703

\bibitem[{{Fontana} {et~al.}(2009){Fontana}, {Santini}, {Grazian},
  {Pentericci}, {Fiore}, {Castellano}, {Giallongo}, {Menci}, {Salimbeni},
  {Cristiani}, {Nonino}, \& {Vanzella}}]{fontana09}
{Fontana}, A., {Santini}, P., {Grazian}, A., {et~al.} 2009, \aap, 501, 15

\bibitem[{{F{\"o}rster Schreiber} {et~al.}(2009){F{\"o}rster Schreiber},
  {Genzel}, {Bouch{\'e}}, {Cresci}, {Davies}, {Buschkamp}, {Shapiro},
  {Tacconi}, {Hicks}, {Genel}, {Shapley}, {Erb}, {Steidel}, {Lutz},
  {Eisenhauer}, {Gillessen}, {Sternberg}, {Renzini}, {Cimatti}, {Daddi},
  {Kurk}, {Lilly}, {Kong}, {Lehnert}, {Nesvadba}, {Verma}, {McCracken},
  {Arimoto}, {Mignoli}, \& {Onodera}}]{fs09}
{F{\"o}rster Schreiber}, N.~M., {Genzel}, R., {Bouch{\'e}}, N., {et~al.} 2009,
  \apj, 706, 1364

\bibitem[{{Fumagalli} {et~al.}(2013){Fumagalli}, {Labbe}, {Patel}, {Franx},
  {van Dokkum}, {Brammer}, {da Cunha}, {Forster Schreiber}, {Kriek}, {Quadri},
  {Rix}, {Wake}, {Whitaker}, {Lundgren}, {Marchesini}, {Maseda}, {Momcheva},
  {Nelson}, {Pacifici}, \& {Skelton}}]{fumagalli13}
{Fumagalli}, M., {Labbe}, I., {Patel}, S.~G., {et~al.} 2013, ArXiv e-prints

\bibitem[{{Fumagalli} {et~al.}(2012){Fumagalli}, {Patel}, {Franx}, {Brammer},
  {van Dokkum}, {da Cunha}, {Kriek}, {Lundgren}, {Momcheva}, {Rix}, {Schmidt},
  {Skelton}, {Whitaker}, {Labbe}, \& {Nelson}}]{fumagalli12}
{Fumagalli}, M., {Patel}, S.~G., {Franx}, M., {et~al.} 2012, \apjl, 757, L22

\bibitem[{{Galametz} {et~al.}(2013){Galametz}, {Grazian}, {Fontana},
  {Ferguson}, {Ashby}, {Barro}, {Castellano}, {Dahlen}, {Donley}, {Faber},
  {Grogin}, {Guo}, {Huang}, {Kocevski}, {Koekemoer}, {Lee}, {McGrath}, {Peth},
  {Willner}, {Almaini}, {Cooper}, {Cooray}, {Conselice}, {Dickinson}, {Dunlop},
  {Fazio}, {Foucaud}, {Gardner}, {Giavalisco}, {Hathi}, {Hartley}, {Koo},
  {Lai}, {de Mello}, {McLure}, {Lucas}, {Paris}, {Pentericci}, {Santini},
  {Simpson}, {Sommariva}, {Targett}, {Weiner}, {Wuyts}, \& {the CANDELS
  Team}}]{galametz13}
{Galametz}, A., {Grazian}, A., {Fontana}, A., {et~al.} 2013, \apjs, 206, 10

\bibitem[{{Genzel} {et~al.}(2008){Genzel}, {Burkert}, {Bouch{\'e}}, {Cresci},
  {F{\"o}rster Schreiber}, {Shapley}, {Shapiro}, {Tacconi}, {Buschkamp},
  {Cimatti}, {Daddi}, {Davies}, {Eisenhauer}, {Erb}, {Genel}, {Gerhard},
  {Hicks}, {Lutz}, {Naab}, {Ott}, {Rabien}, {Renzini}, {Steidel}, {Sternberg},
  \& {Lilly}}]{genzel08}
{Genzel}, R., {Burkert}, A., {Bouch{\'e}}, N., {et~al.} 2008, \apj, 687, 59

\bibitem[{{Gobat} {et~al.}(2012){Gobat}, {Strazzullo}, {Daddi}, {Onodera},
  {Renzini}, {B{\'e}thermin}, {Dickinson}, {Carollo}, \& {Cimatti}}]{gobat12}
{Gobat}, R., {Strazzullo}, V., {Daddi}, E., {et~al.} 2012, \apjl, 759, L44

\bibitem[{{Gonzalez} {et~al.}(2012){Gonzalez}, {Bouwens}, {llingworth},
  {Labbe}, {Oesch}, {Franx}, \& {Magee}}]{gonzalez12}
{Gonzalez}, V., {Bouwens}, R., {llingworth}, G., {et~al.} 2012, ArXiv e-prints

\bibitem[{{Grogin} {et~al.}(2011){Grogin}, {Kocevski}, {Faber}, {Ferguson},
  {Koekemoer}, {Riess}, {Acquaviva}, {Alexander}, {Almaini}, {Ashby}, {Barden},
  {Bell}, {Bournaud}, {Brown}, {Caputi}, {Casertano}, {Cassata}, {Castellano},
  {Challis}, {Chary}, {Cheung}, {Cirasuolo}, {Conselice}, {Roshan Cooray},
  {Croton}, {Daddi}, {Dahlen}, {Dav{\'e}}, {de Mello}, {Dekel}, {Dickinson},
  {Dolch}, {Donley}, {Dunlop}, {Dutton}, {Elbaz}, {Fazio}, {Filippenko},
  {Finkelstein}, {Fontana}, {Gardner}, {Garnavich}, {Gawiser}, {Giavalisco},
  {Grazian}, {Guo}, {Hathi}, {H{\"a}ussler}, {Hopkins}, {Huang}, {Huang},
  {Jha}, {Kartaltepe}, {Kirshner}, {Koo}, {Lai}, {Lee}, {Li}, {Lotz}, {Lucas},
  {Madau}, {McCarthy}, {McGrath}, {McIntosh}, {McLure}, {Mobasher},
  {Moustakas}, {Mozena}, {Nandra}, {Newman}, {Niemi}, {Noeske}, {Papovich},
  {Pentericci}, {Pope}, {Primack}, {Rajan}, {Ravindranath}, {Reddy}, {Renzini},
  {Rix}, {Robaina}, {Rodney}, {Rosario}, {Rosati}, {Salimbeni}, {Scarlata},
  {Siana}, {Simard}, {Smidt}, {Somerville}, {Spinrad}, {Straughn}, {Strolger},
  {Telford}, {Teplitz}, {Trump}, {van der Wel}, {Villforth}, {Wechsler},
  {Weiner}, {Wiklind}, {Wild}, {Wilson}, {Wuyts}, {Yan}, \& {Yun}}]{candelsgro}
{Grogin}, N.~A., {Kocevski}, D.~D., {Faber}, S.~M., {et~al.} 2011, \apjs, 197,
  35

\bibitem[{{Guo} {et~al.}(2011){Guo}, {Giavalisco}, {Cassata}, {Ferguson},
  {Dickinson}, {Renzini}, {Koekemoer}, {Grogin}, {Papovich}, {Tundo},
  {Fontana}, {Lotz}, \& {Salimbeni}}]{guo11}
{Guo}, Y., {Giavalisco}, M., {Cassata}, P., {et~al.} 2011, \apj, 735, 18

\bibitem[{{Guo} {et~al.}(2012){Guo}, {Giavalisco}, {Ferguson}, {Cassata}, \&
  {Koekemoer}}]{guo12b}
{Guo}, Y., {Giavalisco}, M., {Ferguson}, H.~C., {Cassata}, P., \& {Koekemoer},
  A.~M. 2012, \apj, 757, 120

\bibitem[{{Hopkins} {et~al.}(2006){Hopkins}, {Hernquist}, {Cox}, {Di Matteo},
  {Robertson}, \& {Springel}}]{hopkins06}
{Hopkins}, P.~F., {Hernquist}, L., {Cox}, T.~J., {et~al.} 2006, \apjs, 163, 1

\bibitem[{{Hopkins} {et~al.}(2008){Hopkins}, {Hernquist}, {Cox}, \& {Kere{\v
  s}}}]{hopkins08a}
{Hopkins}, P.~F., {Hernquist}, L., {Cox}, T.~J., \& {Kere{\v s}}, D. 2008,
  \apjs, 175, 356

\bibitem[{{Ilbert} {et~al.}(2010){Ilbert}, {Salvato}, {Le Floc'h}, {Aussel},
  {Capak}, {McCracken}, {Mobasher}, {Kartaltepe}, {Scoville}, {Sanders},
  {Arnouts}, {Bundy}, {Cassata}, {Kneib}, {Koekemoer}, {Le F{\`e}vre}, {Lilly},
  {Surace}, {Taniguchi}, {Tasca}, {Thompson}, {Tresse}, {Zamojski}, {Zamorani},
  \& {Zucca}}]{ilbert10}
{Ilbert}, O., {Salvato}, M., {Le Floc'h}, E., {et~al.} 2010, \apj, 709, 644

\bibitem[{{Karim} {et~al.}(2011){Karim}, {Schinnerer},
  {Mart{\'{\i}}nez-Sansigre}, {Sargent}, {van der Wel}, {Rix}, {Ilbert},
  {Smol{\v c}i{\'c}}, {Carilli}, {Pannella}, {Koekemoer}, {Bell}, \&
  {Salvato}}]{karim11}
{Karim}, A., {Schinnerer}, E., {Mart{\'{\i}}nez-Sansigre}, A., {et~al.} 2011,
  \apj, 730, 61

\bibitem[{{Kartaltepe} {et~al.}(2012){Kartaltepe}, {Dickinson}, {Alexander},
  {Bell}, {Dahlen}, {Elbaz}, {Faber}, {Lotz}, {McIntosh}, {Wiklind}, {Altieri},
  {Aussel}, {Bethermin}, {Bournaud}, {Charmandaris}, {Conselice}, {Cooray},
  {Dannerbauer}, {Dav{\'e}}, {Dunlop}, {Dekel}, {Ferguson}, {Grogin}, {Hwang},
  {Ivison}, {Kocevski}, {Koekemoer}, {Koo}, {Lai}, {Leiton}, {Lucas}, {Lutz},
  {Magdis}, {Magnelli}, {Morrison}, {Mozena}, {Mullaney}, {Newman}, {Pope},
  {Popesso}, {van der Wel}, {Weiner}, \& {Wuyts}}]{kartaltepe12}
{Kartaltepe}, J.~S., {Dickinson}, M., {Alexander}, D.~M., {et~al.} 2012, \apj,
  757, 23

\bibitem[{{Kauffmann} {et~al.}(2003){Kauffmann}, {Heckman}, {White}, {Charlot},
  {Tremonti}, {Peng}, {Seibert}, {Brinkmann}, {Nichol}, {SubbaRao}, \&
  {York}}]{kauffman03}
{Kauffmann}, G., {Heckman}, T.~M., {White}, S.~D.~M., {et~al.} 2003, \mnras,
  341, 54

\bibitem[{{Kaviraj} {et~al.}(2013{\natexlab{a}}){Kaviraj}, {Cohen}, {Ellis},
  {Peirani}, {Windhorst}, {O'Connell}, {Silk}, {Whitmore}, {Hathi}, {Ryan},
  {Dopita}, {Frogel}, \& {Dekel}}]{kaviraj13a}
{Kaviraj}, S., {Cohen}, S., {Ellis}, R.~S., {et~al.} 2013{\natexlab{a}},
  \mnras, 428, 925

\bibitem[{{Kaviraj} {et~al.}(2012){Kaviraj}, {Cohen}, {Ellis}, {Peirani},
  {Windhorst}, {O'Connell}, {Silk}, {Whitmore}, {Hathi}, {Ryan}, {Dopita},
  {Frogel}, \& {Dekel}}]{kaviraj12}
---. 2012, ArXiv e-prints

\bibitem[{{Kaviraj} {et~al.}(2013{\natexlab{b}}){Kaviraj}, {Cohen},
  {Windhorst}, {Silk}, {O'Connell}, {Dopita}, {Dekel}, {Hathi}, {Straughn}, \&
  {Rutkowski}}]{kaviraj13b}
{Kaviraj}, S., {Cohen}, S., {Windhorst}, R.~A., {et~al.} 2013{\natexlab{b}},
  \mnras, 429, L40

\bibitem[{{Kennicutt}(1998)}]{ken98}
{Kennicutt}, Jr., R.~C. 1998, \araa, 36, 189

\bibitem[{{Kere{\v s}} {et~al.}(2005){Kere{\v s}}, {Katz}, {Weinberg}, \&
  {Dav{\'e}}}]{keres05}
{Kere{\v s}}, D., {Katz}, N., {Weinberg}, D.~H., \& {Dav{\'e}}, R. 2005,
  \mnras, 363, 2

\bibitem[{{Kirkpatrick} {et~al.}(2013){Kirkpatrick}, {Pope}, {Charmandaris},
  {Daddi}, {Elbaz}, {Hwang}, {Pannella}, {Scott}, {Altieri}, {Aussel}, {Coia},
  {Dannerbauer}, {Dasyra}, {Dickinson}, {Kartaltepe}, {Leiton}, {Magdis},
  {Magnelli}, {Popesso}, \& {Valtchanov}}]{kirkpatric13}
{Kirkpatrick}, A., {Pope}, A., {Charmandaris}, V., {et~al.} 2013, \apj, 763,
  123

\bibitem[{{Koekemoer} {et~al.}(2011){Koekemoer}, {Faber}, {Ferguson}, {Grogin},
  {Kocevski}, {Koo}, {Lai}, {Lotz}, {Lucas}, {McGrath}, {Ogaz}, {Rajan},
  {Riess}, {Rodney}, {Strolger}, {Casertano}, {Castellano}, {Dahlen},
  {Dickinson}, {Dolch}, {Fontana}, {Giavalisco}, {Grazian}, {Guo}, {Hathi},
  {Huang}, {van der Wel}, {Yan}, {Acquaviva}, {Alexander}, {Almaini}, {Ashby},
  {Barden}, {Bell}, {Bournaud}, {Brown}, {Caputi}, {Cassata}, {Challis},
  {Chary}, {Cheung}, {Cirasuolo}, {Conselice}, {Roshan Cooray}, {Croton},
  {Daddi}, {Dav{\'e}}, {de Mello}, {de Ravel}, {Dekel}, {Donley}, {Dunlop},
  {Dutton}, {Elbaz}, {Fazio}, {Filippenko}, {Finkelstein}, {Frazer}, {Gardner},
  {Garnavich}, {Gawiser}, {Gruetzbauch}, {Hartley}, {H{\"a}ussler},
  {Herrington}, {Hopkins}, {Huang}, {Jha}, {Johnson}, {Kartaltepe},
  {Khostovan}, {Kirshner}, {Lani}, {Lee}, {Li}, {Madau}, {McCarthy},
  {McIntosh}, {McLure}, {McPartland}, {Mobasher}, {Moreira}, {Mortlock},
  {Moustakas}, {Mozena}, {Nandra}, {Newman}, {Nielsen}, {Niemi}, {Noeske},
  {Papovich}, {Pentericci}, {Pope}, {Primack}, {Ravindranath}, {Reddy},
  {Renzini}, {Rix}, {Robaina}, {Rosario}, {Rosati}, {Salimbeni}, {Scarlata},
  {Siana}, {Simard}, {Smidt}, {Snyder}, {Somerville}, {Spinrad}, {Straughn},
  {Telford}, {Teplitz}, {Trump}, {Vargas}, {Villforth}, {Wagner}, {Wandro},
  {Wechsler}, {Weiner}, {Wiklind}, {Wild}, {Wilson}, {Wuyts}, \&
  {Yun}}]{candelskoe}
{Koekemoer}, A.~M., {Faber}, S.~M., {Ferguson}, H.~C., {et~al.} 2011, \apjs,
  197, 36

\bibitem[{{Kravtsov} {et~al.}(1997){Kravtsov}, {Klypin}, \& {Khokhlov}}]{arm}
{Kravtsov}, A.~V., {Klypin}, A.~A., \& {Khokhlov}, A.~M. 1997, \apjs, 111, 73

\bibitem[{{Kriek} {et~al.}(2009{\natexlab{a}}){Kriek}, {van Dokkum}, {Franx},
  {Illingworth}, \& {Magee}}]{kriek09}
{Kriek}, M., {van Dokkum}, P.~G., {Franx}, M., {Illingworth}, G.~D., \&
  {Magee}, D.~K. 2009{\natexlab{a}}, \apjl, 705, L71

\bibitem[{{Kriek} {et~al.}(2009{\natexlab{b}}){Kriek}, {van Dokkum},
  {Labb{\'e}}, {Franx}, {Illingworth}, {Marchesini}, \& {Quadri}}]{fast}
{Kriek}, M., {van Dokkum}, P.~G., {Labb{\'e}}, I., {et~al.} 2009{\natexlab{b}},
  \apj, 700, 221

\bibitem[{{Kriek} {et~al.}(2011){Kriek}, {van Dokkum}, {Whitaker}, {Labb{\'e}},
  {Franx}, \& {Brammer}}]{kriek11}
{Kriek}, M., {van Dokkum}, P.~G., {Whitaker}, K.~E., {et~al.} 2011, \apj, 743,
  168

\bibitem[{{Krist}(1995)}]{tinytim}
{Krist}, J. 1995, in Astronomical Society of the Pacific Conference Series,
  Vol.~77, Astronomical Data Analysis Software and Systems IV, ed. R.~A.
  {Shaw}, H.~E. {Payne}, \& J.~J.~E. {Hayes}, 349

\bibitem[{{Krogager} {et~al.}(2013){Krogager}, {Zirm}, {Toft}, {Man}, \&
  {Brammer}}]{krogager13}
{Krogager}, J.-K., {Zirm}, A.~W., {Toft}, S., {Man}, A., \& {Brammer}, G. 2013,
  ArXiv e-prints

\bibitem[{{K{\"u}mmel} {et~al.}(2009){K{\"u}mmel}, {Walsh}, {Pirzkal},
  {Kuntschner}, \& {Pasquali}}]{axe}
{K{\"u}mmel}, M., {Walsh}, J.~R., {Pirzkal}, N., {Kuntschner}, H., \&
  {Pasquali}, A. 2009, \pasp, 121, 59

\bibitem[{{Lacy} {et~al.}(2004){Lacy}, {Storrie-Lombardi}, {Sajina},
  {Appleton}, {Armus}, {Chapman}, {Choi}, {Fadda}, {Fang}, {Frayer},
  {Heinrichsen}, {Helou}, {Im}, {Marleau}, {Masci}, {Shupe}, {Soifer},
  {Surace}, {Teplitz}, {Wilson}, \& {Yan}}]{lacy04}
{Lacy}, M., {Storrie-Lombardi}, L.~J., {Sajina}, A., {et~al.} 2004, \apjs, 154,
  166

\bibitem[{{Laidler} {et~al.}(2006){Laidler}, {Grogin}, {Clubb}, {Ferguson},
  {Papovich}, {Dickinson}, {Idzi}, {MacDonald}, {Ouchi}, \& {Mobasher}}]{tfit}
{Laidler}, V.~G., {Grogin}, N., {Clubb}, K., {et~al.} 2006, in Astronomical
  Society of the Pacific Conference Series, Vol. 351, Astronomical Data
  Analysis Software and Systems XV, ed. C.~{Gabriel}, C.~{Arviset}, D.~{Ponz},
  \& S.~{Enrique}, 228

\bibitem[{{Lee} \& {Yi}(2013)}]{lee13}
{Lee}, J. \& {Yi}, S.~K. 2013, \apj, 766, 38

\bibitem[{{Lee} {et~al.}(2009){Lee}, {Idzi}, {Ferguson}, {Somerville},
  {Wiklind}, \& {Giavalisco}}]{lee09}
{Lee}, S.-K., {Idzi}, R., {Ferguson}, H.~C., {et~al.} 2009, \apjs, 184, 100

\bibitem[{{Magdis} {et~al.}(2012){Magdis}, {Daddi}, {B{\'e}thermin}, {Sargent},
  {Elbaz}, {Pannella}, {Dickinson}, {Dannerbauer}, {da Cunha}, {Walter},
  {Rigopoulou}, {Charmandaris}, {Hwang}, \& {Kartaltepe}}]{magdis12}
{Magdis}, G.~E., {Daddi}, E., {B{\'e}thermin}, M., {et~al.} 2012, \apj, 760, 6

\bibitem[{{Magnelli} {et~al.}(2013){Magnelli}, {Popesso}, {Berta}, {Pozzi},
  {Elbaz}, {Lutz}, {Dickinson}, {Altieri}, {Andreani}, {Aussel},
  {B{\'e}thermin}, {Bongiovanni}, {Cepa}, {Charmandaris}, {Chary}, {Cimatti},
  {Daddi}, {F{\"o}rster Schreiber}, {Genzel}, {Gruppioni}, {Harwit}, {Hwang},
  {Ivison}, {Magdis}, {Maiolino}, {Murphy}, {Nordon}, {Pannella}, {P{\'e}rez
  Garc{\'{\i}}a}, {Poglitsch}, {Rosario}, {Sanchez-Portal}, {Santini}, {Scott},
  {Sturm}, {Tacconi}, \& {Valtchanov}}]{magnelli13}
{Magnelli}, B., {Popesso}, P., {Berta}, S., {et~al.} 2013, \aap, 553, A132

\bibitem[{{Maraston} {et~al.}(2010){Maraston}, {Pforr}, {Renzini}, {Daddi},
  {Dickinson}, {Cimatti}, \& {Tonini}}]{maraston10}
{Maraston}, C., {Pforr}, J., {Renzini}, A., {et~al.} 2010, \mnras, 407, 830

\bibitem[{{McIntosh} {et~al.}(2013){McIntosh}, {Wagner}, {Cooper}, {Bell},
  {Keres}, {van den Bosch}, {Gallazzi}, {Haines}, {Mann}, {Pasquali}, \&
  {Christian}}]{mcintosh13}
{McIntosh}, D.~H., {Wagner}, C., {Cooper}, A., {et~al.} 2013, ArXiv e-prints

\bibitem[{{Mullaney} {et~al.}(2012){Mullaney}, {Daddi}, {B{\'e}thermin},
  {Elbaz}, {Juneau}, {Pannella}, {Sargent}, {Alexander}, \&
  {Hickox}}]{mullaney12b}
{Mullaney}, J.~R., {Daddi}, E., {B{\'e}thermin}, M., {et~al.} 2012, \apjl, 753,
  L30

\bibitem[{{Muzzin} {et~al.}(2013){Muzzin}, {Marchesini}, {Stefanon}, {Franx},
  {McCracken}, {Milvang-Jensen}, {Dunlop}, {Fynbo}, {Le Fevre}, {Brammer}, \&
  {Labbe}}]{muzzin13smf}
{Muzzin}, A., {Marchesini}, D., {Stefanon}, M., {et~al.} 2013, ArXiv e-prints

\bibitem[{{Naab} {et~al.}(2009){Naab}, {Johansson}, \& {Ostriker}}]{naab09b}
{Naab}, T., {Johansson}, P.~H., \& {Ostriker}, J.~P. 2009, \apjl, 699, L178

\bibitem[{{Naab} {et~al.}(2007){Naab}, {Johansson}, {Ostriker}, \&
  {Efstathiou}}]{naab07}
{Naab}, T., {Johansson}, P.~H., {Ostriker}, J.~P., \& {Efstathiou}, G. 2007,
  \apj, 658, 710

\bibitem[{{Newman} {et~al.}(2013{\natexlab{a}}){Newman}, {Ellis}, {Andreon},
  {Treu}, {Raichoor}, \& {Trinchieri}}]{newman13}
{Newman}, A.~B., {Ellis}, R.~S., {Andreon}, S., {et~al.} 2013{\natexlab{a}},
  ArXiv e-prints

\bibitem[{{Newman} {et~al.}(2012){Newman}, {Ellis}, {Bundy}, \&
  {Treu}}]{newman12}
{Newman}, A.~B., {Ellis}, R.~S., {Bundy}, K., \& {Treu}, T. 2012, \apj, 746,
  162

\bibitem[{{Newman} {et~al.}(2013{\natexlab{b}}){Newman}, {Buschkamp}, {Genzel},
  {Forster Schreiber}, {Kurk}, {Sternberg}, {Gnat}, {Rosario}, {Mancini},
  {Lilly}, {Renzini}, {Burkert}, {Carollo}, {Cresci}, {Davies}, {Eisenhauer},
  {Genel}, {Shapiro Griffin}, {Hicks}, {Lutz}, {Naab}, {Peng}, {Tacconi},
  {Wuyts}, {Zamorani}, {Vergani}, \& {Weiner}}]{snewman13}
{Newman}, S.~F., {Buschkamp}, P., {Genzel}, R., {et~al.} 2013{\natexlab{b}},
  ArXiv e-prints

\bibitem[{{Noeske} {et~al.}(2007){Noeske}, {Weiner}, {Faber}, {Papovich},
  {Koo}, {Somerville}, {Bundy}, {Conselice}, {Newman}, {Schiminovich}, {Le
  Floc'h}, {Coil}, {Rieke}, {Lotz}, {Primack}, {Barmby}, {Cooper}, {Davis},
  {Ellis}, {Fazio}, {Guhathakurta}, {Huang}, {Kassin}, {Martin}, {Phillips},
  {Rich}, {Small}, {Willmer}, \& {Wilson}}]{mainseq}
{Noeske}, K.~G., {Weiner}, B.~J., {Faber}, S.~M., {et~al.} 2007, \apjl, 660,
  L43

\bibitem[{{Oser} {et~al.}(2010){Oser}, {Ostriker}, {Naab}, {Johansson}, \&
  {Burkert}}]{oser10}
{Oser}, L., {Ostriker}, J.~P., {Naab}, T., {Johansson}, P.~H., \& {Burkert}, A.
  2010, \apj, 725, 2312

\bibitem[{{Pacifici} {et~al.}(2012){Pacifici}, {Charlot}, {Blaizot}, \&
  {Brinchmann}}]{paci12}
{Pacifici}, C., {Charlot}, S., {Blaizot}, J., \& {Brinchmann}, J. 2012, \mnras,
  421, 2002

\bibitem[{{Pacifici} {et~al.}(2013){Pacifici}, {Kassin}, {Weiner}, {Charlot},
  \& {Gardner}}]{paci13}
{Pacifici}, C., {Kassin}, S.~A., {Weiner}, B., {Charlot}, S., \& {Gardner},
  J.~P. 2013, \apjl, 762, L15

\bibitem[{{Papovich} {et~al.}(2011){Papovich}, {Finkelstein}, {Ferguson},
  {Lotz}, \& {Giavalisco}}]{papovich11}
{Papovich}, C., {Finkelstein}, S.~L., {Ferguson}, H.~C., {Lotz}, J.~M., \&
  {Giavalisco}, M. 2011, \mnras, 412, 1123

\bibitem[{{Papovich} {et~al.}(2007){Papovich}, {Rudnick}, {Le Floc'h}, {van
  Dokkum}, {Rieke}, {Taylor}, {Armus}, {Gawiser}, {Huang}, {Marcillac}, \&
  {Franx}}]{papovich07}
{Papovich}, C., {Rudnick}, G., {Le Floc'h}, E., {et~al.} 2007, \apj, 668, 45

\bibitem[{{Patel} {et~al.}(2012{\natexlab{a}}){Patel}, {Holden}, {Kelson},
  {Franx}, {van der Wel}, \& {Illingworth}}]{patel12}
{Patel}, S.~G., {Holden}, B.~P., {Kelson}, D.~D., {et~al.} 2012{\natexlab{a}},
  \apjl, 748, L27

\bibitem[{{Patel} {et~al.}(2012{\natexlab{b}}){Patel}, {van Dokkum}, {Franx},
  {Quadri}, {Muzzin}, {Marchesini}, {Williams}, {Holden}, \&
  {Stefanon}}]{patel13}
{Patel}, S.~G., {van Dokkum}, P.~G., {Franx}, M., {et~al.} 2012{\natexlab{b}},
  ArXiv e-prints

\bibitem[{{Peng} {et~al.}(2002){Peng}, {Ho}, {Impey}, \& {Rix}}]{galfit}
{Peng}, C.~Y., {Ho}, L.~C., {Impey}, C.~D., \& {Rix}, H.-W. 2002, \aj, 124, 266

\bibitem[{{P{\'e}rez-Gonz{\'a}lez} {et~al.}(2013){P{\'e}rez-Gonz{\'a}lez},
  {Cava}, {Barro}, {Villar}, {Cardiel}, {Ferreras},
  {Rodr{\'{\i}}guez-Espinosa}, {Alonso-Herrero}, {Balcells}, {Cenarro}, {Cepa},
  {Charlot}, {Cimatti}, {Conselice}, {Daddi}, {Donley}, {Elbaz}, {Espino},
  {Gallego}, {Gobat}, {Gonz{\'a}lez-Mart{\'{\i}}n}, {Guzm{\'a}n},
  {Hern{\'a}n-Caballero}, {Mu{\~n}oz-Tu{\~n}{\'o}n}, {Renzini},
  {Rodr{\'{\i}}guez-Zaur{\'{\i}}n}, {Tresse}, {Trujillo}, \&
  {Zamorano}}]{shards}
{P{\'e}rez-Gonz{\'a}lez}, P.~G., {Cava}, A., {Barro}, G., {et~al.} 2013, \apj,
  762, 46

\bibitem[{{P{\'e}rez-Gonz{\'a}lez} {et~al.}(2010){P{\'e}rez-Gonz{\'a}lez},
  {Egami}, {Rex}, {Rawle}, {Kneib}, {Richard}, {Johansson}, {Altieri}, {Blain},
  {Bock}, {Boone}, {Bridge}, {Chung}, {Cl{\'e}ment}, {Clowe}, {Combes}, {Cuby},
  {Dessauges-Zavadsky}, {Dowell}, {Espino-Briones}, {Fadda}, {Fiedler},
  {Gonzalez}, {Horellou}, {Ilbert}, {Ivison}, {Jauzac}, {Lutz}, {Pell{\'o}},
  {Pereira}, {Rieke}, {Rodighiero}, {Schaerer}, {Smith}, {Valtchanov}, {Walth},
  {van der Werf}, {Werner}, \& {Zemcov}}]{pg10}
{P{\'e}rez-Gonz{\'a}lez}, P.~G., {Egami}, E., {Rex}, M., {et~al.} 2010, \aap,
  518, L15

\bibitem[{{P{\'e}rez-Gonz{\'a}lez}
  {et~al.}(2008{\natexlab{a}}){P{\'e}rez-Gonz{\'a}lez}, {Rieke}, {Villar},
  {Barro}, {Blaylock}, {Egami}, {Gallego}, {Gil de Paz}, {Pascual}, {Zamorano},
  \& {Donley}}]{pg08}
{P{\'e}rez-Gonz{\'a}lez}, P.~G., {Rieke}, G.~H., {Villar}, V., {et~al.}
  2008{\natexlab{a}}, \apj, 675, 234

\bibitem[{{P{\'e}rez-Gonz{\'a}lez}
  {et~al.}(2008{\natexlab{b}}){P{\'e}rez-Gonz{\'a}lez}, {Trujillo}, {Barro},
  {Gallego}, {Zamorano}, \& {Conselice}}]{pg08b}
{P{\'e}rez-Gonz{\'a}lez}, P.~G., {Trujillo}, I., {Barro}, G., {et~al.}
  2008{\natexlab{b}}, \apj, 687, 50

\bibitem[{{Pforr} {et~al.}(2012){Pforr}, {Maraston}, \& {Tonini}}]{pforr12}
{Pforr}, J., {Maraston}, C., \& {Tonini}, C. 2012, \mnras, 422, 3285

\bibitem[{{Poggianti} {et~al.}(2013){Poggianti}, {Moretti}, {Calvi},
  {D'Onofrio}, {Valentinuzzi}, {Fritz}, \& {Renzini}}]{poggianti13}
{Poggianti}, B.~M., {Moretti}, A., {Calvi}, R., {et~al.} 2013, \apj, 777, 125

\bibitem[{{Reddy} {et~al.}(2010){Reddy}, {Erb}, {Pettini}, {Steidel}, \&
  {Shapley}}]{reddy10}
{Reddy}, N.~A., {Erb}, D.~K., {Pettini}, M., {Steidel}, C.~C., \& {Shapley},
  A.~E. 2010, \apj, 712, 1070

\bibitem[{{Robertson} {et~al.}(2006){Robertson}, {Bullock}, {Cox}, {Di Matteo},
  {Hernquist}, {Springel}, \& {Yoshida}}]{robertson06}
{Robertson}, B., {Bullock}, J.~S., {Cox}, T.~J., {et~al.} 2006, \apj, 645, 986

\bibitem[{{Rodighiero} {et~al.}(2010){Rodighiero}, {Cimatti}, {Gruppioni},
  {Popesso}, {Andreani}, {Altieri}, {Aussel}, {Berta}, {Bongiovanni},
  {Brisbin}, {Cava}, {Cepa}, {Daddi}, {Dominguez-Sanchez}, {Elbaz}, {Fontana},
  {F{\"o}rster Schreiber}, {Franceschini}, {Genzel}, {Grazian}, {Lutz},
  {Magdis}, {Magliocchetti}, {Magnelli}, {Maiolino}, {Mancini}, {Nordon},
  {Perez Garcia}, {Poglitsch}, {Santini}, {Sanchez-Portal}, {Pozzi},
  {Riguccini}, {Saintonge}, {Shao}, {Sturm}, {Tacconi}, {Valtchanov},
  {Wetzstein}, \& {Wieprecht}}]{rodi10b}
{Rodighiero}, G., {Cimatti}, A., {Gruppioni}, C., {et~al.} 2010, \aap, 518, L25

\bibitem[{{Rodighiero} {et~al.}(2011){Rodighiero}, {Daddi}, {Baronchelli},
  {Cimatti}, {Renzini}, {Aussel}, {Popesso}, {Lutz}, {Andreani}, {Berta},
  {Cava}, {Elbaz}, {Feltre}, {Fontana}, {F{\"o}rster Schreiber},
  {Franceschini}, {Genzel}, {Grazian}, {Gruppioni}, {Ilbert}, {Le Floch},
  {Magdis}, {Magliocchetti}, {Magnelli}, {Maiolino}, {McCracken}, {Nordon},
  {Poglitsch}, {Santini}, {Pozzi}, {Riguccini}, {Tacconi}, {Wuyts}, \&
  {Zamorani}}]{rodi11}
{Rodighiero}, G., {Daddi}, E., {Baronchelli}, I., {et~al.} 2011, \apjl, 739,
  L40

\bibitem[{{Salim} {et~al.}(2009){Salim}, {Dickinson}, {Michael Rich},
  {Charlot}, {Lee}, {Schiminovich}, {P{\'e}rez-Gonz{\'a}lez}, {Ashby},
  {Papovich}, {Faber}, {Ivison}, {Frayer}, {Walton}, {Weiner}, {Chary},
  {Bundy}, {Noeske}, \& {Koekemoer}}]{salim09}
{Salim}, S., {Dickinson}, M., {Michael Rich}, R., {et~al.} 2009, \apj, 700, 161

\bibitem[{{Salvato} {et~al.}(2009){Salvato}, {Hasinger}, {Ilbert}, {Zamorani},
  {Brusa}, {Scoville}, {Rau}, {Capak}, {Arnouts}, {Aussel}, {Bolzonella},
  {Buongiorno}, {Cappelluti}, {Caputi}, {Civano}, {Cook}, {Elvis}, {Gilli},
  {Jahnke}, {Kartaltepe}, {Impey}, {Lamareille}, {Le Floc'h}, {Lilly},
  {Mainieri}, {McCarthy}, {McCracken}, {Mignoli}, {Mobasher}, {Murayama},
  {Sasaki}, {Sanders}, {Schiminovich}, {Shioya}, {Shopbell}, {Silverman},
  {Smol{\v c}i{\'c}}, {Surace}, {Taniguchi}, {Thompson}, {Trump}, {Urry}, \&
  {Zamojski}}]{salvato09}
{Salvato}, M., {Hasinger}, G., {Ilbert}, O., {et~al.} 2009, \apj, 690, 1250

\bibitem[{{Salvato} {et~al.}(2011){Salvato}, {Ilbert}, {Hasinger}, {Rau},
  {Civano}, {Zamorani}, {Brusa}, {Elvis}, {Vignali}, {Aussel}, {Comastri},
  {Fiore}, {Le Floc'h}, {Mainieri}, {Bardelli}, {Bolzonella}, {Bongiorno},
  {Capak}, {Caputi}, {Cappelluti}, {Carollo}, {Contini}, {Garilli}, {Iovino},
  {Fotopoulou}, {Fruscione}, {Gilli}, {Halliday}, {Kneib}, {Kakazu},
  {Kartaltepe}, {Koekemoer}, {Kovac}, {Ideue}, {Ikeda}, {Impey}, {Le Fevre},
  {Lamareille}, {Lanzuisi}, {Le Borgne}, {Le Brun}, {Lilly}, {Maier},
  {Manohar}, {Masters}, {McCracken}, {Messias}, {Mignoli}, {Mobasher}, {Nagao},
  {Pello}, {Puccetti}, {Perez-Montero}, {Renzini}, {Sargent}, {Sanders},
  {Scodeggio}, {Scoville}, {Shopbell}, {Silvermann}, {Taniguchi}, {Tasca},
  {Tresse}, {Trump}, \& {Zucca}}]{salvato11}
{Salvato}, M., {Ilbert}, O., {Hasinger}, G., {et~al.} 2011, \apj, 742, 61

\bibitem[{{Santini} {et~al.}(2009){Santini}, {Fontana}, {Grazian}, {Salimbeni},
  {Fiore}, {Fontanot}, {Boutsia}, {Castellano}, {Cristiani}, {de Santis},
  {Gallozzi}, {Giallongo}, {Menci}, {Nonino}, {Paris}, {Pentericci}, \&
  {Vanzella}}]{santini09}
{Santini}, P., {Fontana}, A., {Grazian}, A., {et~al.} 2009, \aap, 504, 751

\bibitem[{{Santini} {et~al.}(2012){Santini}, {Rosario}, {Shao}, {Lutz},
  {Maiolino}, {Alexander}, {Altieri}, {Andreani}, {Aussel}, {Bauer}, {Berta},
  {Bongiovanni}, {Brandt}, {Brusa}, {Cepa}, {Cimatti}, {Daddi}, {Elbaz},
  {Fontana}, {F{\"o}rster Schreiber}, {Genzel}, {Grazian}, {Le Floc'h},
  {Magnelli}, {Mainieri}, {Nordon}, {P{\'e}rez Garcia}, {Poglitsch}, {Popesso},
  {Pozzi}, {Riguccini}, {Rodighiero}, {Salvato}, {Sanchez-Portal}, {Sturm},
  {Tacconi}, {Valtchanov}, \& {Wuyts}}]{santini12b}
{Santini}, P., {Rosario}, D.~J., {Shao}, L., {et~al.} 2012, \aap, 540, A109

\bibitem[{{Saracco} {et~al.}(2010){Saracco}, {Longhetti}, \&
  {Gargiulo}}]{saracco10}
{Saracco}, P., {Longhetti}, M., \& {Gargiulo}, A. 2010, \mnras, 408, L21

\bibitem[{{Schaerer} {et~al.}(2011){Schaerer}, {de Barros}, \&
  {Stark}}]{schaerer11}
{Schaerer}, D., {de Barros}, S., \& {Stark}, D.~P. 2011, \aap, 536, A72

\bibitem[{{Shen} {et~al.}(2003){Shen}, {Mo}, {White}, {Blanton}, {Kauffmann},
  {Voges}, {Brinkmann}, \& {Csabai}}]{shen03}
{Shen}, S., {Mo}, H.~J., {White}, S.~D.~M., {et~al.} 2003, \mnras, 343, 978

\bibitem[{{Silva} {et~al.}(2004){Silva}, {Maiolino}, \& {Granato}}]{silva04}
{Silva}, L., {Maiolino}, R., \& {Granato}, G.~L. 2004, \mnras, 355, 973

\bibitem[{{Smail} {et~al.}(1997){Smail}, {Ivison}, \& {Blain}}]{smail97}
{Smail}, I., {Ivison}, R.~J., \& {Blain}, A.~W. 1997, \apjl, 490, L5

\bibitem[{{Somerville} {et~al.}(2012){Somerville}, {Gilmore}, {Primack}, \&
  {Dom{\'{\i}}nguez}}]{somerville12}
{Somerville}, R.~S., {Gilmore}, R.~C., {Primack}, J.~R., \& {Dom{\'{\i}}nguez},
  A. 2012, \mnras, 423, 1992

\bibitem[{{Somerville} {et~al.}(2008){Somerville}, {Hopkins}, {Cox},
  {Robertson}, \& {Hernquist}}]{somerville08}
{Somerville}, R.~S., {Hopkins}, P.~F., {Cox}, T.~J., {Robertson}, B.~E., \&
  {Hernquist}, L. 2008, \mnras, 391, 481

\bibitem[{{Springel} \& {Hernquist}(2005)}]{springel05b}
{Springel}, V. \& {Hernquist}, L. 2005, \apjl, 622, L9

\bibitem[{{Springel} {et~al.}(2005){Springel}, {White}, {Jenkins}, {Frenk},
  {Yoshida}, {Gao}, {Navarro}, {Thacker}, {Croton}, {Helly}, {Peacock}, {Cole},
  {Thomas}, {Couchman}, {Evrard}, {Colberg}, \& {Pearce}}]{springel05}
{Springel}, V., {White}, S.~D.~M., {Jenkins}, A., {et~al.} 2005, \nat, 435, 629

\bibitem[{{Stark} {et~al.}(2009){Stark}, {Ellis}, {Bunker}, {Bundy}, {Targett},
  {Benson}, \& {Lacy}}]{stark09}
{Stark}, D.~P., {Ellis}, R.~S., {Bunker}, A., {et~al.} 2009, \apj, 697, 1493

\bibitem[{{Stefanon} {et~al.}(2013){Stefanon}, {Marchesini}, {Rudnick},
  {Brammer}, \& {Whitaker}}]{stefanon13}
{Stefanon}, M., {Marchesini}, D., {Rudnick}, G.~H., {Brammer}, G.~B., \&
  {Whitaker}, K.~E. 2013, \apj, 768, 92

\bibitem[{{Stern} {et~al.}(2005){Stern}, {Eisenhardt}, {Gorjian}, {Kochanek},
  {Caldwell}, {Eisenstein}, {Brodwin}, {Brown}, {Cool}, {Dey}, {Green},
  {Jannuzi}, {Murray}, {Pahre}, \& {Willner}}]{stern05}
{Stern}, D., {Eisenhardt}, P., {Gorjian}, V., {et~al.} 2005, \apj, 631, 163

\bibitem[{{Szomoru} {et~al.}(2011){Szomoru}, {Franx}, {Bouwens}, {van Dokkum},
  {Labb{\'e}}, {Illingworth}, \& {Trenti}}]{szo11}
{Szomoru}, D., {Franx}, M., {Bouwens}, R.~J., {et~al.} 2011, \apjl, 735, L22

\bibitem[{{Szomoru} {et~al.}(2012){Szomoru}, {Franx}, \& {van Dokkum}}]{szo12}
{Szomoru}, D., {Franx}, M., \& {van Dokkum}, P.~G. 2012, \apj, 749, 121

\bibitem[{{Targett} {et~al.}(2013){Targett}, {Dunlop}, {Cirasuolo}, {McLure},
  {Bruce}, {Fontana}, {Galametz}, {Paris}, {Dav{\'e}}, {Dekel}, {Faber},
  {Ferguson}, {Grogin}, {Kartaltepe}, {Kocevski}, {Koekemoer}, {Kurczynski},
  {Lai}, \& {Lotz}}]{targett13}
{Targett}, T.~A., {Dunlop}, J.~S., {Cirasuolo}, M., {et~al.} 2013, \mnras, 432,
  2012

\bibitem[{{Thomas} {et~al.}(2005){Thomas}, {Maraston}, {Bender}, \& {Mendes de
  Oliveira}}]{thomas05}
{Thomas}, D., {Maraston}, C., {Bender}, R., \& {Mendes de Oliveira}, C. 2005,
  \apj, 621, 673

\bibitem[{{Toft} {et~al.}(2007){Toft}, {van Dokkum}, {Franx}, {Labbe},
  {F{\"o}rster Schreiber}, {Wuyts}, {Webb}, {Rudnick}, {Zirm}, {Kriek}, {van
  der Werf}, {Blakeslee}, {Illingworth}, {Rix}, {Papovich}, \&
  {Moorwood}}]{toft07}
{Toft}, S., {van Dokkum}, P., {Franx}, M., {et~al.} 2007, \apj, 671, 285

\bibitem[{{Tomczak} {et~al.}(2013){Tomczak}, {Quadri}, {Tran}, {Labbe},
  {Straatman}, {Papovich}, {Glazebrook}, {Allen}, {Kacprzak},
  {Kawinwanichakij}, {Kelson}, {McCarthy}, {Mehrtens}, {Monson}, {Persson},
  {Spitler}, {Tilvi}, \& {van Dokkum}}]{tomczak13}
{Tomczak}, A.~R., {Quadri}, R.~F., {Tran}, K.-V.~H., {et~al.} 2013, ArXiv
  e-prints

\bibitem[{{Toomre}(1964)}]{toomre}
{Toomre}, A. 1964, \apj, 139, 1217

\bibitem[{{Tremonti} {et~al.}(2007){Tremonti}, {Moustakas}, \&
  {Diamond-Stanic}}]{tremonti07}
{Tremonti}, C.~A., {Moustakas}, J., \& {Diamond-Stanic}, A.~M. 2007, \apjl,
  663, L77

\bibitem[{{Trujillo} {et~al.}(2007){Trujillo}, {Conselice}, {Bundy}, {Cooper},
  {Eisenhardt}, \& {Ellis}}]{trujillo07}
{Trujillo}, I., {Conselice}, C.~J., {Bundy}, K., {et~al.} 2007, \mnras, 382,
  109

\bibitem[{{Trump} {et~al.}(2013){Trump}, {Konidaris}, {Barro}, {Koo},
  {Kocevski}, {Juneau}, {Weiner}, {Faber}, {McLean}, {Yan},
  {P{\'e}rez-Gonz{\'a}lez}, \& {Villar}}]{trump13}
{Trump}, J.~R., {Konidaris}, N.~P., {Barro}, G., {et~al.} 2013, \apjl, 763, L6

\bibitem[{{Trump} {et~al.}(2011){Trump}, {Weiner}, {Scarlata}, {Kocevski},
  {Bell}, {McGrath}, {Koo}, {Faber}, {Laird}, {Mozena}, {Rangel}, {Yan},
  {Yesuf}, {Atek}, {Dickinson}, {Donley}, {Dunlop}, {Ferguson}, {Finkelstein},
  {Grogin}, {Hathi}, {Juneau}, {Kartaltepe}, {Koekemoer}, {Nandra}, {Newman},
  {Rodney}, {Straughn}, \& {Teplitz}}]{trump11b}
{Trump}, J.~R., {Weiner}, B.~J., {Scarlata}, C., {et~al.} 2011, \apj, 743, 144

\bibitem[{{van der Wel} {et~al.}(2012){van der Wel}, {Bell}, {H{\"a}ussler},
  {McGrath}, {Chang}, {Guo}, {McIntosh}, {Rix}, {Barden}, {Cheung}, {Faber},
  {Ferguson}, {Galametz}, {Grogin}, {Hartley}, {Kartaltepe}, {Kocevski},
  {Koekemoer}, {Lotz}, {Mozena}, {Peth}, \& {Peng}}]{vdw12}
{van der Wel}, A., {Bell}, E.~F., {H{\"a}ussler}, B., {et~al.} 2012, \apjs,
  203, 24

\bibitem[{{van der Wel} {et~al.}(2011){van der Wel}, {Rix}, {Wuyts}, {McGrath},
  {Koekemoer}, {Bell}, {Holden}, {Robaina}, \& {McIntosh}}]{vdw11a}
{van der Wel}, A., {Rix}, H.-W., {Wuyts}, S., {et~al.} 2011, \apj, 730, 38

\bibitem[{{van Dokkum} \& {Brammer}(2010)}]{dokkum10b}
{van Dokkum}, P.~G. \& {Brammer}, G. 2010, \apjl, 718, L73

\bibitem[{{van Dokkum} {et~al.}(2008){van Dokkum}, {Franx}, {Kriek}, {Holden},
  {Illingworth}, {Magee}, {Bouwens}, {Marchesini}, {Quadri}, {Rudnick},
  {Taylor}, \& {Toft}}]{dokkum08}
{van Dokkum}, P.~G., {Franx}, M., {Kriek}, M., {et~al.} 2008, \apjl, 677, L5

\bibitem[{{Wang} {et~al.}(2012){Wang}, {Huang}, {Faber}, {Fang}, {Wuyts},
  {Fazio}, {Yan}, {Dekel}, {Guo}, {Ferguson}, {Grogin}, {Lotz}, {Weiner},
  {McGrath}, {Kocevski}, {Hathi}, {Lucas}, {Koekemoer}, {Kong}, \&
  {Gu}}]{wang12}
{Wang}, T., {Huang}, J.-S., {Faber}, S.~M., {et~al.} 2012, \apj, 752, 134

\bibitem[{{Whitaker} {et~al.}(2012{\natexlab{a}}){Whitaker}, {Kriek}, {van
  Dokkum}, {Bezanson}, {Brammer}, {Franx}, \& {Labb{\'e}}}]{whitaker12}
{Whitaker}, K.~E., {Kriek}, M., {van Dokkum}, P.~G., {et~al.}
  2012{\natexlab{a}}, \apj, 745, 179

\bibitem[{{Whitaker} {et~al.}(2011){Whitaker}, {Labb{\'e}}, {van Dokkum},
  {Brammer}, {Kriek}, {Marchesini}, {Quadri}, {Franx}, {Muzzin}, {Williams},
  {Bezanson}, {Illingworth}, {Lee}, {Lundgren}, {Nelson}, {Rudnick}, {Tal}, \&
  {Wake}}]{whitaker11}
{Whitaker}, K.~E., {Labb{\'e}}, I., {van Dokkum}, P.~G., {et~al.} 2011, \apj,
  735, 86

\bibitem[{{Whitaker} {et~al.}(2012{\natexlab{b}}){Whitaker}, {van Dokkum},
  {Brammer}, \& {Franx}}]{whitaker12b}
{Whitaker}, K.~E., {van Dokkum}, P.~G., {Brammer}, G., \& {Franx}, M.
  2012{\natexlab{b}}, \apjl, 754, L29

\bibitem[{{Whitaker} {et~al.}(2013){Whitaker}, {van Dokkum}, {Brammer},
  {Momcheva}, {Skelton}, {Franx}, {Kriek}, {Labbe}, {Fumagalli}, {Lundgren},
  {Nelson}, {Patel}, \& {Rix}}]{whitaker13}
{Whitaker}, K.~E., {van Dokkum}, P.~G., {Brammer}, G., {et~al.} 2013, ArXiv
  e-prints

\bibitem[{{Wild} {et~al.}(2010){Wild}, {Heckman}, \& {Charlot}}]{wild10}
{Wild}, V., {Heckman}, T., \& {Charlot}, S. 2010, \mnras, 405, 933

\bibitem[{{Williams} {et~al.}(2013){Williams}, {Giavalisco}, {Cassata},
  {Tundo}, {Wiklind}, {Guo}, {Lee}, {Barro}, {Wuyts}, {Bell}, {Conselice},
  {Dekel}, {Faber}, {Ferguson}, {Grogin}, {Hathi}, {Huang}, {Kocevski},
  {Koekemoer}, {Koo}, {Ravindranath}, \& {Salimbeni}}]{williams13}
{Williams}, C.~C., {Giavalisco}, M., {Cassata}, P., {et~al.} 2013, ArXiv
  e-prints

\bibitem[{{Williams} {et~al.}(2011){Williams}, {Quadri}, \&
  {Franx}}]{williams11}
{Williams}, R.~J., {Quadri}, R.~F., \& {Franx}, M. 2011, \apjl, 738, L25

\bibitem[{{Williams} {et~al.}(2010){Williams}, {Quadri}, {Franx}, {van Dokkum},
  {Toft}, {Kriek}, \& {Labb{\'e}}}]{williams10}
{Williams}, R.~J., {Quadri}, R.~F., {Franx}, M., {et~al.} 2010, \apj, 713, 738

\bibitem[{{Windhorst} {et~al.}(2011){Windhorst}, {Cohen}, {Hathi}, {McCarthy},
  {Ryan}, {Yan}, {Baldry}, {Driver}, {Frogel}, {Hill}, {Kelvin}, {Koekemoer},
  {Mechtley}, {O'Connell}, {Robotham}, {Rutkowski}, {Seibert}, {Straughn},
  {Tuffs}, {Balick}, {Bond}, {Bushouse}, {Calzetti}, {Crockett}, {Disney},
  {Dopita}, {Hall}, {Holtzman}, {Kaviraj}, {Kimble}, {MacKenty}, {Mutchler},
  {Paresce}, {Saha}, {Silk}, {Trauger}, {Walker}, {Whitmore}, \&
  {Young}}]{windhorst11}
{Windhorst}, R.~A., {Cohen}, S.~H., {Hathi}, N.~P., {et~al.} 2011, \apjs, 193,
  27

\bibitem[{{Wuyts} {et~al.}(2010){Wuyts}, {Cox}, {Hayward}, {Franx},
  {Hernquist}, {Hopkins}, {Jonsson}, \& {van Dokkum}}]{wuyts10}
{Wuyts}, S., {Cox}, T.~J., {Hayward}, C.~C., {et~al.} 2010, \apj, 722, 1666

\bibitem[{{Wuyts} {et~al.}(2012){Wuyts}, {F{\"o}rster Schreiber}, {Genzel},
  {Guo}, {Barro}, {Bell}, {Dekel}, {Faber}, {Ferguson}, {Giavalisco}, {Grogin},
  {Hathi}, {Huang}, {Kocevski}, {Koekemoer}, {Koo}, {Lotz}, {Lutz}, {McGrath},
  {Newman}, {Rosario}, {Saintonge}, {Tacconi}, {Weiner}, \& {van der
  Wel}}]{wuyts12}
{Wuyts}, S., {F{\"o}rster Schreiber}, N.~M., {Genzel}, R., {et~al.} 2012, \apj,
  753, 114

\bibitem[{{Wuyts} {et~al.}(2011{\natexlab{a}}){Wuyts}, {F{\"o}rster Schreiber},
  {Lutz}, {Nordon}, {Berta}, {Altieri}, {Andreani}, {Aussel}, {Bongiovanni},
  {Cepa}, {Cimatti}, {Daddi}, {Elbaz}, {Genzel}, {Koekemoer}, {Magnelli},
  {Maiolino}, {McGrath}, {P{\'e}rez Garc{\'{\i}}a}, {Poglitsch}, {Popesso},
  {Pozzi}, {Sanchez-Portal}, {Sturm}, {Tacconi}, \& {Valtchanov}}]{wuyts11a}
{Wuyts}, S., {F{\"o}rster Schreiber}, N.~M., {Lutz}, D., {et~al.}
  2011{\natexlab{a}}, \apj, 738, 106

\bibitem[{{Wuyts} {et~al.}(2011{\natexlab{b}}){Wuyts}, {F{\"o}rster Schreiber},
  {van der Wel}, {Magnelli}, {Guo}, {Genzel}, {Lutz}, {Aussel}, {Barro},
  {Berta}, {Cava}, {Graci{\'a}-Carpio}, {Hathi}, {Huang}, {Kocevski},
  {Koekemoer}, {Lee}, {Le Floc'h}, {McGrath}, {Nordon}, {Popesso}, {Pozzi},
  {Riguccini}, {Rodighiero}, {Saintonge}, \& {Tacconi}}]{wuyts11b}
{Wuyts}, S., {F{\"o}rster Schreiber}, N.~M., {van der Wel}, A., {et~al.}
  2011{\natexlab{b}}, \apj, 742, 96

\bibitem[{{Wuyts} {et~al.}(2007){Wuyts}, {Labb{\'e}}, {Franx}, {Rudnick}, {van
  Dokkum}, {Fazio}, {F{\"o}rster Schreiber}, {Huang}, {Moorwood}, {Rix},
  {R{\"o}ttgering}, \& {van der Werf}}]{wuyts07}
{Wuyts}, S., {Labb{\'e}}, I., {Franx}, M., {et~al.} 2007, \apj, 655, 51

\bibitem[{{Xue} {et~al.}(2011){Xue}, {Luo}, {Brandt}, {Bauer}, {Lehmer},
  {Broos}, {Schneider}, {Alexander}, {Brusa}, {Comastri}, {Fabian}, {Gilli},
  {Hasinger}, {Hornschemeier}, {Koekemoer}, {Liu}, {Mainieri}, {Paolillo},
  {Rafferty}, {Rosati}, {Shemmer}, {Silverman}, {Smail}, {Tozzi}, \&
  {Vignali}}]{chandra4m}
{Xue}, Y.~Q., {Luo}, B., {Brandt}, W.~N., {et~al.} 2011, \apjs, 195, 10

\bibitem[{{Zahid} {et~al.}(2013){Zahid}, {Torrey}, {Kudritzki}, {Kewley},
  {Dave}, \& {Geller}}]{zahid13}
{Zahid}, H.~J., {Torrey}, P., {Kudritzki}, R., {et~al.} 2013, ArXiv e-prints

\end{thebibliography}

\end{document}